\documentclass{jfm}

\usepackage{graphicx}
\usepackage{newtxtext}
\usepackage{newtxmath}
\usepackage{natbib}
\usepackage{hyperref}
\usepackage{wrapfig}

\usepackage[lined,boxed,linesnumbered,ruled]{algorithm2e}

\hypersetup{
    colorlinks = true,
    urlcolor   = blue,
    citecolor  = black,
}

\newcommand{\RomanNumeralCaps}[1]
\linenumbers

\title{Dynamics-augmented cluster-based network model}

\author{
Chang Hou \aff{1},
Nan Deng \aff{1} \corresp{\email{dengnan@hit.edu.cn}}
\and 
Bernd R. Noack \aff{1,2} \corresp{\email{bernd.noack@hit.edu.cn}}
}

\affiliation{
\aff{1}School of Mechanical Engineering and Automation, Harbin Institute of Technology, Shenzhen
518055, People's Republic of China
\aff{2}Guangdong Provincial Key Laboratory of Intelligent Morphing Mechanisms and Adaptive Robotics,
Harbin Institute of Technology, Shenzhen
518055, People's Republic of China}
\begin{document}
\maketitle

\begin{abstract}

\begin{wrapfigure}{r}{0.4\textwidth}
\centering
\includegraphics[width=5.2cm]{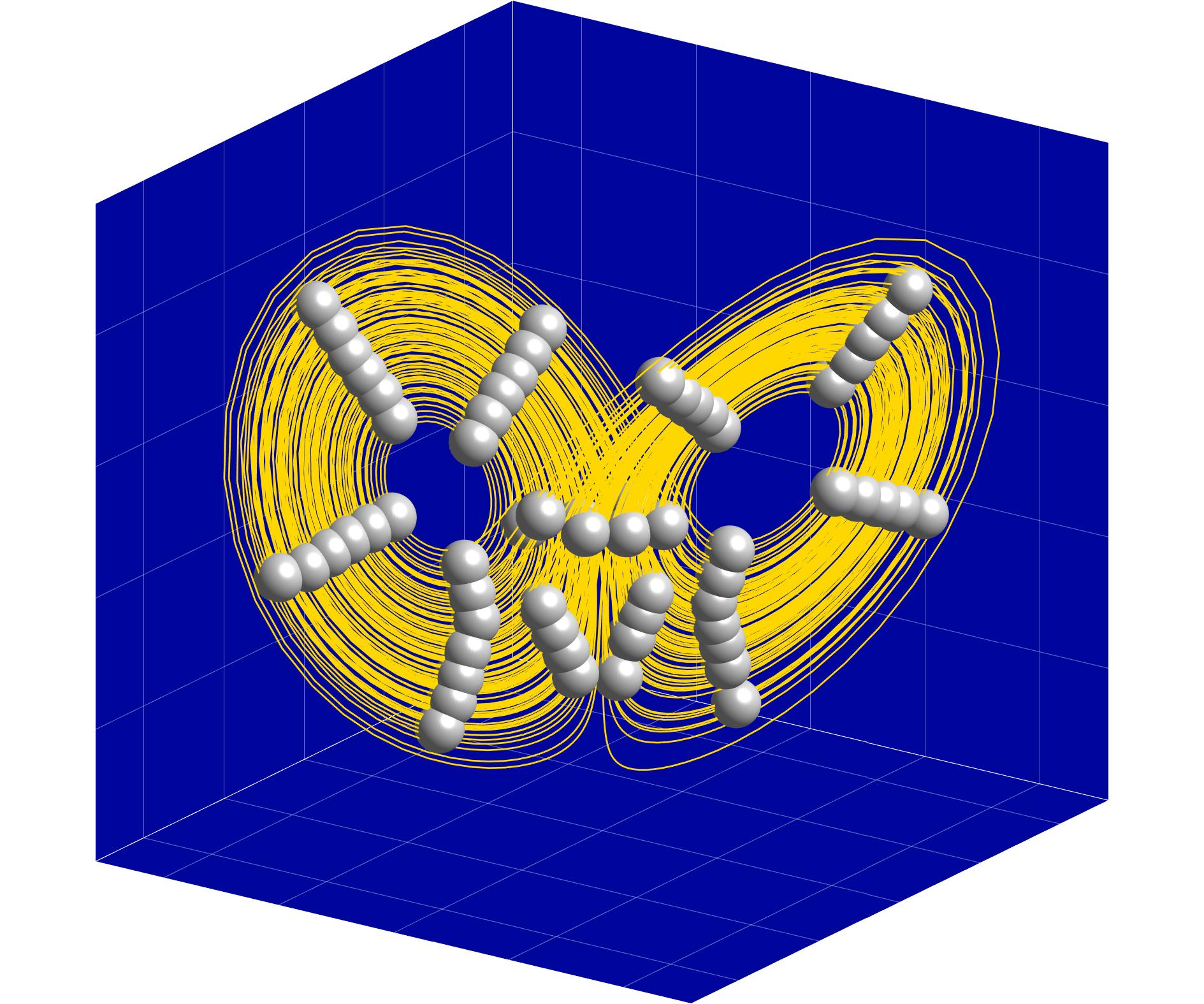}
\end{wrapfigure}

In this study, we propose a novel data-driven reduced-order model for complex dynamics, including nonlinear, multi-attractor, multi-frequency, and multiscale behaviours.
The starting point is a fully automatable cluster-based network model (CNM) (Li \emph{et al.} \emph{J. Fluid Mech.} vol. 906, 2021, A21)
which kinematically coarse-grains the state with clusters and dynamically predicts the transitions in a network model.
In the proposed dynamics-augmented CNM (dCNM), the prediction error is reduced with trajectory-based clustering using the same number of centroids.
The dCNM is first exemplified for the Lorenz system and then demonstrated for the three-dimensional sphere wake featuring periodic, quasi-periodic and chaotic flow regimes.
For both plants, the dCNM significantly outperforms the CNM in resolving the multi-frequency and multiscale dynamics.
This increased prediction accuracy is obtained by stratification of the state space aligned with the direction of the trajectories.
Thus, the dCNM has numerous potential applications to a large spectrum of shear flows, even for complex dynamics.
\end{abstract}

\begin{keywords}
Wakes/Jets: Wakes,
Nonlinear dynamic systems: Low-Dimensional Models
% Authors should not enter keywords on the manuscript, as these must be chosen by the author during the online submission process and will then be added during the typesetting process (see \href{https://www.cambridge.org/core/journals/journal-of-fluid-mechanics/information/list-of-keywords}{Keyword PDF} for the full list).  Other classifications will be added at the same time.
\end{keywords}

% {\bf MSC Codes }  {\it(Optional)} Please enter your MSC Codes here

%%%%%%%%%%%%%%%%%%%%%%%%%%%%%%%%%%%%%%%%%%%%%%%%%%%%%%%%%%%%%%%
\section{Introduction}
\label{sec:headings}
Advancements in computational capabilities and flow measurement technologies are producing a rapidly increasing amount of high-fidelity flow data.
The coherent spatio-temporal structures of the flow data enable data-driven reduced-order models (ROMs).
In terms of kinematics, ROMs furnish simplified descriptions that enrich our understanding of fundamental flow processes \citep{holmes2012turbulence}, facilitated by increasingly powerful machine learning methods \citep{brunton2020machine}.
ROMs may also allow the prediction of future states with acceptable accuracy. %, as they capture the essential dynamics from the reduced-order representations.
In the context of flow control, ROMs are serving as efficient tools for designing and testing control strategies, replacing costly high-fidelity simulations with an acceptable trade-off in accuracy \citep{bergmann2008optimal}.

% First-principle-based ROMs, Galerkin & the mean-field model
First-principle-based ROMs have historically been the foundation of the ROM community, as only a limited number of large data sets were available at that time.
The Galerkin framework is one of the most classic methods in this category.
By projecting the Navier-Stokes equations onto a low-dimensional subspace, the Galerkin model elegantly describes the original dynamics, exhibiting self-amplified amplitude-limited dynamics \citep{stuart1971nonlinear,busse1991numerical,noack1994global}.
\citet{landau1944problem} and \citet{stuart1958non} pioneered the mean-field model, a major progress in first-principle-based ROMs that provides insight into flow instabilities and bifurcation theory.
For instance, in the case of a supercritical Hopf bifurcation, mean-field models have been applied to the vortex shedding behind a cylinder \citep{strykowski1990formation,schumm1994self,noack2003hierarchy} and high Reynolds number turbulent wake flow \citep{bourgeois2013generalized}.
For more complex flows undergoing successive bifurcations, including both Pitchfork and Hopf bifurcations, weakly nonlinear mean-field analysis is applied to the wake of axisymmetric bodies \citep{fabre2008bifurcations}, the wake of a disk \citep{meliga2009global} and the fluidic pinball \citep{deng2020low}.
Furthermore, in the field of resolvent analysis, 
the mean-field theory also contributes by decomposing the system into time-resolved linear dynamics and a feedback term involving quadratic nonlinearity \citep{mckeon2004further, gomez2016reduced, rigas2017one}.

%% Data-driven ROMs
In contrast to a first principle ROM,
a data-driven version is based on a low-dimensional representation of flow snapshots.
Proper orthogonal decomposition (POD) is a commonly used example.
POD begins with the eigenvalue or singular value decomposition of the correlation matrix, yielding a low-dimensional subspace comprising leading orthogonal eigenvectors.
This subspace provides an ``optimal" Galerkin expansion with minimal average residual in the energy norm.
Since \citet{aubry1988dynamics} introduced the groundbreaking POD-Galerkin model for unforced turbulent boundary layers, numerous POD models have emerged for various configurations.
Examples include POD models for channel flow \citep{podvin1998low,podvin2009proper}, the wake of a two-dimensional square cylinder \citet{bergmann2009enablers}, laminar and turbulent vortex shedding \citep{iollo2000stability}, and flow past a circular cylinder with dynamic subgrid-scale model and variational multiscale model \citet{iollo2000stability}. 
There are also various variations of the POD model, e.g.\ integrating the actuation terms into the projection system for control design \citep{bergmann2008optimal,luchtenburg2009generalized} and balanced POD \citep{rowley2005model}, which is derived from a POD approximation to the product of controllability and observability Gramians to obtain an approximately balanced truncation \citep{moore1981principal}.
%% Automated data-driven ROMs
Increasingly powerful machine learning methods can make data-driven ROMs more automated. 
Examples include
the sparse identification of nonlinear dynamics (SINDy) aim at human interpretable models \citep{brunton2016discovering}, 
ROMs with artificial neural networks \citep{san2018extreme,san2019artificial,zhu2019machine,KouJQ2021pas}, 
turbulence modelling and flow estimation with multi-input multi-output by deep neural networks \citep{kutz2017deep, li2022Machine-learned},
and manifold learning methods \citep{farzamnik2023snapshots}.

%% CROMs
In this work, we focus on automated data-driven modelling.
The starting point is cluster-based ROMs (CROMs), pioneered by \citet{burkardt2006centroidal}.
Clustering is an unsupervised classification of patterns into groups commonly used in data science \citep{jain1988algorithms, jain1999data, jain2010data}, 
it is popular in data mining, document retrieval, image segmentation, and feature detection \citep{kim2022cluster}.
The foundation of the CROM lies in the cluster-based Markov model (CMM)
proposed by \citet{kaiser2014cluster}, which combines a cluster analysis of an ensemble of snapshots and a Markov model for transitions between different flow states reduced by clustering.
The CMM has provided a valuable physical understanding of the mixing layer, Ahmed body wakes \citep{kaiser2014cluster}, combustion-related mixing \citep{cao2014cluster}, and supersonic mixing layer \citep{li2020cluster}.
\citet{nair2019cluster} applied the cluster-based model to feedback control for drag reduction and first introduced a directed network for dynamical modelling.
Building on this concept, \citet{fernex2021cluster} and \citet{li2021cluster} further proposed the cluster-based network model (CNM) with improved long-timescale resolution.
Instead of the “stroboscopic” view of the CMM, the CNM focuses on non-trivial transitions.
The dynamics are restricted to a simple network model between the cluster centroids, like a deterministic-stochastic flight schedule which allows only a few possible flights with corresponding probabilities and flight times consistent with the data set.
Networks of complex dynamic systems have gained great interest, forming an increasingly important interdisciplinary field known as network science \citep{watts1998collective,albert2002statistical,borner2007network,barabasi2013network}.
Network-based approaches are often used in fluid flows \citep{nair2015network,hadjighasem2016spectral,taira2016network,yeh2021network,taira2022network}, in conjunction with clustering analysis \citep{bollt2001combinatorial,schlueter2017coherent,murayama2018characterization,krueger2019quantitative}.
The critical structures that modify the dynamical system can be identified by the intra- and inter-cluster interactions using community detection \citep{meena2018network,meena2021identifying}.

\begin{figure}
    \centering
    \includegraphics[width=11cm]{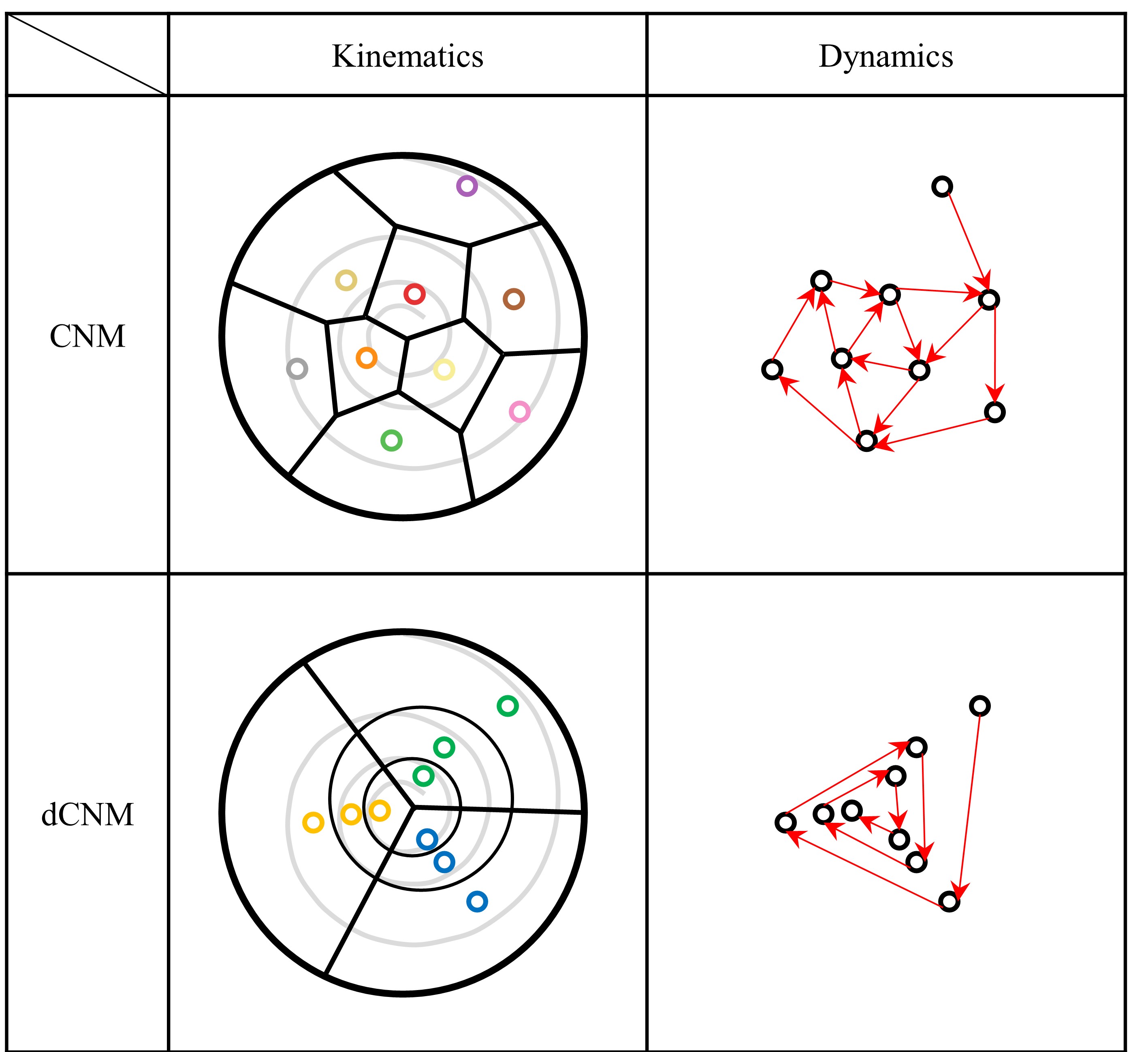}
    \caption{Principle sketches: The CNM and the dCNM are illustrated using an inward spiral trajectory in a two-dimensional state space with the same number of centroids.
    The thick solid lines denote cluster divisions, and the thin solid lines represent sub-cluster divisions. 
    The centroids are represented by coloured dots, and their colours represent their cluster affiliation.
    The CNM centroids are derived from snapshot averages within each cluster and show uniform geometric coverage, whereas the dCNM centroids incorporate dynamic features and exhibit a weighted distribution. 
    Consequently, dCNM accurately reconstructs the cycle-to-cycle variations and also ensures precise transition sequencing.
    %Principle sketches: The CNM versus the dCNM, exemplified by an inward spiralling trajectory in a two-dimensional state space with the same number of centroids. 
    %In the comparison of kinematics, the thick solid lines represent the cluster divisions, and the thin solid lines represent the sub-%cluster divisions.
    %The centroids are represented by coloured dots, with their colours representing their cluster affiliation.
    %In the comparison of dynamics, the red solid lines with an arrowhead represent the transition between centroids.
    %The CNM centroids are derived from the average of snapshots in each cluster, with good geometric coverage.
    %While dCNM centroids are assigned with the incorporation of dynamics, therefore exhibiting a weighted distribution.
    %As a consequence, the CNM only reconstructs coarse large-scale dynamics, which occasionally leads to misestimations. 
    %Conversely, the dCNM can accurately reconstruct the cycle-to-cycle variations and meanwhile guarantee a precise sequence.
    %With multi-stage clustering, the dCNM is less sensitive to the number of clusters; thus, a smaller number of clusters can be acceptable.
    }
\label{Method}
\end{figure}

%% Weakness and solutions 
CROMs are fully automated, robust, and physically interpretable, while the model accuracy is strongly related to the clustering process.
The state space is equivalently discretised in the above-mentioned CROMs, leading to a lack of dynamic coverage. 
For example, the CNM struggles to capture multiscale behaviours such as the oscillations near attractors and the amplitude variations between trajectories.
To address this issue, an effective solution is to employ dynamics-augmented clustering to determine the centroid distribution.
Inspired by the hierarchical clustering \citep{deng2022cluster} and the network sparsification \citep{nair2015network}, we propose a dynamics-augmented cluster-based network model (dCNM) with an improved resolution of complex dynamics.
In this case, the time-resolved dynamics are reflected by the evolution of trajectory segments after the state space is clustered.
These segments are automatically identified from cluster transitions and are represented by centroids obtained through segment averaging. 
A second-stage clustering further refines the centroids, eliminating the network redundancy and also deepening the comprehension of underlying physical mechanisms.
The proposed dCNM can systematically identify complex dynamics involved in the case of multi-attractor, multi-frequency, and multiscale dynamic systems.
Figure~\ref{Method} provides a comparative illustration of CNM and dCNM in terms of kinematics and dynamics, exemplified by an inward spiral trajectory in a two-dimensional state space.

%% Configuration
The dCNM is first applied to the \citet{lorenz1963deterministic} system as an illustrative example. 
The Lorenz attractor is notable for the ``butterfly effect'', showcasing the chaotic dynamics governed by only three ordinary differential equations.
%The dCNM is initially applied to the Lorenz system \citep{lorenz1963deterministic} as an illustrative example and subsequently demonstrated on the quasi-periodic and chaotic wake of a three-dimensional sphere.
%The Lorenz system, governed by only three ordinary differential equations, is notable for the ``butterfly effect'', showcasing chaotic dynamics on two attractors.
%
Subsequently, we demonstrate the dCNM on the sphere wake of the periodic, quasi-periodic, and chaotic flow regimes.
The sphere wake is a well-investigated benchmark configuration, serving as a prototype flow of bluff body wakes commonly encountered in many modern applications, for instance, the design of drones and air taxis.
Despite the simple geometry, the sphere wake can experience a series of bifurcations with increasing Reynolds number.
Along the route to turbulence, the flow system exhibits steady, periodic, quasi-periodic, and chaotic flow regimes.
The transient and post-transient flow dynamics, characterised by multi-frequency and multiscale behaviours, provide a challenging testing ground for reduced-order modelling.
%The quasi-periodic and chaotic scenario, characterised by multi-frequency and multiscale behaviours, provides a challenging testing ground for reduced-order modelling.

%%%%%%%%%%%%%%%%%%%%%%%%%%%%%%
This manuscript is organised as follows:
In \S~\ref{sec:methods}, the clustering algorithm and the different perspectives on the dCNM strategy are described. 
In \S~\ref{sec:Lorenz}, the dCNM is illustrated on the Lorenz system, and in \S~\ref{sec:sphere}, it is demonstrated on the sphere wake of three flow regimes: the periodic flow, the quasi-periodic flow and the chaotic flow.
%In \S~\ref{sec:Lorenz}, the dCNM is exemplified on the Lorenz system, and in \S~\ref{sec:sphere} it is demonstrated on the sphere wake of quasi-periodic and chaotic flow regimes.
In \S~\ref{sec:conclusion}, the main findings and improvements are summarised, and future directions are suggested.

%%%%%%%%%%%%%%%%%%%%%%%%%%%%%%%%%%%%%%%%%%%%%%%%%%%%%%%%%%%%%%%
\section{Dynamics-augmented cluster-based network model}
\label{sec:methods}
In this section, we detail the process of the dynamics-augmented cluster-based network model.
In \S~\ref{sec2.1}, the $k$-means++ clustering algorithm and its demonstration on the state space are introduced.
The second-stage clustering on the trajectory segments is further discussed in \S~\ref{sec2.2}.
In \S~\ref{sec2.3}, the transition characteristics are described, 
and in \S~\ref{sec2.4}, different criteria are introduced to evaluate the performance of the proposed model.
The variables used in this section are listed in table~\ref{tab:variables}.
\begin{table}
\centering
\def~{\hphantom{0}}
  \begin{tabular}{ll}
        Variables   &   Description\\[4pt]
        $\boldsymbol{u}^m$    &   Time-resolved snapshots\\[4pt]
        $M$   &   Number of snapshots\\[4pt]  
        \multicolumn{2}{l}{ ----------------- Clustering the state space ----------------- }\\[4pt]  
        $K$   &   Number of clusters\\[4pt] 
        $\mathcal{C}_{k}$, $\mathcal{C}_{i}$   &   Clusters obtained by the state space clustering\\[4pt]
        $\chi_{k}^{m}$   &   Characteristic function of the state space clustering\\[4pt]
        $M_{k}$   &   Number of snapshots in cluster $\mathcal{C}_{k}$\\[4pt]
        $\chi_{ik}^m$   &   Characteristic function of transition from $\mathcal{C}_{k}$ to $\mathcal{C}_{i}$\\[4pt]
        $\boldsymbol{c}_{k}$   &   Centroids of clusters \\[4pt]
        $n_{ik}$   &   Number of transitions from $\mathcal{C}_{k}$ to $\mathcal{C}_{i}$\\[4pt]
        $n_{k}$   &   Total number of transitions from $\mathcal{C}_{k}$\\[4pt]
        $n_{\mathrm{traj}}$   &   Total number of transitions of the data set\\[4pt]
        $Q_{ik}$   &   Cluster transition probability from $\mathcal{C}_{k}$ to $\mathcal{C}_{i}$\\[4pt]
        $T_{ik}$   &   Cluster transition time from $\mathcal{C}_{k}$ to $\mathcal{C}_{i}$\\[4pt]   
        $\mathbf{Q}$   &   Cluster transition probability matrix\\[4pt]
        $\mathbf{T}$   &   Cluster transition time matrix\\[4pt]
        $\boldsymbol{R}^u$   &   Cluster deviation on snapshots\\[4pt]
        \multicolumn{2}{l}{ ----------------- Clustering the trajectory segments ----------------- }\\[4pt]  
        $\mathcal{T}_{(kl)}$   &   The $l$-th trajectory segment in $\mathcal{C}_{k}$\\[4pt]
        $\chi_{(kl)}^{m}$   &   Characteristic function of the second-stage clustering\\[4pt]
        $M_{(kl)}$   &   Number of snapshots in trajectory segment $\mathcal{T}_{(kl)}$\\[4pt]
        $\boldsymbol{c}_{(kl)}$, $\boldsymbol{c}_{(ij)}$   &   Centroids of trajectory segments \\[4pt]
        $\boldsymbol{L} = [L_{1}, \ldots, L_{K}]^{\intercal}$   &   Number of sub-clusters for the second-stage clustering\\[4pt]
        $n_{(ij) (kl)}$   &   Number of transitions from $\boldsymbol{c}_{(kl)}$ only to $\boldsymbol{c}_{(ij)}$\\[4pt]
        $Q_{(ij) (kl)}$   &   Centroid transition probability from $\boldsymbol{c}_{(kl)}$ to $\boldsymbol{c}_{(ij)}$\\[4pt]
        $\mathbf{Q}_{ik}$   &   Centroid transition probability matrix\\[4pt]
        $\mathcal{Q}_k$   &   Centroid transition probability tensor\\[4pt]
        $\boldsymbol{R^{\mathcal{T}}}$   &   Cluster deviation on trajectory segments
  \end{tabular}
\caption{Table of variables.
Subscripts $k$ and $i$ are related to the level of clusters from the state space clustering, and subscripts $l$ and $j$ are related to the level of trajectory segments.}
\label{tab:variables}
\end{table}
%

%%%%%%%%%%%%%%%%%%%%%%%%%%%%%%%%%%%%%%%%%%
\subsection{Clustering the state space}
\label{sec2.1}
The dynamics-augmented clustering procedure is divided into two steps. 
Initially, the state space is clustered, yielding coarse-grained state transition dynamics with trajectory segments composed of time-continuous snapshots within each cluster.
Subsequently, we cluster these trajectory segments, utilising centroids derived from the average of each segment. 
This step optimises the centroid distribution and eliminates the redundancy of the trajectory segments.

The first-stage clustering discretises the high-dimensional state space by grouping the snapshots.
We first define a Hilbert space $\mathscr{L}^{2}(\Omega)$, in which the inner product of vector fields in the domain $\Omega$ is given by a square-integrable function:
\begin{equation}
(\boldsymbol{u}, \boldsymbol{v})_{\Omega} = \int_{\Omega}  \mathrm{d} \boldsymbol{x} \, \boldsymbol{u} \cdot \boldsymbol{v},
\label{eq1}
\end{equation}
where $\boldsymbol{u}$ and $\boldsymbol{v}$ represent snapshots of this vector field, also known as observations in the machine learning context. 
The corresponding norm is defined as:
\begin{equation}
\|\boldsymbol{u}\|_{\Omega}:=\sqrt{(\boldsymbol{u}, \boldsymbol{u})_{\Omega}}.
\label{eq2}
\end{equation}
The distance $D$ between two snapshots can be calculated as follows:
\begin{equation}
D(\boldsymbol{u},\boldsymbol{v}) = \| \boldsymbol{u} - \boldsymbol{v} \|_{\Omega}.
\label{eq3}
\end{equation}

The unsupervised $k$-means++ algorithm \citep{macqueen1967classification, lloyd1982least, arthur2007k} is used for clustering.
It operates automatically, devoid of assumptions or data categorisation.
Serving as the foundation of cluster analysis, this algorithm partitions a set of $M$ time-resolved snapshots $\boldsymbol{u}^m$, where $m=1 \dots M$, into $K$ clusters $\mathcal{C}_k$, where $k = 1 \dots K$. 
Each cluster corresponds to a centroidal Voronoi cell, with the centroid defined as the average of the snapshots within the same cluster. 
The algorithm comprises the following steps:
\begin{enumerate}
\item Initialisation: $K$ centroids $\boldsymbol{c}_{k}$, where $k = 1 \dots K$, are randomly selected.
In contrast to the $k$-means algorithm, $k$-means++ optimises the placement of these centroids to prevent sensitivity to initial conditions.
\item Assignment: Each snapshot $\boldsymbol{u}^{m}$ is allocated to the nearest centroid by
$\underset{k}{\rm arg \min} \ D(\boldsymbol{u}^{m},\boldsymbol{c}_{k})$. 
The characteristic function is used to mark their affiliation, and it is defined as follows:
\begin{equation}
\chi_{k}^{m}:=\left\{\begin{array}{ll}
1, & \text { if } \boldsymbol{u}^{m} \in \mathcal{C}_{k} \\
0, & \text { otherwise }
\end{array}\right.
\label{eq4}
\end{equation}
\item Update: Each centroid is recalculated by averaging all the snapshots belonging to the corresponding cluster as follows:
\begin{equation}
\boldsymbol{c}_{k}=\frac{1}{M_{k}} \sum_{\boldsymbol{u}^{m} \in \mathcal{C}_{k}} \boldsymbol{u}^{m}=\frac{1}{M_{k}} \sum_{m=1}^{M} \chi_{k}^{m} \boldsymbol{u}^{m} ,
\label{eq5}
\end{equation}
where 
\begin{equation}
M_{k}=\sum_{m=1}^{M} \chi_{k}^{m}.
\label{eq6}
\end{equation}
\item Iteration: The Assignment and Update steps are repeated until convergence is reached. 
Convergence means that the centroids do not move or stabilise below a certain threshold. 
The algorithm minimises the intra-cluster variance and maximises the inter-cluster variance. 
The intra-cluster variance is computed as follows:
\begin{equation}
J\left(\boldsymbol{c}_{1}, \ldots, \boldsymbol{c}_{K}\right) = \sum_{k=1}^{K} \sum_{m=1}^{M} \chi_{k}^{m}
\left\| \boldsymbol{u}^{m}-\boldsymbol{c}_{k} \right\|_{\Omega}^{2} .
\label{eq7}
\end{equation}
Each iteration reduces the value of the criterion $J$ until convergence is reached.
\end {enumerate}

The cluster probability distribution $\boldsymbol{P} = [P_1, \dots,  P_K]$ is determined by $P_k = M_{k}/M$ for each cluster $\mathcal{C}_k$, and satisfies the normalisation condition $\sum_{k=1}^{K}P_k = 1$.

The geometric properties of the clusters are quantified for further analysis.
The cluster standard deviation on the snapshots $R_{k}^{u}$ measures the cluster size, following \citet{kaiser2014cluster}, as:
\begin{equation}
R_{k}^{u} = \sqrt{ \frac{1}{M_{k}} \sum_{m=1}^{M} \chi_{k}^{m}
\left\| \boldsymbol{u}^{m}-\boldsymbol{c}_{k} \right\|_{\Omega}^{2}} .
\label{eq8}
\end{equation}

The time-resolved snapshots should be equidistantly sampled and cover a statistically representative time window of the coherent structure evolution. 
As a rule of thumb, at least ten periods of the dominant frequency are needed to obtain reasonably accurate statistical moments and at least $K$ snapshots per characteristic period to capture an accurate temporal evolution.

%%%%%%%%%%%%%%%%%%%%%%%%%%%%%%%%%%%%%%%%%%
\subsection{Clustering the trajectory segments}
\label{sec2.2}
After the state space is discretized, the trajectory is also divided into segments. 
We use the cluster transition information to identify the trajectory segments that pass through a cluster.

Based on the temporal information from the given data set, the nonlinear dynamics between snapshots are modelled as linear transitions between clusters, known as the classic CNM \citep{fernex2021cluster, li2021cluster}. 
We infer the probability of cluster transition from the data as follows:
\begin{equation}
Q_{ik}=\frac{n_{ik}}{n_{k}}, \quad i, k=1, \ldots, K,
\label{eq9}
\end{equation}
where $Q_{ik}$ is the direct cluster transition probability from cluster $\mathcal{C}_{k}$ to $\mathcal{C}_{i}$ and $n_{ik}$ is the number of transitions from $\mathcal{C}_{k}$ only to $\mathcal{C}_{i}$:
\begin{equation}
n_{ik}=\sum_{m=1}^{M} \chi_{ik}^m,
\label{eq10}
\end{equation}
where
\begin{equation}
\chi_{ik}^m=\left\{\begin{array}{ll}
1, & \text { if } \boldsymbol{u}^{m} \in \mathcal{C}_{k}\; \& \; \boldsymbol{u}^{m+1} \in \mathcal{C}_{i} \\
0, & \text { otherwise }
\end{array}\right.
\label{eq11}
\end{equation}
$n_{k}$ is the total number of transitions from $\mathcal{C}_{k}$ regardless of the destination cluster:
\begin{equation}
n_{k}=\sum_{i=1}^{K} n_{ik}, \quad i, k=1, \ldots, K.
\label{eq12}
\end{equation}

If $Q_{ik} \neq 0$, it can be inferred that in cluster $\mathcal{C}_{k}$ there exists at least one trajectory segment that is bound for cluster $\mathcal{C}_{i}$.
We assign distinct labels to each trajectory segment corresponding to all destination clusters, denoted as $\mathcal{T}_{(kl)}$, where $k$ and $l$ represent the $l$-th segment in $\mathcal{C}_{k}$.
Therefore the snapshots are marked according to their trajectory affiliations by a characteristic function:
\begin{equation}
\chi_{(kl)}^{m} = \left\{\begin{array}{ll}
1, & \text { if } \boldsymbol{u}^{m} \in \mathcal{T}_{(kl)} \\
0, & \text { otherwise }
\end{array}\right.
\label{eq13}
\end{equation}
where $k$ represents the cluster affiliation, and $l$ represents the trajectory segment affiliation.
The total number of trajectory segments in $\mathcal{C}_{k}$ equals $n_{k}$. 
Note that the final trajectory segment of the data set will not be considered as it will not lead to any destination cluster and is usually incomplete.
The total number of trajectory segments in the data set can be obtained by the sum of $n_{k}$ as follows:
\begin{equation}
n_{\mathrm{traj}} = \sum_{k=1}^{K} n_{k}.
\label{eq14}
\end{equation}

Analogous trajectory segments within the same cluster will be merged in the subsequent clustering stage. 
Operations on the trajectories can often be costly. 
Efficiency in clustering can be achieved by mapping the operations performed on trajectory segments to their corresponding averages, i.e., the trajectory segment centroids, given their topological relationship.
Additionally, the propagation of our model relies on centroids, rendering the trajectory information essentially unnecessary.
We define the centroids $\boldsymbol{c}_{(kl)}$ as the average of snapshots belonging to the same trajectory segment:
\begin{equation}
\boldsymbol{c}_{(kl)}=\frac{1}{M_{(kl)}} \sum_{\boldsymbol{u}^{m} \in \mathcal{T}_{(kl)}}\boldsymbol{u}^{m}=\frac{1}{M_{(kl)}} \sum_{m=1}^{M} \chi_{(kl)}^{m} \boldsymbol{u}^{m},
\label{eq15}
\end{equation}
where
\begin{equation}
M_{(kl)}= \sum_{m=1}^{M} \chi_{(kl)}^{m}.
\label{eq16}
\end{equation}

The subsequent question pertains to how to determine the number of sub-clusters.
The allocation of sub-clusters within each cluster can be automatically learnt from the data.
To maintain the spatial resolution, more sub-clusters should be assigned to clusters with a larger transverse size. 
%To achieve a better spatial resolution, more sub-clusters should be assigned to clusters with more significant trajectory dispersion. 
We first introduce a transverse cluster size vector $\boldsymbol{R}^{\mathcal{T}}$, which is defined by the standard deviation of the $n_{k}$ centroids $\boldsymbol{c}_{(kl)}$ with respect to the cluster centroid $\boldsymbol{c}_{k}$ as follows:
%We first introduce the standard deviation vector of the trajectory segments $\boldsymbol{R}^{\mathcal{T}}$
%
\begin{equation}
R_{{k}}^{\mathcal{T}} = \sqrt{\frac{1}{n_{k}} \sum_{l=1}^{n_{k}}\left\|\boldsymbol{c}_{(kl)}-\boldsymbol{c}_{k}\right\|_{\Omega}^{2}}.
\label{eq17}
\end{equation}
Next, we denote the number of sub-clusters as $L_{k}$ for clustering the centroids in cluster $\mathcal{C}_{k}$. 
A $K$-dimensional vector $\boldsymbol{L} = [L_{1}, \dots, L_{K}]^{\intercal}$ records the numbers of sub-clusters in each cluster, with $L_{k}$ determined by:
\begin{equation}
L_{k} = \min (\lfloor \hat{R}^{\mathcal{T}}_{k} n_{\mathrm{traj}} (1 - \beta) \rfloor +1, n_{k}).
\label{eq18}
\end{equation}
Here, the vector $\boldsymbol{R}^{\mathcal{T}}$ is normalised with the sum $\sum_{k=1}^{K} R_{{k}}^{\mathcal{T}}$, denoted as $\boldsymbol{\hat{R}}^{\mathcal{T}}$, which ensures a suitable distribution of sub-clusters for the ensemble of $n_{\mathrm{traj}}$ trajectories.
To increase the flexibility of the model, we introduce a sparsification controller $\beta \in [0, 1]$ in this clustering process.
For the extreme value of $\beta=1$, all the centroids are merged into one centroid, 
and the dCNM is identical to a classic CNM, with the maximum sparsification. 
For the other extreme $\beta=0$, the dCNM is minimally sparsified according to the transverse cluster size.
For periodic or quasi-periodic systems, the dCNM with a large $\beta$ can capture most of the dynamics, while for complex systems such as chaotic systems, a small $\beta$ may be needed.
In addition, the minimum function prevents the possibility that the left-hand side of the equation exceeds the number of centroids $n_{k}$ when $\beta$ is too small, causing the second-stage clustering to not be performed.
The choice of $\beta$ is discussed in Appendix~\ref{app:c}.

The refined centroids are obtained by averaging a series of centroids related to analogous trajectory segments.
The redundancy of the $n_{\mathrm{traj}}$ centroids is mitigated, and the corresponding transition network becomes sparse.
The $k$-means++ algorithm is also used in the second-stage clustering. 
It will iteratively update the centroids $\boldsymbol{c}_{(kl)}$ and the characteristic function $\chi_{(kl)}^{m}$ until convergence or the maximum number of iterations is reached. 
The overall clustering process of the dCNM is summarised in Algorithm~\ref{alg1}.
\begin{algorithm}[t]
\SetAlgoLined
    \KwIn{$\boldsymbol{u}^m$: Snapshots;\\
          $K$: Number of clusters;\\
          $\beta$: Sparsification index $(0 \le \beta \le 1)$;}% input
    \KwOut{$\boldsymbol{c}_{(kl)}$: Refined centroids;\\
           $R_{k}^{u}$, $R_{k}^{\mathcal{T}}$: Geometric properties;\\
           $\chi_{(kl)}^{m}$: Characteristic function}% output
    Apply $k$-means++ algorithm with $K$ clusters to $\boldsymbol{u}^m$\\
    Save the characteristic function as $\chi_{k}^{m}$\\
    \For{$k \leftarrow 1$ \KwTo $K$}{
        \For{$i \leftarrow 1$ \KwTo $K$}{
            Compute the transition probability $Q_{ik}$\\
            \If {$Q_{ik} \neq 0$}{
                Locate the time-continuous snapshots in cluster $\mathcal{C}_{k}$ on each trajectory segment to $\mathcal{C}_{i}$\\
                Save the characteristic function $\chi_{(kl)}^{m}$ accordingly\
            }
        }
        Compute and save the centroids $\boldsymbol{c}_{(kl)}$ by $\chi_{(kl)}^{m}$, compute the geometric properties $R_{k}^{u}$ and $R_{k}^{\mathcal{T}}$\
    }    
    Compute $\boldsymbol{L}$ by $R_{{k}}^{\mathcal{T}}$ and $\beta$\\
    \For{$k \leftarrow 1$ \KwTo $K$}{
        Locate the centroids $\boldsymbol{c}_{(kl)}$ in cluster $\mathcal{C}_{k}$\\
        Apply $k$-means++ algorithm with $L_{k}$ clusters directly to $\boldsymbol{c}_{(kl)}$\\
        Update the characteristic function $\chi_{(kl)}^{m}$ and the centroids $\boldsymbol{c}_{(kl)}$.\

    }
 \caption{Pseudocode for the dynamics-augmented clustering procedure}\label{alg1}
\end{algorithm}
%

%%%%%%%%%%%%%%%%%%%%%%%%%%%%%%%%%%%%%%%%%%
\subsection{Characterising the transition dynamics}
\label{sec2.3}
We use the centroids obtained from \S~\ref{sec2.2} as the nodes of the network and the linear transitions between these centroids as the edges of the network. 
First, we introduce two transition properties: the centroid transition probability $Q_{(ij) (kl)}$ and the transition time $T_{ik}$.

\begin{figure}
    \centering
    \includegraphics[width=5cm]{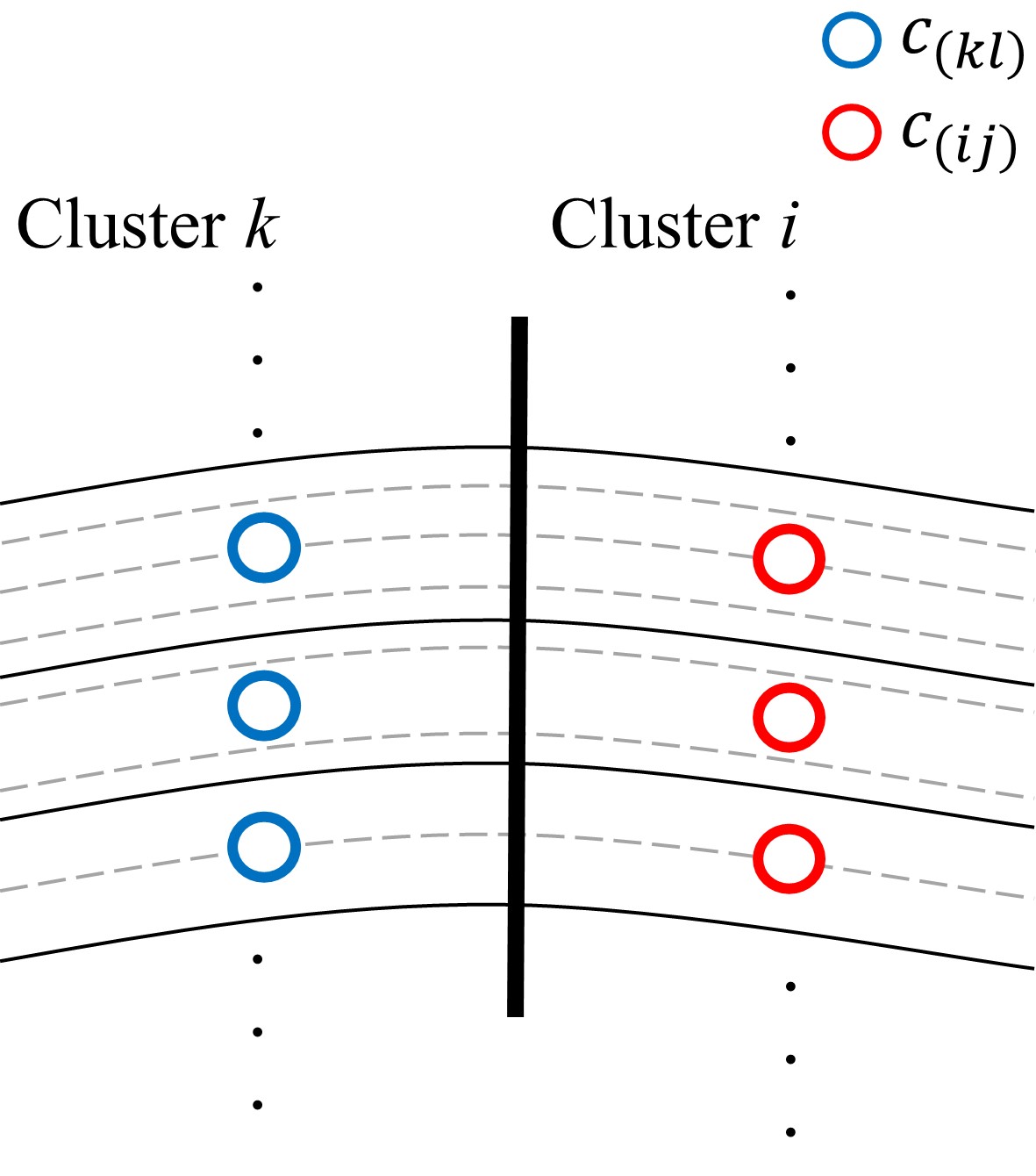}
    \caption{Illustration of the subscripts in the refined centroid transitions.
    After the state space is clustered, only one subscript is needed to distinguish the different clusters, such as $\mathcal{C}_{k}$ and $\mathcal{C}_{i}$. 
    After the trajectory segments are clustered, two subscripts are needed to represent the refined centroids, such as $\boldsymbol{c}_{(kl)}$ in $\mathcal{C}_{k}$ and $\boldsymbol{c}_{(ij)}$ in $\mathcal{C}_{i}$.}
\label{Centroid_transition}
\end{figure}

Figure~\ref{Centroid_transition} illustrates the definition of the subscripts in the centroid transition probability $Q_{(ij) (kl)}$, which can contain all possible transitions between the refined centroids of clusters $\mathcal{C}_{k}$ and $\mathcal{C}_{i}$.
Considering the transitions between these centroids, we define $Q_{(ij) (kl)}$ as:
\begin{equation}
    Q_{(ij) (kl)} = \frac{n_{(ij) (kl)}}{n_{k}}, \quad i, k =1, \ldots, K, \quad j = 1, \ldots, L_{i}, \quad l = 1, \ldots, L_{k},
\label{eq19}
\end{equation}
where $n_{(ij) (kl)}$ is the number of transitions from $\boldsymbol{c}_{(kl)}$ only to $\boldsymbol{c}_{(ij)}$. 
This definition differs from that of the CNM, which uses the cluster transition $Q_{ik}$ in~\ref{eq9} to define the probability. 
In fact, we can compute $Q_{ik}$ by summing up $Q_{(ij) (kl)}$ as follows:
\begin{equation}
    Q_{ik}=\sum_{j=1}^{L_{i}} \sum_{l=1}^{L_{k}} Q_{(ij) (kl)}.
\label{eq20}
\end{equation}

\begin{figure}
    \centerline{\includegraphics[width=7cm]{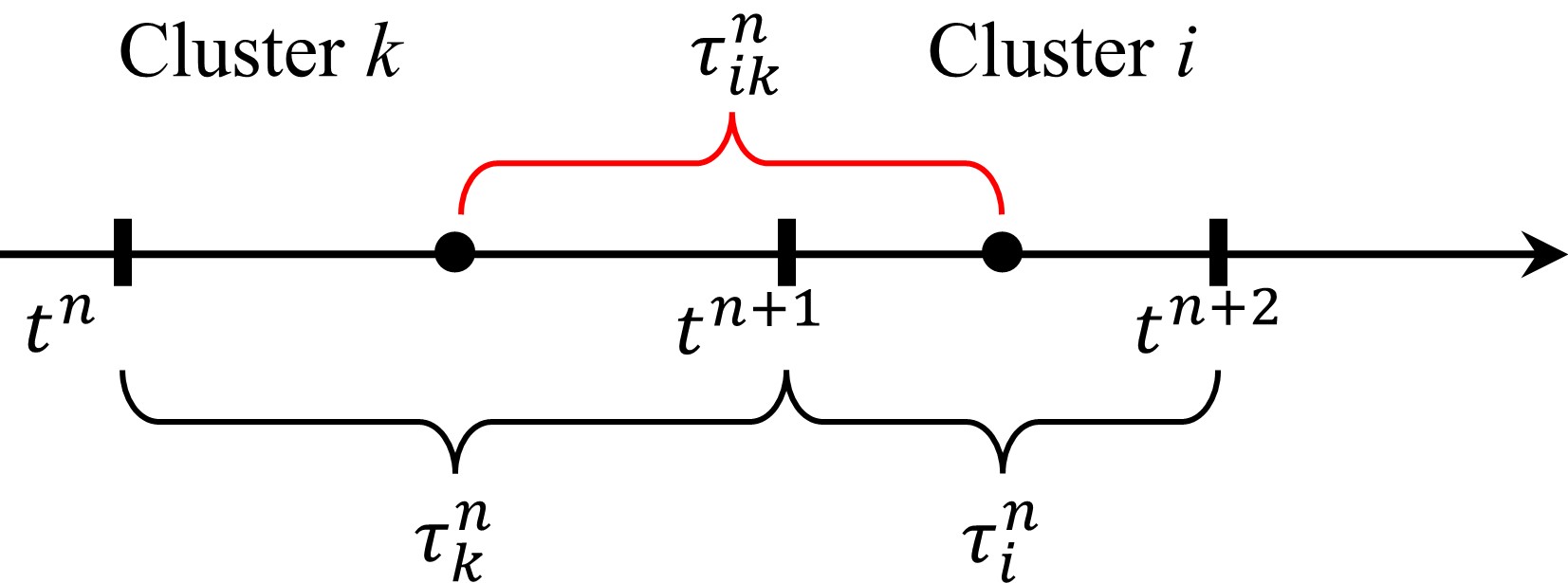}}
    \caption{Individual transition time $\tau^{n}_{ik}$ for the transition from cluster $\mathcal{C}_{k}$ to $\mathcal{C}_{i}$}
\label{TransitionTime}
\end{figure}

The definition of the transition time $T_{ik}$ is identical to the CNM, as shown in figure~\ref{TransitionTime}.
This property is not further investigated in the present work, as the transition time crossing the same clusters varies little in most dynamic systems.

Let $t^{n}$ be the instant when the first snapshot enters, and $t^{n+1}$ be the instant when the last snapshot leaves on one trajectory segment passing through cluster $\mathcal{C}_{k}$.
The residence time $\tau_{k}^{n}$ is the duration of staying in cluster $\mathcal{C}_{k}$ on this segment, which is given by:
\begin{equation}
    \tau_{k}^{n}=t^{n+1}-t^{n}.
\label{eq21}
\end{equation}
For an individual transition from $\mathcal{C}_{k}$ to $\mathcal{C}_{i}$, the transition time is defined as $\tau_{ik}^{n}$, which can be obtained by the average of the residence times from both clusters:
\begin{equation}
    \tau_{ik}^{n}=(\tau_{k}^n + \tau_{i}^n)/2.
\label{eq22}
\end{equation}
By averaging $\tau_{ik}^{n}$ for all the individual transitions from $\mathcal{C}_{k}$ to $\mathcal{C}_{i}$, the transition time can be expressed as follows:
\begin{equation}
    T_{ik}=\frac{\sum_{n=1}^{n_{ik}} \tau_{ik}^{n}}{n_{ik}}.
\label{eq23}
\end{equation}

The essential dynamics can also be summarised into single entities as in the CNM, since the cluster-level information is still retained in the current model.
For completeness, we introduce the cluster transition probability matrix $\mathbf{Q}$ and the cluster transition time matrix $\mathbf{T}$ as:
\begin{eqnarray}
\begin{array}{l}
    \mathbf{Q}=Q_{ik} \in \mathbb{R}^{K \times K}, \quad i, k=1, \ldots, K\\
    \mathbf{T}=T_{ik} \in \mathbb{R}^{K \times K}, \quad i, k=1, \ldots, K.
\end{array}
\label{eq24}
\end{eqnarray}
The cluster indices are reordered in both matrices to enhance readability.
$\mathcal{C}_1$ is the cluster with the highest distribution probability, 
$\mathcal{C}_2$ is the cluster with the highest transition probability leaving from $\mathcal{C}_1$, 
and $\mathcal{C}_3$ is the cluster with the highest transition probability leaving from $\mathcal{C}_2$, so on and so forth.
If the cluster with the highest probability is already assigned, we choose the cluster with the second highest probability. 
If all the clusters with nonzero transition probabilities are already assigned, we choose the next cluster with the highest distribution probability among the rest.

By analogy with $\mathbf{Q}$, the centroid transition probability $Q_{(ij) (kl)}$ for given affiliations of the departure cluster $k$ and destination cluster $i$ can form a centroid transition matrix $\mathbf{Q}_{ik}$ that captures all possible centroid dynamics between the two clusters:
\begin{equation}
\mathbf{Q}_{ik} = Q_{(ij) (kl)} \in \mathbb{R}^{L_{i} \times L_{k}}, \quad j = 1, \ldots, L_{i}, \quad l = 1, \ldots, L_{k}.
\label{eq25}
\end{equation}
Moreover, to summarise the centroid transition dynamics, the centroid transition probability $Q_{(ij) (kl)}$ for a given affiliation $k$ of only the departure cluster can form a centroid transition tensor $\mathcal{Q}_k$ that captures all the possible centroid dynamics from this cluster, as:
\begin{equation}
\mathcal{Q}_{k} = Q_{(ij) (kl)} \in \mathbb{R}^{K \times L_{i} \times L_{k}}, \quad i =1, \ldots, K, \quad j = 1, \ldots, L_{i}, \quad l = 1, \ldots, L_{k}.
\label{eq26}
\end{equation}

The dCNM propagates the state motion based on the centroids $\boldsymbol{c}_{(kl)}$ for the reconstruction. 
To determine the transition dynamics, we first use $\mathcal{Q}_{k}$ to find the centroid transitions from the initial centroid $\boldsymbol{c}_{(kl)}$ to the destination $\boldsymbol{c}_{(ij)}$. 
As the destination centroids are determined, the cluster-level dynamics are determined correspondingly.
Then, $\mathbf{T}$ is used to identify the related transition time. 

We assume a linear state propagation between the two centroids $\boldsymbol{c}_{(kl)}$ and $\boldsymbol{c}_{(ij)}$ obtained from the tensors, as follows:
\begin{equation}
\boldsymbol{u}^m(t) = \alpha_{ik}(t) \boldsymbol{c}_{(ij)} + \left[ 1-\alpha_{ik}(t) \right] \boldsymbol{c}_{(kl)}, \quad \alpha_{ik}=\frac{t-t_{k}}{T_{ik}}.
\label{eq27}
\end{equation}
Here $t_{k}$ is the time when the centroid $\boldsymbol{c}_{(kl)}$ is left. 
Note that we can use splines \citep{fernex2021cluster} or add the trajectory supporting points \citep{hou2022tCNM} to interpolate the motion between the centroids for smoother trajectories.

Intriguingly, we observe that the trajectory-based clustering of the dCNM enhances the resolution of the cluster transitions.
Now each centroid only has a limited number of destination centroids, often within the same cluster.
This minimises the likelihood of selecting the wrong destination cluster based solely on the cluster transition probability matrix, as is the case in classic CNM. 
Consequently, it becomes feasible to accurately resolve long-term cluster transitions without the need for historical information.
It can be argued that dCNM effectively constrains cluster transitions, leading to outcomes similar to those obtained with the higher-order CNM \citep{fernex2021cluster}. 
This improvement is attained by replacing higher-order indexing with higher-dimensionality dual indexing.
Specifically, the dual indexing also results in a substantial reduction in the model complexity.
While the complexity of the high-order CNM is defined as $K^{\tilde{L}}$, where $K$ is the number of clusters and $\tilde{L}$ is the order, the model complexity of the dCNM is expressed as $\sum_{{k}=1}^{K} {L}_{k}$, which is a significantly lower value, particularly when $\tilde{L}$ is relatively large.
In terms of computational efficiency, dCNM with $\beta = 0.80$ reduces the computational time by $40\%$ as compared to CNM with the same number of centroids. 
This improvement is primarily attributed to the hierarchical clustering.
The computational load of the first-stage clustering on the state space is reduced by a small number of clusters $K$.
The second-stage clustering on the trajectory segments accounts only for $20\%$ of the total computation time.

%%%%%%%%%%%%%%%%%%%%%%%%%%%%%%%%%%%%%%%%%%
\subsection{Validation}
\label{sec2.4}
The auto-correlation function and the representation error are used for validation.
We examine the prediction errors for cluster-based models considering both spatial and temporal perspectives. 
The spatial error arises from the inadequate representation by cluster centroids, as evidenced by the representation error and the auto-correlation function.
The temporal error arises due to the imprecise reconstruction of intricate snapshot transition dynamics. This can be observed directly through the temporal evolution of snapshot affiliations and, to some extent, through the auto-correlation function.

The auto-correlation function is a practical tool for evaluating ROMs, as it can statistically reflect the prediction errors.
Additionally, the auto-correlation function circumvents the problem of directly comparing two trajectories with finite prediction horizons, which may suffer from phase mismatch \citep{fernex2021cluster}. 
This is particularly relevant for chaotic dynamics, whereby minor differences in initial conditions can lead to divergent trajectories, making the direct comparison of time series meaningless.
The unbiased auto-correlation function of the state vector \citep{protas2015optimal} is given by:
\begin{equation}
R(\tau )=\frac{1}{T-\tau } \int_{0}^{T-\tau} (\boldsymbol{u}(x,t) , \boldsymbol{u}(x,t+\tau ))_\Omega\mathrm {d}t, \quad \tau\in \left [ 0,T \right ].
\label{eq31}
\end{equation}
In this study, $R(\tau)$ will be normalised by $R(0)$ \citep{deng2022cluster}.
This function can also infer the spectral behaviour by computing the fluctuation energy at the vanishing delay.

The representation error can be numerically computed as:
\begin{equation}
    E_{r}=\frac{1}{M} \sum_{m=1}^{M} D_{\mathcal{T}}^{m},
\label{eq29}
\end{equation}
where $D_{\mathcal{T}}^{m}$ is the minimum distance from the snapshot $\boldsymbol{u}^m$ to the states on the reconstructed trajectory $\mathcal{T}$:
\begin{equation}
    D_{\mathcal{T}}^{m}=\min _{\boldsymbol{u}^{n} \in \mathcal{T}}\left\|\boldsymbol{u}^{m}-\boldsymbol{u}^{n}\right\|_{\Omega}.
\label{eq30}
\end{equation}
%

%%%%%%%%%%%%%%%%%%%%%%%%%%%%%%%%%%%%%%%%%%%%%%%%%%%%%%%%%%%%%%%
\section{Lorenz system as an illustrative example}
\label{sec:Lorenz}

In this section, we apply the dCNM to the \citet{lorenz1963deterministic} system to illustrate its superior spatial resolution in handling multiscale dynamics.
We also compare it with the CNM \citep{li2021cluster,fernex2021cluster} of the same rank as a reference.

The Lorenz system is a three-dimensional autonomous system with non-periodic, deterministic, and dissipative dynamics that exhibit exponential divergence and convergence to strange fractal attractors. 
The system is governed by three coupled nonlinear differential equations:
\begin{eqnarray}
\begin{array}{l}
\mathrm{d}x/\mathrm{d}t =\sigma(y-x),\\
\mathrm{d}y/\mathrm{d}t =x(\rho-z)-y,\\
\mathrm{d}z/\mathrm{d}t =x y-\beta z.
\end{array}
\label{eq32}
\end{eqnarray}
The system parameters are set as $\sigma = 10$, $\rho = 28$, and $\beta = 8/3$.
These equations emulate the Rayleigh-B\'enard convection. 
The trajectory of the system revolves around two weakly unstable oscillatory fixed points, forming two sets of attractors, that are loosely called ``ears''. 
These two ears have similar but not identical shapes, with the left ear being rounder and thicker in the toroidal region. 
The region where the ears overlap is called the branching region. 
The Lorenz system has two main types of dynamics. 
One is that the inner loop in each ear varies and oscillates for several cycles. 
The other is that the inner loop may randomly switch from one ear to another in the branching region and resume oscillatory motion.

We numerically integrate the system using the fourth-order explicit Runge-Kutta method.
The time-resolved $10000$ snapshots data with $\boldsymbol{u}^m=[x, y, z]^{\intercal}$ are collected at a sampling time step of $\Delta t = 0.015$ with an initial condition of $[-3, 0, 31]^{\intercal}$ \citep{fernex2021cluster}.
This time step corresponds to approximately one-fiftieth of a typical cycle period.
The first $5\%$ of the snapshots are neglected to reserve only the post-transient dynamics.

\begin{figure}
    \centerline{\includegraphics[width=12cm]{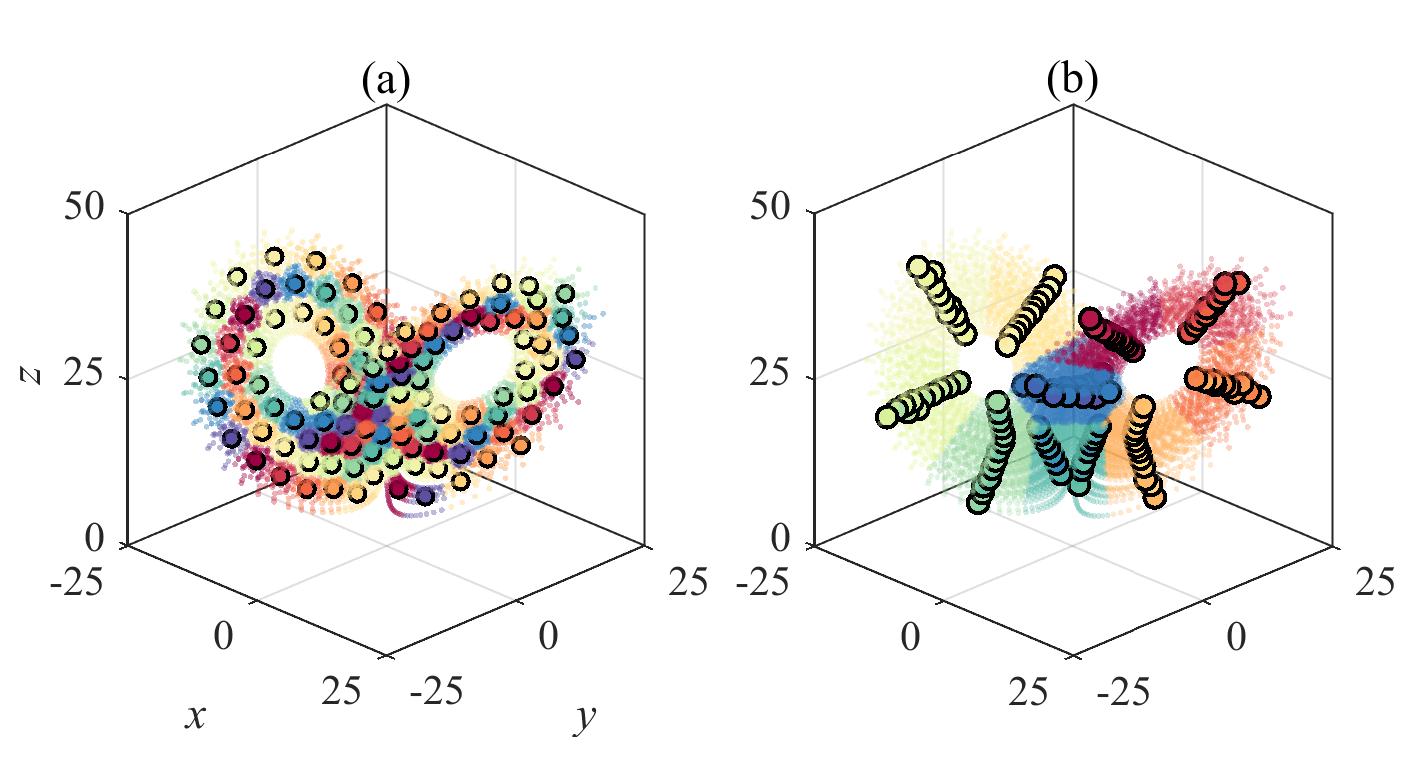}}
    \caption[width=\linewidth]{Phase portrait of the clustered Lorenz system from the CNM and dCNM. 
    The small dots represent the snapshots, and the large dots represent the centroids.
    Snapshots and centroids with the same colour belong to the same cluster.
    As a comparison, the CNM result in (a) is shown with the same number of centroids as the corresponding dCNM result.
    The dCNM result in (b) is shown with $K = 10$ and $\beta = 0.90$.}
\label{Lorenz_clustering}
\end{figure}

Figure~\ref{Lorenz_clustering} shows the phase portrait of the clustered Lorenz system from the CNM and dCNM. 
We set $K =10$ for the state space clustering of the dCNM, which is consistent with previous studies \citep{kaiser2014cluster,li2021cluster}. 
This number is large enough for the further subdivision of transition dynamics and is also small enough to obtain a simple structure for understanding. 
The sparsification index $\beta$ is chosen with large numbers as $\beta=0.90$ to allow for a distinct visualisation of the centroids.
In addition, since the trajectory in each ``ear'' is confined to a two-dimensional surface, a high value of $\beta$ is deemed suitable.
The normalized transverse cluster size vector $\boldsymbol{\hat{R}}^{\mathcal{T}} = [0.1163,0.1262,0.1164,0.0921,0.0943,0.0908,0.1116,0.0840,0.0866,0.0817]^{\intercal}$ corresponds to the number of sub-clusters $\boldsymbol{L} = [13,15,13,11,11,11,13,10,10,10]^{\intercal}$.

The two models exhibit notable differences in centroid distribution. 
CNM clustering relies solely on the spatial topology in the phase space, evenly dividing the entire attractor and dispersing centroids uniformly throughout the phase portrait. 
It can be inferred that increasing the number of centroids under this uniform distribution does not lead to substantial changes, merely resulting in a denser centroid distribution.
This uniform distribution possesses certain disadvantages regarding the dynamics. 
First, it unnecessarily complicates the transition rhythm as the deterministic large-scale transition may be fragmented into several stochastic transitions. 
Second, even with many centroids, it fails to capture the increasing oscillation amplitude between the loops in one ear, as the uniform distribution provides only a limited number of centroid orbits.
The same result occurs for the branching region where these limited numbers of centroids usually oversimplify the switch between ears.
In contrast, the distribution of the dCNM centroids resembles a weighted reallocation. 
For the Lorenz system, the state space is stratified along the trajectory direction, leading to a concentrated distribution of the dCNM centroids in the radial direction of the attractor and the branching region, which correspond to the system's primary dynamics.
Additionally, varying quantities of the centroids can be observed in the radial direction in the toroidal region, depending on its thickness. 
In thinner toroidal regions with smaller variations between trajectory segments, the second-stage clustering assigns fewer sub-clusters and, consequently, builds fewer centroids.

\begin{figure}
    \centerline{\includegraphics[width=12cm]{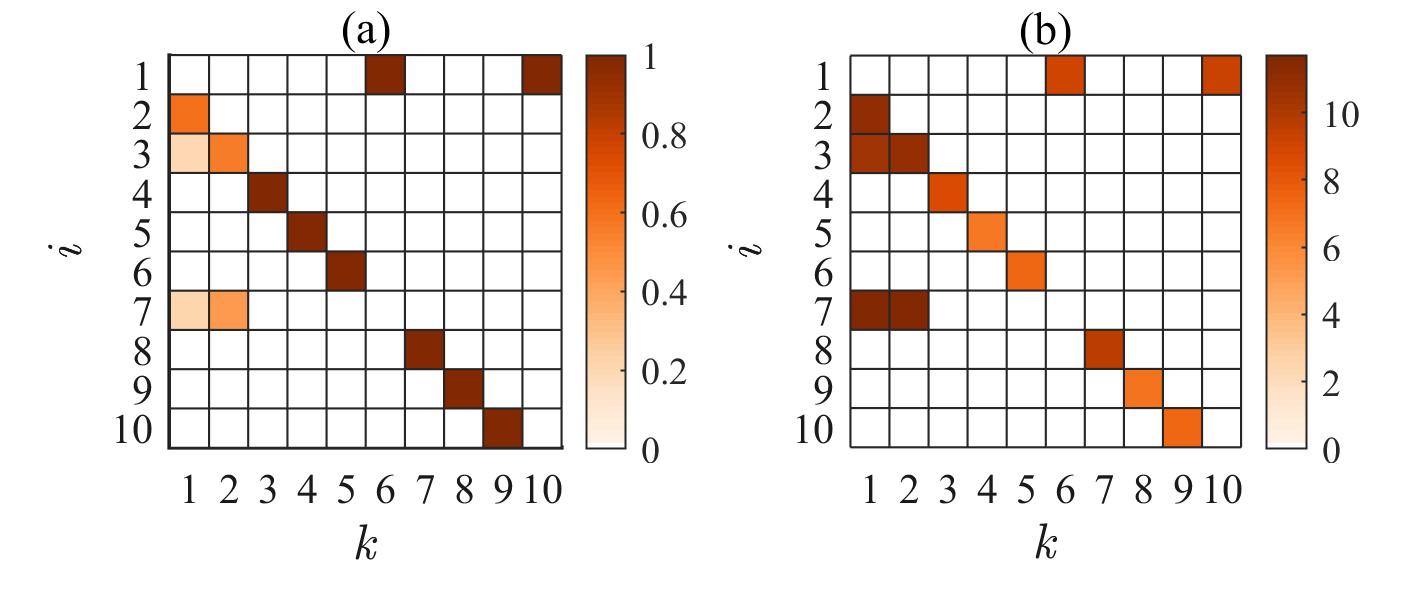}}
    \caption{Transition matrices of the Lorenz system. 
    The colour bar indicates the values of the terms.
    (a) Transition probability matrix $\mathbf{Q}$. 
    (b) Transition time matrix $\mathbf{T}$.}
\label{Lorenz_matrixes}
\end{figure}

The cluster transition matrices, which are a distinctive feature of cluster modelling, are preserved because the dCNM maintains the coarse-grained transitions at the cluster level.
Figure~\ref{Lorenz_matrixes} illustrates the cluster transition probability matrix $\mathbf{Q}$ and the corresponding transition time matrix $\mathbf{T}$ to illustrate the significant dynamics of the Lorenz system.
It is worth noting that in the case of the CNM with an equivalent number of centroids, the matrices become considerably larger, which diminishes their readability and interpretability.
The matrices reveal three distinct cluster groups.
The first group comprises clusters $\mathcal{C}_{1}$ and $\mathcal{C}_{2}$, which resolve the branching region and exhibit similar transition probabilities to clusters $\mathcal{C}_{3}$ and $\mathcal{C}_{7}$. 
The branching region is further linked to different ears and is crucial to the attractor oscillation.
Clusters $\mathcal{C}_{1}$ and $\mathcal{C}_{2}$ can be referred to as flipper clusters \citep{kaiser2014cluster}, representing a switch between the different groups.
The equivalent transition probability from $\mathcal{C}_{2}$ is consistent with the random jumping behaviour of the two ears.
The other two groups demonstrate an inner-group circulation corresponding to the main components of the two ears,  exemplified by the cluster chains $\mathcal{C}_{3} \to \mathcal{C}_{4} \to \mathcal{C}_{5} \to \mathcal{C}_{6}$ and $\mathcal{C}_{7} \to \mathcal{C}_{8} \to \mathcal{C}_{9} \to \mathcal{C}_{10}$.
These chains exhibit deterministic transition probabilities that resolve the cyclic behaviour. 
In the second-stage clustering, these two groups are further categorised into numerous centroid orbits.
Moreover, the transition time matrix resolves the variance in the transition times, with significantly shorter transition times observed in the cyclic groups compared to transitions involving the flipper clusters.

\begin{figure}
    \centerline{\includegraphics[width=12cm]{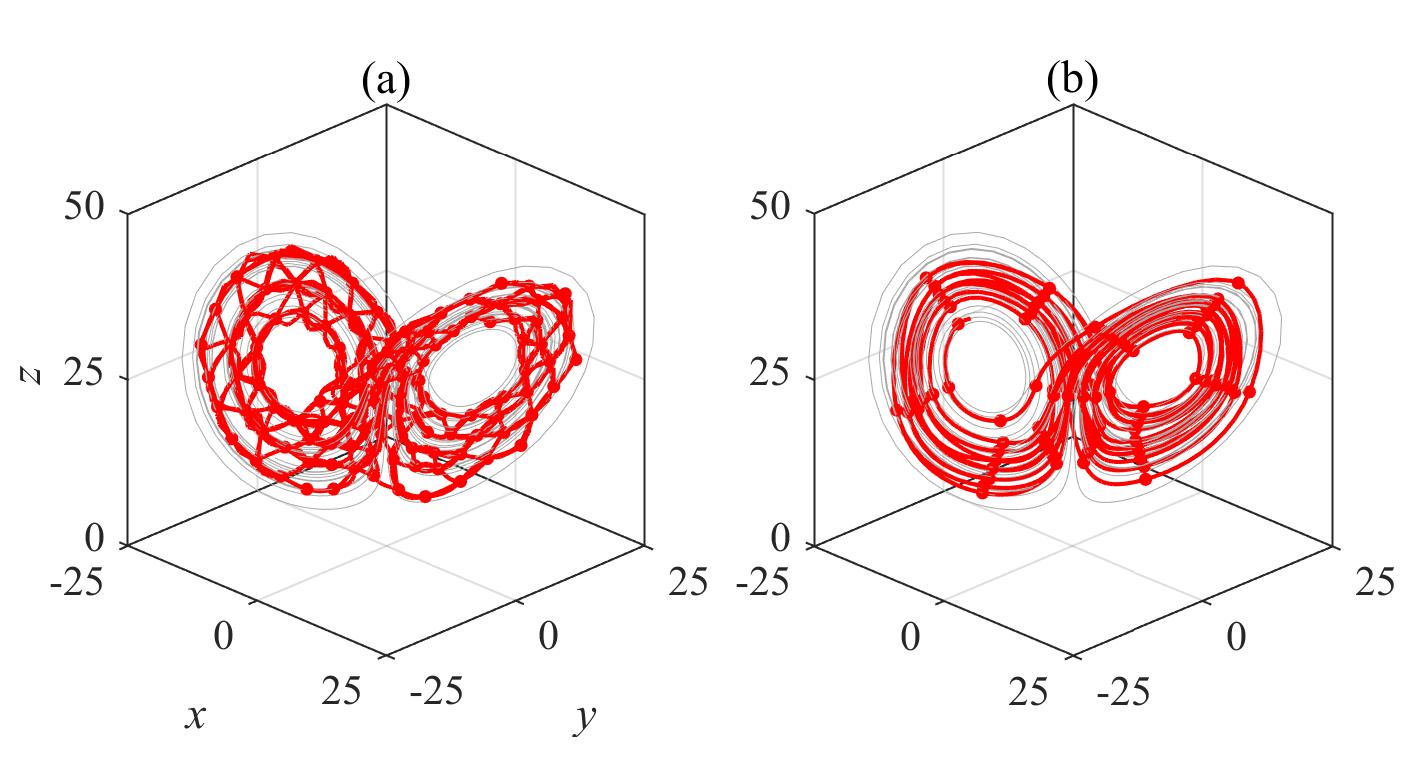}}
    \caption{Trajectory of the Lorenz system.
    The thin grey curve represents the original trajectory, the thick red curve represents the reconstructed trajectory, and the red dots represent the centroids.
    (a) The CNM reconstruction and (b) the dCNM reconstruction are performed with the same parameters as in figure~\ref{Lorenz_clustering}.}
\label{Lorenz_reconstruction}
\end{figure}

The original and reconstructed trajectories in the phase space are directly compared.
We focus solely on the spatial resolution, disregarding phase mismatches during temporal evolution.
Figure~\ref{Lorenz_reconstruction} shows the original Lorenz system and the reconstruction by the CNM and dCNM with the same parameters as in figure~\ref{Lorenz_clustering}.
To ensure clarity, we select a time window from $t=0$ to $t=30$ for the trajectories and employ spline interpolation for a smooth reconstruction. 
Inaccurate or non-physical centroid transitions, along with incomplete dynamic coverage, can lead to substantial deformations in the reconstructed trajectory. 
As expected, the dCNM provides a more accurate reconstruction than the CNM. 
The CNM uses a finite number of centroid orbits to represent oscillating attractors, converting slow and continuous amplitude growth into limited and abrupt amplitude jumps.
Furthermore, the CNM may group one continuous snapshot loop into clusters belonging to different centroid orbits, often when these clusters are adjacent to each other. 
This can lead to unnecessary orbit-crossing centroid transitions and result in nonphysical radial jumps in the reconstructed trajectory.
In contrast, the dCNM provides more comprehensive dynamic coverage, resolving more cyclic behaviour with additional centroid orbits. 
Dual indexing also guarantees accurate centroid transitions.
The radial jumps are eliminated, as departing centroids can only transition to destination centroids within the same centroid orbits. 
Consequently, oscillations are effectively resolved by the centroid orbits, and transitions between them are constrained by densely distributed centroids in the branching region, ensuring a smoothly varied oscillation.

\begin{figure}
    \centerline{\includegraphics[width=12cm]{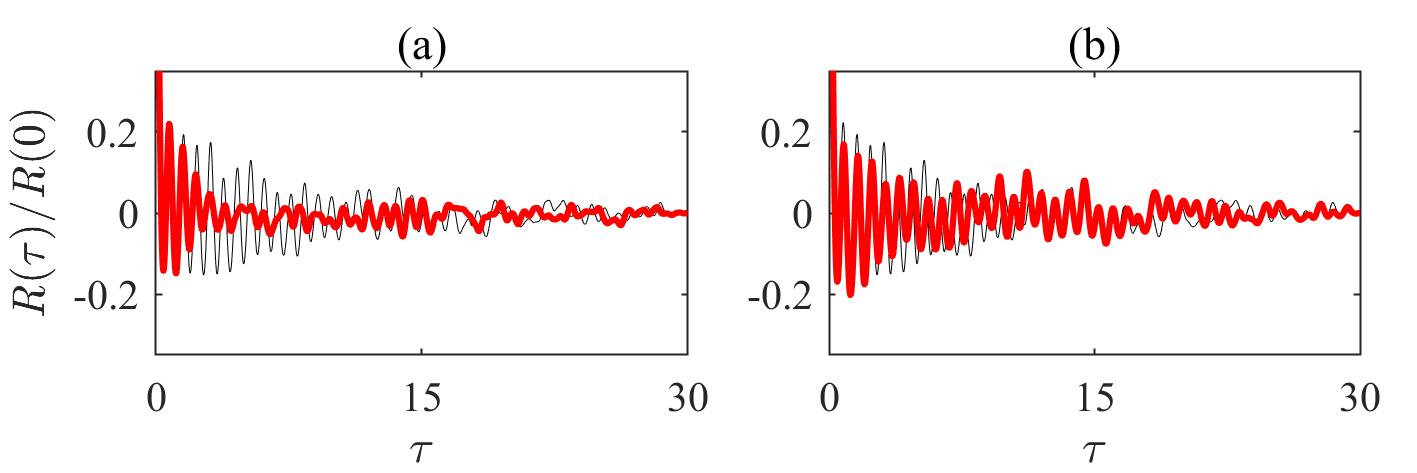}}
    \caption{Auto-correlation function for $\tau \in [0, 30)$ of the Lorenz system.
    The thin black curves represent the original data set, and the thick red curves represent the models: (a) CNM and (b)dCNM.}
\label{Lorenz_autocorrelation}
\end{figure}

The auto-correlation function is computed to reflect the model accuracy, as shown in figure~\ref{Lorenz_autocorrelation}.
In the original data set, the normalised auto-correlation function $R(\tau) / R(0)$ vanishes smoothly as $\tau$ increases, and the variance between the periodic behaviour can be clearly observed.
However, the CNM reconstruction captures only the first four periods of oscillation dynamics.
As $\tau$ increases, there is a sudden amplitude decay accompanied by a phase mismatch.
This can be attributed to amplitude jumps between the centroid loops and commonly occurring orbit-crossing transitions.
In contrast, the dCNM reconstruction accurately captures both the amplitude and frequency of the oscillation dynamics, demonstrating robust and precise long-timescale behaviours.

%%%%%%%%%%%%%%%%%%%%%%%%%%%%%%%%%%%%%%%%%%%%%%%%%%%%%%%%%%%%%%%
\section{Dynamics-augmented modelling of the sphere wake}
\label{sec:sphere}
In this section, we demonstrate the dCNM for the transient and post-transient flow dynamics of the sphere wake. 
The numerical method for obtaining the flow field data set and the flow characteristics is presented in \S~\ref{sec4.1}. 
The performance of the dCNM for the periodic, quasi-periodic and chaotic flow regimes is evaluated in \S~\ref{sec4.2}, \S~\ref{sec4.3} and \S~\ref{sec4.4}, respectively.
The physical interpretation of the modelling strategy is discussed in \S~\ref{sec4.5}.
%In this section, we evaluate the performance of the dCNM for the quasi-periodic and chaotic flow regimes of the sphere wake, which exhibit multi-frequency and multiscale dynamics. 
%The numerical method for obtaining the flow field data set and the flow characteristics is presented in \S~\ref{sec4.1}. 
%The demonstration of the dCNM for the quasi-periodic and chaotic flow regimes is introduced in \S~\ref{sec4.2} and \S~\ref{sec4.3}, respectively.
%The physical interpretation of the modelling strategy is discussed in \S~\ref{sec4.4}.

%%%%%%%%%%%%%%%%%%%%%%%%%%%%%%%%%%%%%%%%%%
\subsection{Numerical methods and flow features}
\label{sec4.1}

\begin{figure}
    \centerline{\includegraphics[width=10cm]{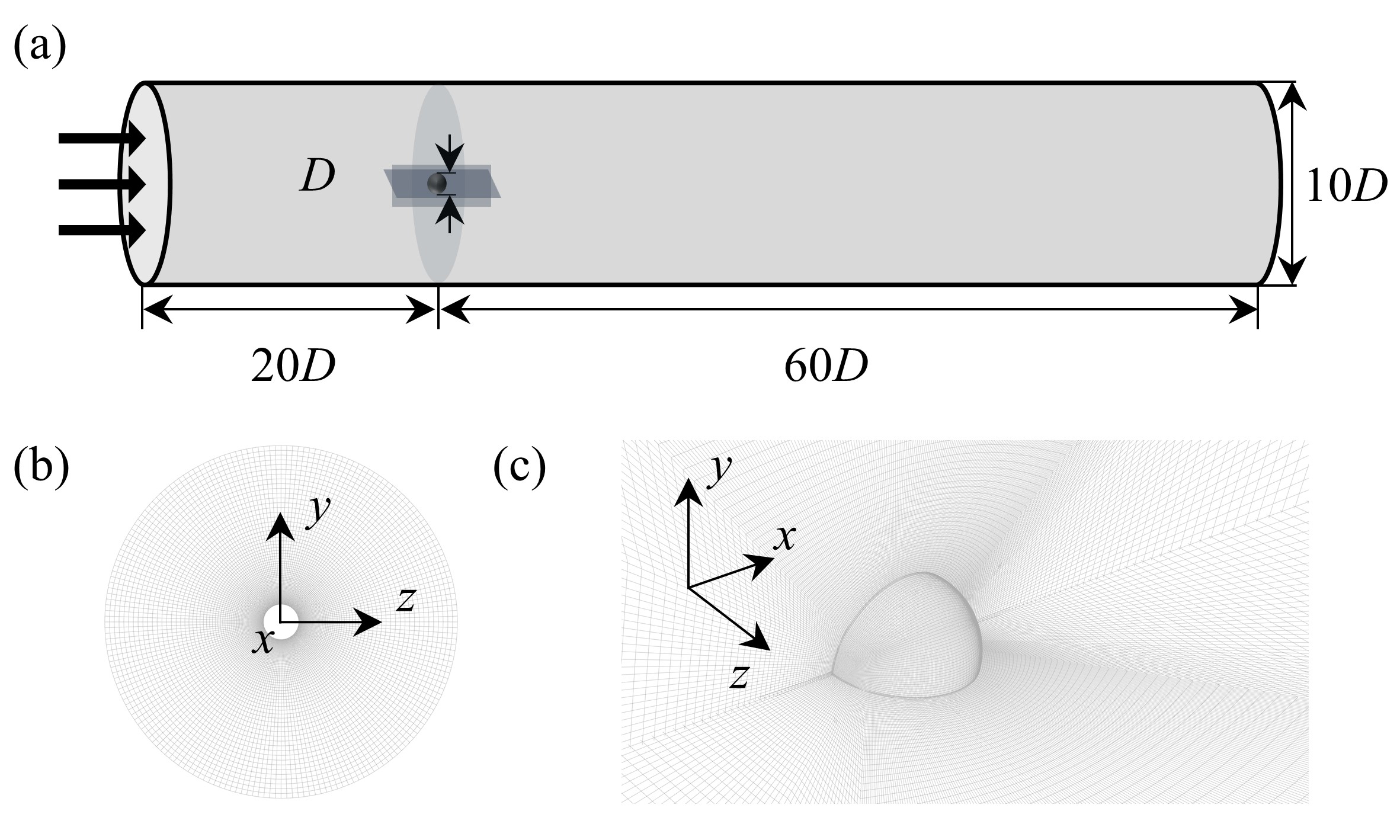}}
    \caption{Numerical sketch of the sphere wake.}
\label{NumericalSketch}
\end{figure}

Numerical simulation is performed to obtain the data set, as shown in figure~\ref{NumericalSketch}.
A sphere with a diameter $D$ is placed in a uniform flow with a streamwise velocity $U_\infty$.
The computational domain takes the form of a cylindrical tube, with its origin at the centre of the sphere and its axial direction along the streamwise direction ($x$-axis).
The dimensions of the domain in the $x$, $y$, and $z$ directions are $80D$, $10D$, and $10D$, respectively.
The inlet is located $20D$ upstream from the sphere.
These specific domain parameters are chosen to minimise any potential distortion arising from the outer boundary conditions while also mitigating computational costs \citep{pan2018wake, lorite2020description}.
The fluid flow is governed by the incompressible Navier–Stokes equations:
\begin{eqnarray}
\begin{array}{l}
\partial{\boldsymbol{u}} / \partial{t} + \boldsymbol{u} \cdot \nabla \boldsymbol{u} + \nabla p- \nabla^{2} \boldsymbol{u} / \Rey = 0,\\
\nabla \cdot \boldsymbol{u}=0,
\end{array}
\label{eq33}
\end{eqnarray}
where $\boldsymbol{u}$ denotes the velocity vector $(u_x, u_y, u_z)$, $p$ is the static pressure, and $\Rey$ is the Reynolds number, which is defined by:
\begin{equation}
\Rey = U_{\infty} D / \nu,
\label{eq34}
\end{equation}
$\nu$ is the kinematic viscosity. 

The net forces on the sphere have three components $F_\alpha$, $\alpha = x, y, z$, and the corresponding force coefficients $C_{\alpha}$ are defined as:
\begin{equation}
C_{\alpha}=\frac{2 F_{\alpha}}{\rho U_\infty^2 S},
\label{eq35}
\end{equation}
where $S=\pi D^{2}/4$ is the projected surface area of the sphere in the streamwise direction.
The total drag force coefficient is $C_D = C_x$.
Since the lift coefficient can have any direction in the $yz$ plane on the axisymmetric sphere, the total lift force coefficient
$C_L$ is given by:
\begin{equation}
C_L=\sqrt{C_{y}^{2}+C_{z}^{2}}.
\label{eq36}
\end{equation}

The flow parameters are non-dimensionalised based on the characteristic length $D$ and the free-stream velocity $U_\infty$.
This implies that the time unit scales are $D/U_\infty$, and the pressure scales are $\rho U_\infty^2$, where $\rho$ is the density.
The Strouhal number $St$ is correspondingly expressed as:
\begin{equation}
St = f,
\label{eq37}
\end{equation}
where $f$ is the characteristic frequency.

ANSYS Fluent $15.0$ is used as the CFD solver for the governing equations with the cell-centred finite volume method (FVM). 
We impose a uniform streamwise velocity $\boldsymbol{u} = [U_\infty, 0, 0]$ at the inlet boundary and an outflow condition at the outlet boundary. 
The outflow condition is set as a Neumann condition for the velocity, $\partial _x \boldsymbol{u} = [0, 0, 0]$, and a Dirichlet condition for the pressure, $p_{\mathrm{out}} = 0$. 
We apply a no-slip boundary condition on the sphere surface and a slip boundary condition on the cylindrical tube walls to prevent wake-wall interpolations. 
The pressure-implicit split-operator (PISO) algorithm is chosen for pressure-velocity coupling.
For the governing equations, the second-order scheme is used for the spatial discretization, and the first-order implicit scheme is used for the temporal term.
To satisfy the Courant–Friedrichs–Levy (CFL) condition, a small integration time step is set as $\Delta t = 0.01$ non-dimensional time unit, such that the Courant number is below $1$ for all simulations. 
For the periodic flow at $\Rey = 300$, the simulation starts in the vicinity of the steady solution and runs for $t = 200$ time units, incorporating the transient and post-transient dynamics.
For the quasi-periodic flow, the simulations are performed for $t = 500$ time units and for the chaotic flow for $t = 700$ time units.
The snapshots are collected at a sampling time step of $\Delta t_s = 0.2$ time units for all the test cases. 
Moreover, we discard the first $200$ time units to eliminate any transient phases for the quasi-periodic and chaotic cases. 
The relevant numerical investigation approach can be found in \citet{johnson1999flow,rajamuni2018transverse}.
For the convergence and validation studies, see Appendix~\ref{app:a}.

The wake of a sphere exhibits different flow regimes as $\Rey$ increases, ultimately transitioning to a chaotic state.
At $\Rey = 20 \sim 24$, flow separation occurs, forming a steady recirculating bubble, as observed in previous studies \citep{sheard2003spheres, eshbal2019measurement}. 
The length of this wake grows linearly with $\ln(\Rey)$. 
When $\Rey$ surpasses $130$ \citep{taneda1956experimental}, the wake bubble starts oscillating in a wave-like manner, while the flow maintains axisymmetry. 
The first Hopf bifurcation takes place at approximately $\Rey \approx 212$ \citep{fabre2008bifurcations}, leading to a loss of axisymmetry and the emergence of a planar-symmetric double-thread wake with two stable and symmetric vortices. 
The orientation of the symmetry plane can vary \citep{johnson1999flow}.
At a subsequent Hopf bifurcation around $\Rey = 270 \sim 272$ \citep{johnson1999flow, fabre2008bifurcations}, the flow becomes time-dependent, initiating periodic vortex shedding with the same symmetry plane as before. 
In the range $272 < \Rey < 420$ \citep{eshbal2019measurement}, periodicity and the symmetry plane diminish, with the vortex shedding becoming quasi-periodic and then fully three-dimensional.
Beyond $\Rey = 420$, shedding becomes irregular and chaotic \citep{ormieres1999transition, eshbal2019measurement, pan2018wake}, due to the azimuthal rotation of the separation point and lateral oscillations of the shedding.

\begin{figure}
    \centerline{\includegraphics[width=10.5cm]{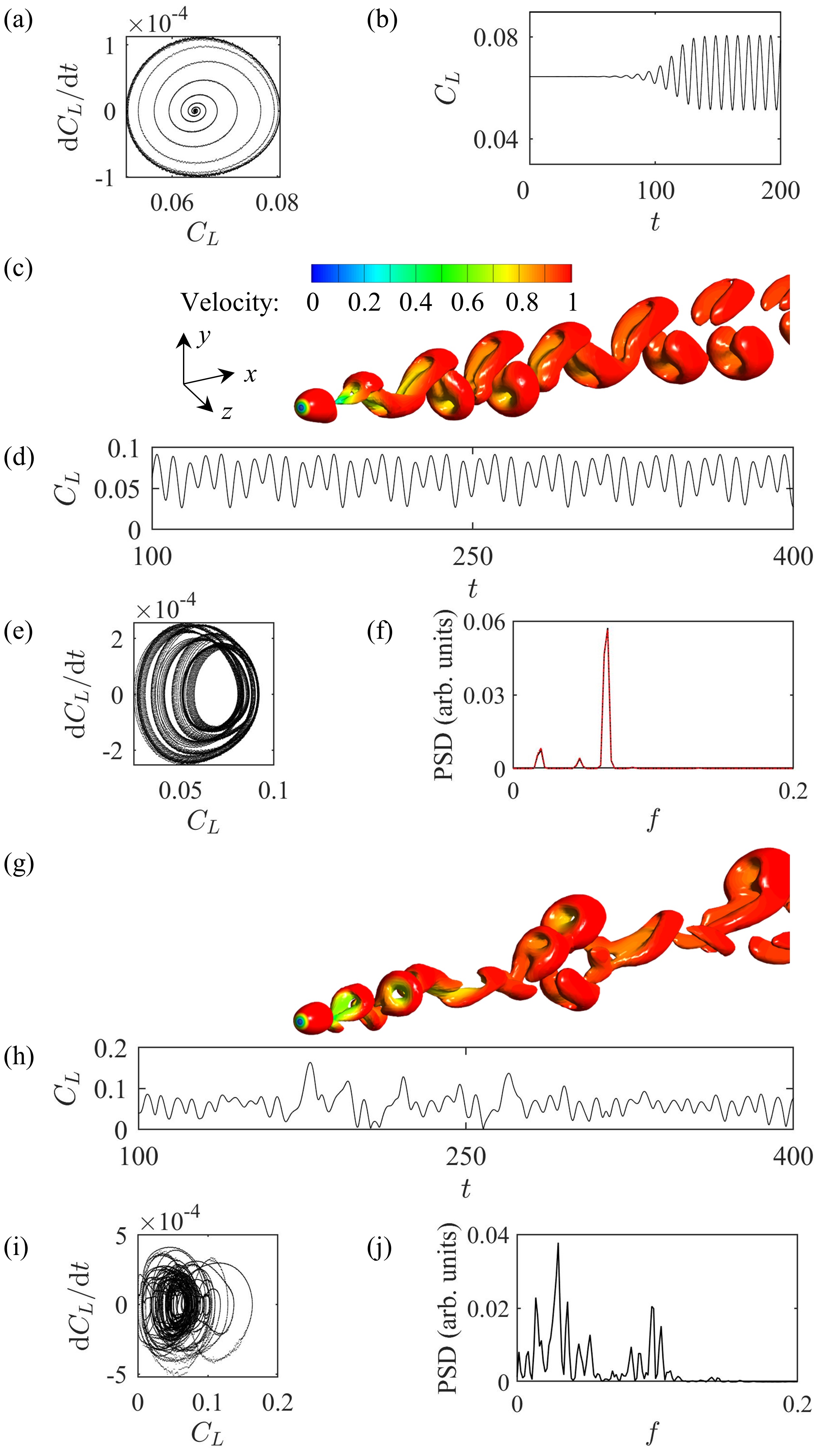}}
    \caption{Flow characteristics of the sphere wake.
    The periodic flow at $\Rey=300$ including the transient and post-transient dynamics is displayed by the
    (a) phase portrait of the lift coefficient $C_L$ and
    (b) temporal evolution of $C_L$.
    The quasi-periodic flow at $\Rey=330$ is displayed by the
    (c) vortex structures, where the vortexes are identified by the $Q$-criteria, and are colour-coded by the non-dimensional velocity $U_{\infty}$,
    (d) temporal evolution of $C_L$,
    (e) phase portrait of $C_L$ and
    (f) power spectral density of $C_L$ on time series of length $T_{\mathrm{traj}} = 100$ (red curve) and $T_{\mathrm{traj}} = 300$ (black curve).
    The chaotic flow at $\Rey=450$ is displayed by the 
    (g) vortex structures,
    (h) temporal evolution of $C_L$,
    (i) phase portrait of $C_L$ and
    (j) power spectral density of $C_L$ on a time series of length $T_{\mathrm{traj}} = 500$.
}
\label{Spherewake}
\end{figure}

In this study, we examine three baseline flow regimes of the sphere wake: periodic flow at $\Rey = 300$, quasi-periodic flow at $\Rey = 330$ and chaotic flow at $\Rey = 450$.
Figure~\ref{Spherewake} illustrates the flow characteristics of these regimes.
Figure~\ref{Spherewake} (a) and (b) depict the lift coefficient $C_L$ of the periodic flow. 
The oscillation of $C_L$ starts after a few time units, with the amplitude gradually increasing in the cycle-to-cycle evolution, eventually forming the limit cycle.
Figure~\ref{Spherewake} (c) shows the instantaneous vortex structures of the quasi-periodic flow, identified by the $Q$-criterion and colour-coded by the non-dimensional velocity $U_{\infty}$.
The vortex shedding forms hairpin vortices with slight variations between successive shedding events, signifying the absence of short-term periodicity while retaining long-term periodic behaviour.
Figure~\ref{Spherewake} (d) and (e) show the temporal evolution of $C_L$ and its phase portrait, respectively.
The amplitude of $C_L$ is strongly associated with the quasi-periodic dynamic, and the modulation also thickens the limit cycle of the oscillator on the phase portrait.
The power spectral density in figure~\ref{Spherewake} (f) indicates two dominant frequencies: a higher frequency linked to natural shedding and a lower frequency associated with amplitude modulation resulting from variations between shedding events.
For chaotic flow, periodicity entirely vanishes, and the flow regime displays the typical features of a chaotic system. 
The hairpin vortexes in figure~\ref{Spherewake} (g) shed irregularly, with varying separation angles and even double spirals. 
The temporal evolution of $C_L$ in figure~\ref{Spherewake} (h) exhibits more complex dynamics, with the phase diagram in figure~\ref{Spherewake} (i) depicting many random loops that no longer exhibit circular patterns.
Furthermore, the power spectral density of $C_L$ in figure~\ref{Spherewake} (j) shows a broad peak, also indicating chaotic features.

We performed a \emph{lossless} POD preprocessing on the snapshots to reduce the computational cost of clustering the three-dimensional flow field data set, as described in Appendix~\ref{app:b}. 
This preprocessing is optional and does not affect the distance measure in the clustering algorithm. 
For consistency, the notation snapshot is maintained in the following for the preprocessed data.

%%%%%%%%%%%%%%%%%%%%%%%%%%%%%%%%%%%%%%%%%%
\subsection{The periodic flow regime at \texorpdfstring{$\Rey = 300$}{Lg}}
\label{sec4.2}
We compare the dCNM to the CNM for the periodic flow regime of the sphere wake at $\Rey = 300$.
The transient and post-transient dynamics are considered, providing insights into the mechanisms for the instability and nonlinear saturation.

\begin{figure}
    \centerline{\includegraphics[width=12cm]{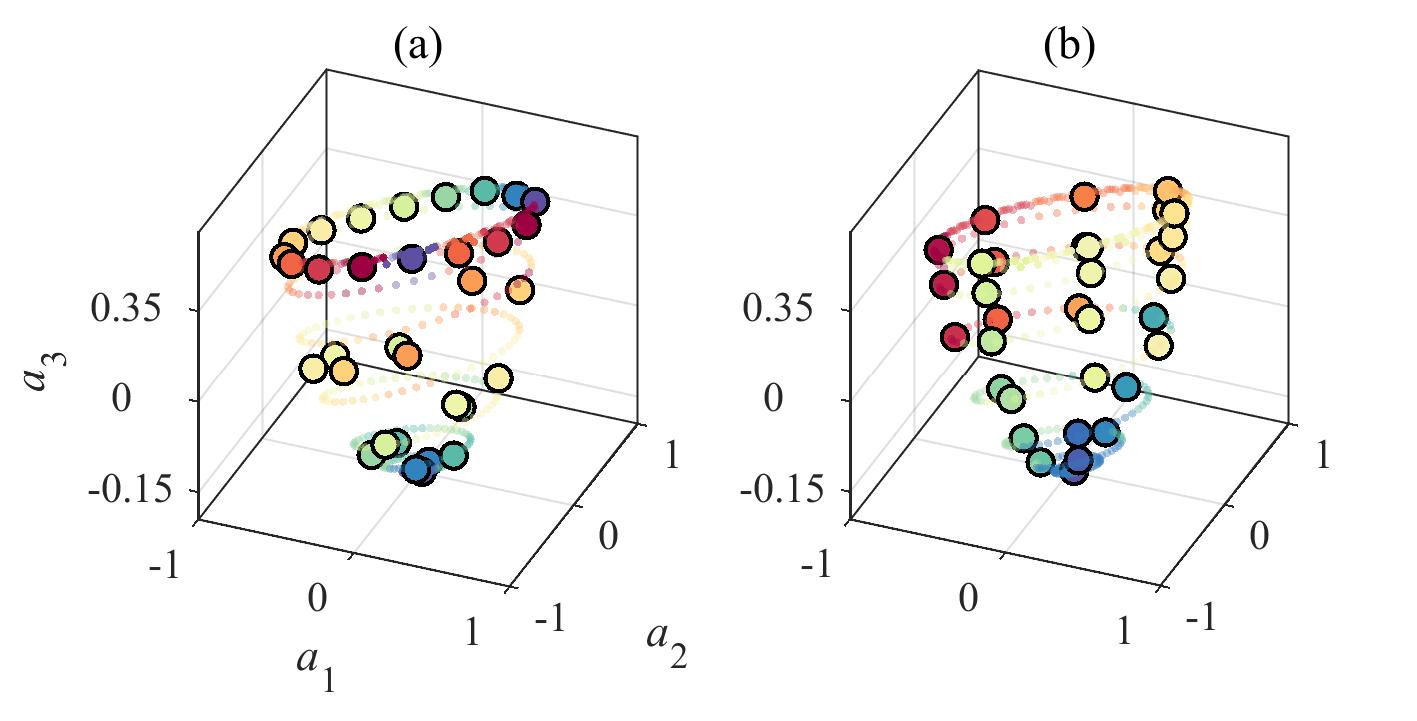}}
    \caption{Three-dimensional visualisation of the clustered periodic flow regime of the sphere wake at $\Rey = 300$. 
    Classical multidimensional scaling (MDS) is applied to the data set to visualise the high-dimensional snapshots and centroids in the subspace.
    The small dots represent the snapshots, and the large dots represent the centroids.
    Snapshots and centroids with the same colour belong to the same cluster.
    For comparison, the CNM result in (a) is shown with the same number of centroids as the corresponding dCNM result.
    The dCNM result in (b) is shown with $K = 10$ and $\beta = 0.50$.}
\label{Re300_clustering}
\end{figure}

The comparison between the CNM and the dCNM clustering for the periodic flow is presented in figure~\ref{Re300_clustering}.
In the dCNM, we set $K=10$ for state space clustering and $\beta = 0.50$ for the sub-clustering.
The value of $\beta$ characterises the trade-off between a small number of sub-clusters and the model accuracy.
The normalised transverse cluster size vertor $\boldsymbol{\hat{R}}^{\mathcal{T}} = [0.1510,0.1101,0.0855,0.1120,0.1069,0.1031,0.0650,0.0934,0.0883,0.0847]^{\intercal}$ corresponds to the number of sub-clusters $\boldsymbol{L} = [3,3,3,4,4,4,3,3,3,3]^{\intercal}$. 
Classical multidimensional scaling (MDS) is applied to project the high-dimensional snapshots and centroids into a three-dimensional subspace $[a_1, a_2, a_3]^\intercal$ for visualisation.
The snapshots form a conical surface in the three-dimensional subspace, where the trajectory spirals up from a fixed point to a periodic motion.
This behaviour is indicative of a Hopf bifurcation, which involves an unstable steady solution and nonlinear saturation to a periodic limit cycle.
Most of the CNM centroids are located on the limit cycle, and only a few resolve the transient phase.
In contrast, the dCNM centroids offer a finer resolution of the amplitude growth.

\begin{figure}
    \centerline{\includegraphics[width=12cm]{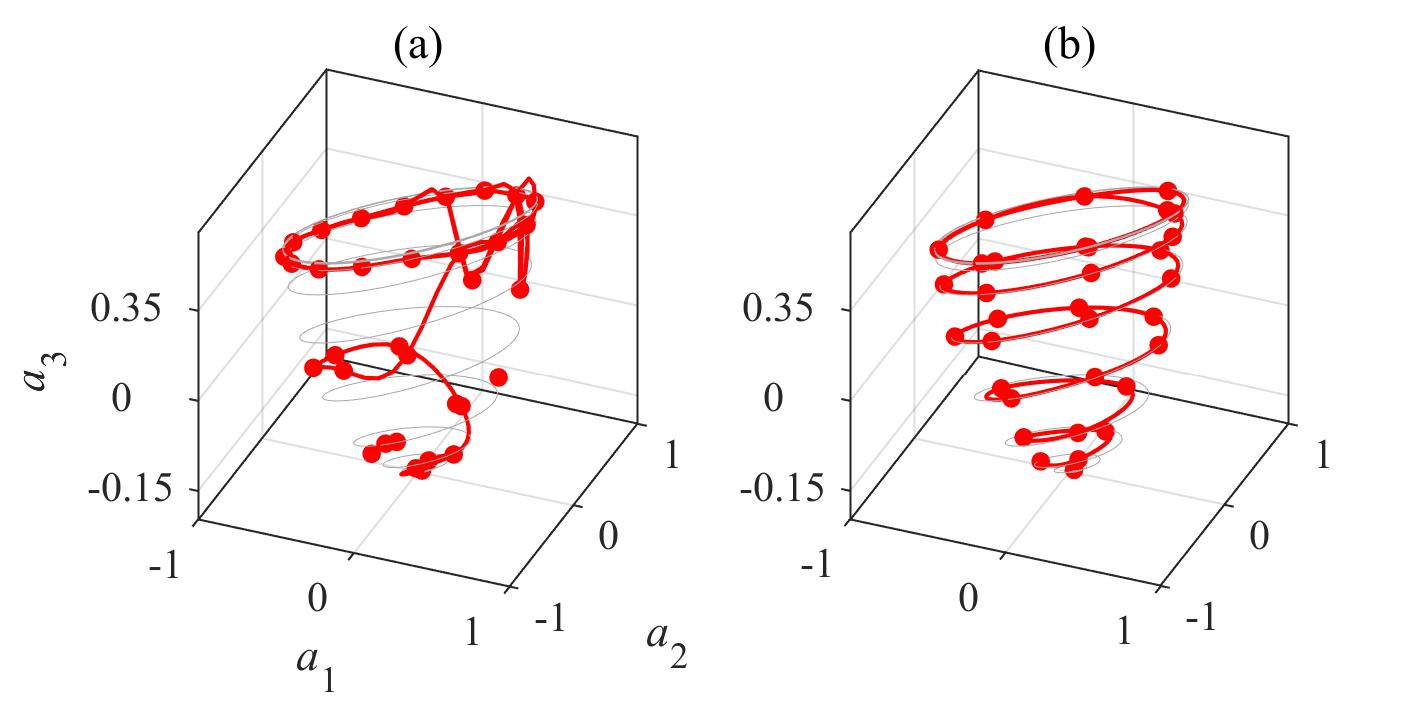}}
        \caption{Trajectory of the periodic flow at $\Rey = 300$.
        The thin grey curve represents the original trajectory, the thick red curve represents the reconstructed trajectory, and the red dots represent the centroids.
        (a) The CNM reconstruction and (b) the dCNM reconstruction are obtained with the same parameters as in figure~\ref{Re300_clustering}.}
\label{Re300_reconstruction}
\end{figure}

The original and reconstructed trajectories of the CNM and dCNM for the periodic flow regime are shown in figure~\ref{Re300_reconstruction}.
The CNM fails to resolve the transient dynamics and only captures the stable limit cycle.
In contrast, the dCNM can effectively resolve both the cyclic behaviour and the growing oscillation amplitude.
Given the deterministic nature of these transient and post-transient dynamics, 
the centroid transition should yield a permutation matrix --- each column and each row of the matrix should only have one element being unity.
However, in the CNM, under-resolved transient dynamics result in a stochastic transition matrix.
The transition uncertainty expressed within a column of the matrix often leads to prediction errors.
Examples are the unphysical jumps between erratic cycles in the transient stage, as displayed in Figure~\ref{Re300_clustering} (a). 
The sub-clusters in dCNM are designed to capture a one-way forward transition from the starting point to the limit cycle, 
ensuring an accurate reconstruction of the deterministic dynamics.

The other extreme is the maximum transition uncertainty, which can be represented by the least informative transition matrix --- a perfectly mixing matrix with elements $Q_{ik} \equiv  1/K$.
The information entropy \citep{shannon1948mathematical} 
\begin{equation}
S(\boldsymbol{Q})=-\sum_{i=1}^{K} \sum_{k=1}^{K} Q_{i k} \ln Q_{i k}
\label{eq38}
\end{equation}
of the permutation matrix vanishes.
In contrast, 
the maximum information entropy  $K\ln K$ is obtained from the perfectly mixing transition matrix
with equal elements  $Q_{ik} \equiv 1/K$.
Here, the knowledge of the current state 
has no predictive value for the future population.
For the current case, 
the reference model has an entropy 
$S_{\mathrm{CNM}} = 7.0586$, 
which is much smaller than the upper bound $33 \ln 33 \approx 115.38$.
The proposed model minimizes entropy,
$S_{\mathrm{dCNM}} = 0$.
Thus our novel clustering measurably increases the prediction accuracy.

%%%%%%%%%%%%%%%%%%%%%%%%%%%%%%%%%%%%%%%%%%
\subsection{The quasi-periodic flow regime at \texorpdfstring{$\Rey = 330$}{Lg}}
\label{sec4.3}
\begin{figure}
    \centerline{\includegraphics[width=12cm]{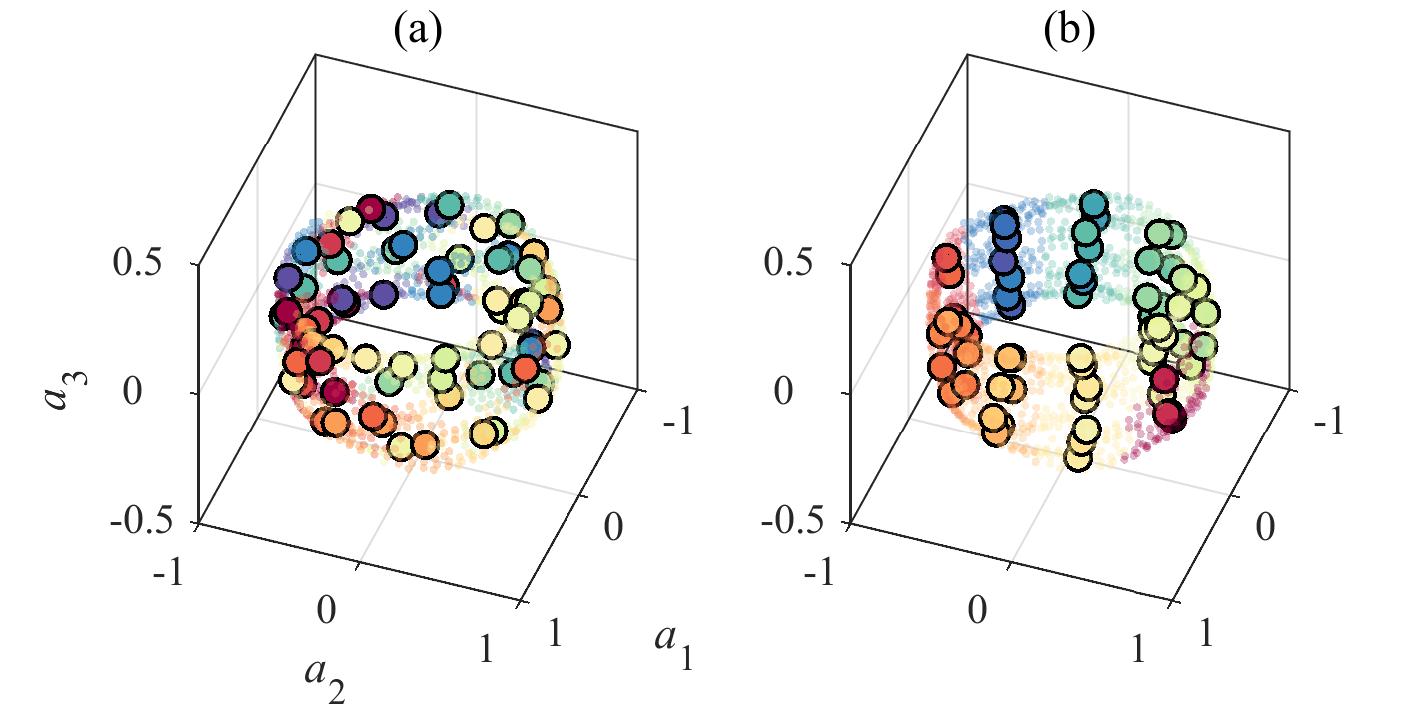}}
    \caption{Same as figure \ref{Re300_clustering}, but for the quasi-periodic flow at $\Rey = 330$. 
    (a) The CNM result with the same number of centroids as the dCNM. 
    (b) The dCNM result with $K = 10$ and $\beta = 0.80$.}
\label{Re330_clustering}
\end{figure}

The clustered quasi-periodic flow of the CNM and dCNM are shown in figure~\ref{Re330_clustering}.
We set $K=10$ for the state space clustering and $\beta = 0.80$ for the subsequent clustering,
which proves adequate for capturing the quasi-periodic dynamics and ensuring clarity in visualisation.
The choice of $\beta$ and other results with different values of $\beta$ are discussed in Appendix~\ref{app:c}.
In this case, the normalized transverse cluster size vector $\boldsymbol{\hat{R}}^{\mathcal{T}} = [0.1043,0.1046,0.1056,0.1068,0.0889,0.1073,0.1061,0.1050,0.1047,0.0668]^{\intercal}$, corresponds to the number of sub-clusters $\boldsymbol{L} = [7,7,7,7,6,7,7,7,7,5]^{\intercal}$.
%This value of $\beta$ proves adequate for capturing the quasi-periodic dynamics and ensuring clarity in visualisation. %Additional results obtained using various $\beta$ values are presented in Appendix~\ref{app:c} for reference.
In the three-dimensional subspace, the snapshots collectively form a hollow cylinder.
The system's dynamics are chiefly governed by two underlying physical phenomena: a cyclic behaviour that synchronises with natural vortex shedding and a quasi-stochastic component responsible for introducing variations between cycles, which is, in turn, synchronised with the oscillator amplitude.

The centroid distribution of the CNM reveals that the clustering algorithm fails to distinguish between the shedding dynamics and inter-cycle variations.
It uniformly groups them based solely on spatial topology.
Nonetheless, the CNM centroids effectively capture the cyclic behaviour, as there exist deterministic transitions between adjacent centroids within an orbit, forming a limit cycle structure akin to the ``ear'' of the Lorenz system.
However, this centroid distribution inadequately models the quasi-stochastic component, as it overlooks the inter-cycle transitions. 
To comprehensively represent this dynamic, clear transitions between the limit cycles are essential.
The clustering process obscures these transitions, causing the quasi-stochastic behaviour to resemble a random walk governed by a fully stochastic process. 
In essence, the clustering process cannot differentiate between the random jumps in the Lorenz system and the quasi-stochastic behaviour in this flow regime. 
This explains why the CNM often struggles with multifrequency problems.
In contrast, the dCNM centroids automatically align along the axial direction of the cylinder with equidistant circumferential spacing, resulting in a greater number of centroid orbits compared to the CNM. 
This enhancement enables the accurate resolution of inter-cycle variations. 
For the quasi-stochastic behaviour, the denser and occasionally overlapping centroids in the axial direction ensure precise spatial representation of the transitions between the limit cycles. 
Additionally, this behaviour can be further constrained by the dual indexing approach for long-timescale periodicity, eliminating random jumps and ensuring accurate transitions between limit cycles.

\begin{figure}
    \centerline{\includegraphics[width=12cm]{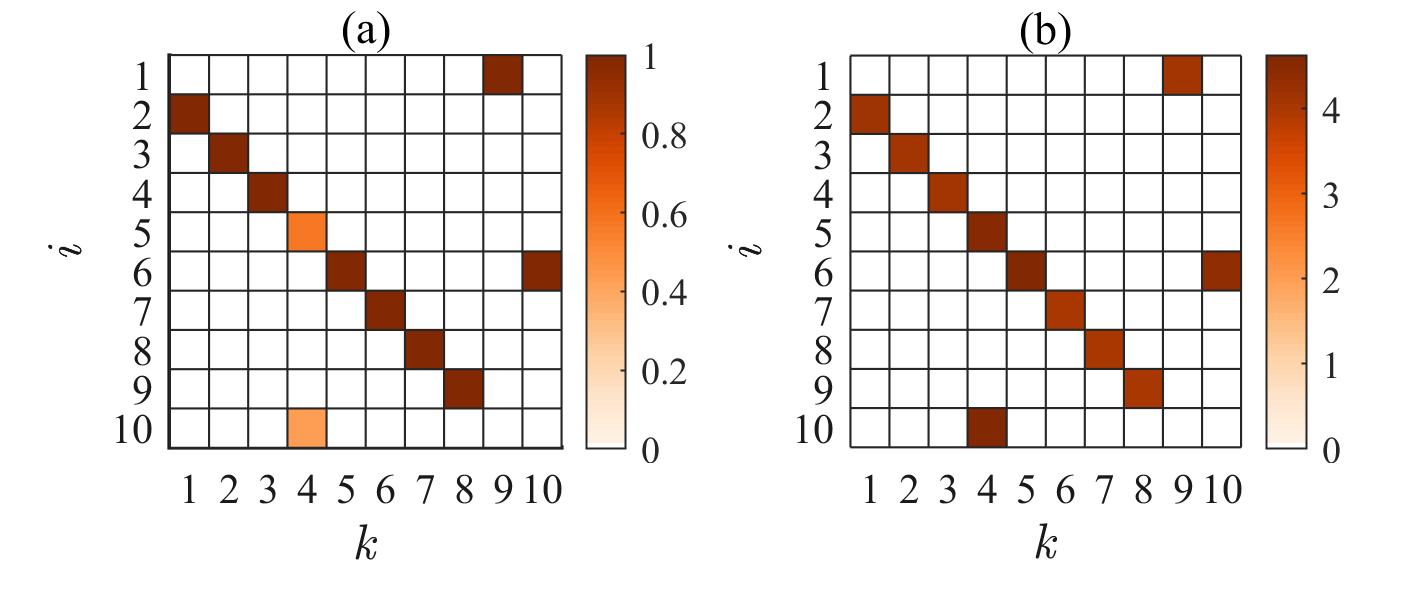}}
    \caption{Same as figure~\ref{Lorenz_matrixes}, but for the quasi-periodic flow at $\Rey = 330$.
    (a) Transition probability matrix $\mathbf{Q}$. 
    (b) Transition time matrix $\mathbf{T}$.}
\label{Re330_matrixes}
\end{figure}

The cluster transition matrices of the quasi-periodic flow regime are illustrated in figure~\ref{Re330_matrixes}.
The quasi-periodic dynamics are evident from $\mathbf{Q}$, which displays dominant transition probabilities corresponding to deterministic cyclic behaviour and minor wandering transitions signifying inter-cycle variations.
Cluster $\mathcal{C}_{4}$ serves as the transition cluster with two destination clusters, $\mathcal{C}_{5}$ and $\mathcal{C}_{10}$.
The two destination clusters have similar transition probabilities since they are visited for comparable times during the quasi-periodic transitions.
The transition cluster bridges the deterministic cluster chains $\mathcal{C}_{1} \to \mathcal{C}_{2} \to \mathcal{C}_{3} \to \mathcal{C}_{4}$ and $\mathcal{C}_{6} \to \mathcal{C}_{7} \to \mathcal{C}_{9} \to \mathcal{C}_{1}$ as two different limit cycles through two short cluster chains: $\mathcal{C}_{4} \to \mathcal{C}_{5} \to \mathcal{C}_{6}$ and $\mathcal{C}_{4} \to \mathcal{C}_{10} \to \mathcal{C}_{6}$. 
These two limit cycles alternate with a fixed order, ultimately forming an extended cluster chain that constitutes the fundamental elements of the long-term periodicity.
However, this characteristic is not effectively portrayed in the transition matrix. 
The purely probabilistic transitions from this matrix can result in arbitrary cluster transitions within the network model, introducing additional transition errors. 
Since the CNM relies on this cluster-level matrix, these transition errors present a notable challenge. 
While the transition tensors $\mathcal{Q}$, which resolve the refined centroid transitions, mitigate this issue, we further discuss the transition tensors and the corresponding centroid transition matrices in Appendix~\ref{app:d}. 
The time matrix $\mathbf{T}$ reveals that the transitions within a cyclic behaviour possess a generally similar time scale, with residence times in adjacent clusters changing smoothly, indicating the presence of a gradually evolving limit cycle.

\begin{figure}
    \centerline{\includegraphics[width=12cm]{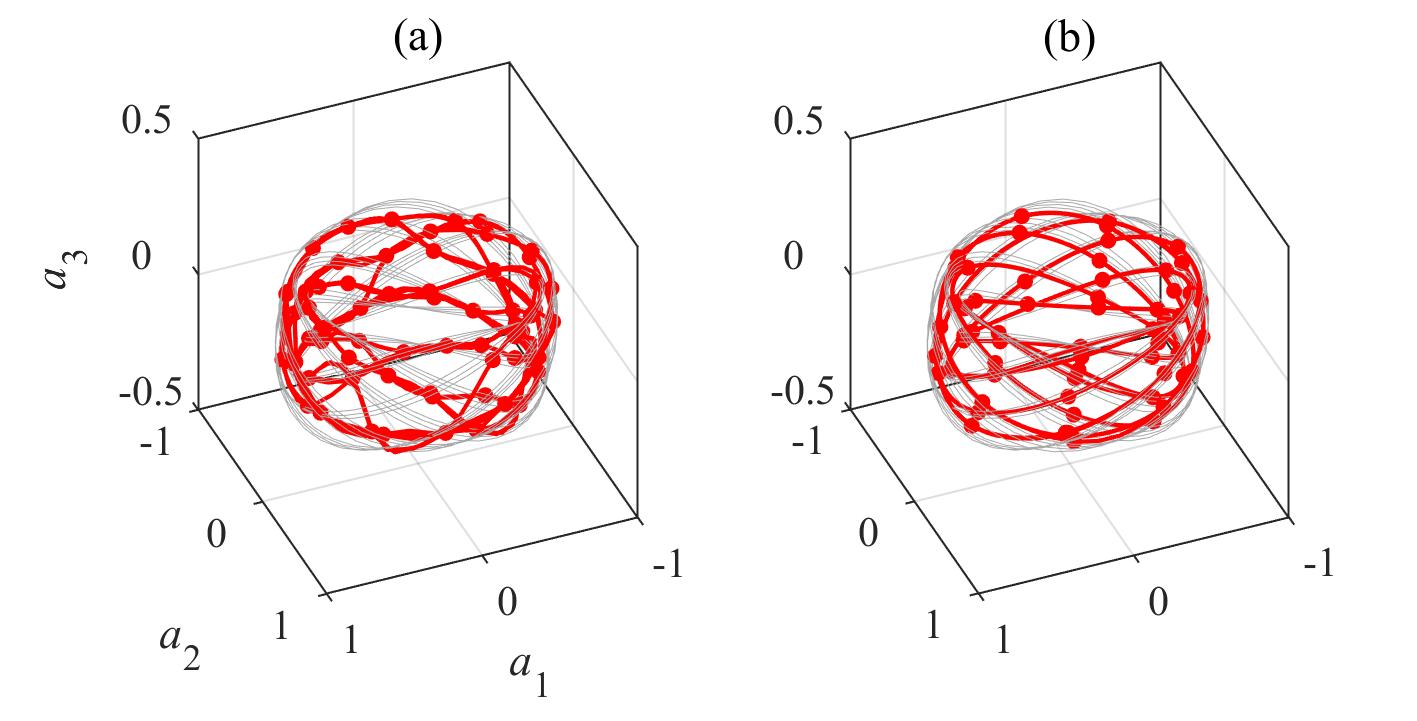}}
        \caption{Same as figure~\ref{Re300_reconstruction}, but for the quasi-periodic flow at $\Rey = 330$.
        (a) The CNM reconstruction and (b) the dCNM reconstruction are obtained with the same parameters as in figure~\ref{Re330_clustering}.}
\label{Re330_reconstruction}
\end{figure}

The original and reconstructed trajectories using the CNM and dCNM for the quasi-periodic flow regime are displayed in figure~\ref{Re330_reconstruction}.
The reconstruction is achieved with the same parameters as in figure~\ref{Re330_clustering}.
As anticipated, the trajectory reconstructed by the CNM undergoes substantial deformation, featuring discontinuous cyclic behaviours and a serrated trajectory. 
Conversely, the dCNM produces cleaner cyclic behaviours with more noticeable variations. 
The reconstructed trajectory accurately replicates the intersecting limit cycles and guides the inter-cycle transition with reduced spatial errors. 
These observations highlight the ability of the dCNM centroids to capture significant dynamics without assuming any prior knowledge of the data set.
A kinematic comparison between the POD reconstruction and the dCNM reconstruction is presented in Appendix~\ref{app:e}.

\begin{figure}
    \centerline{\includegraphics[width=10cm]{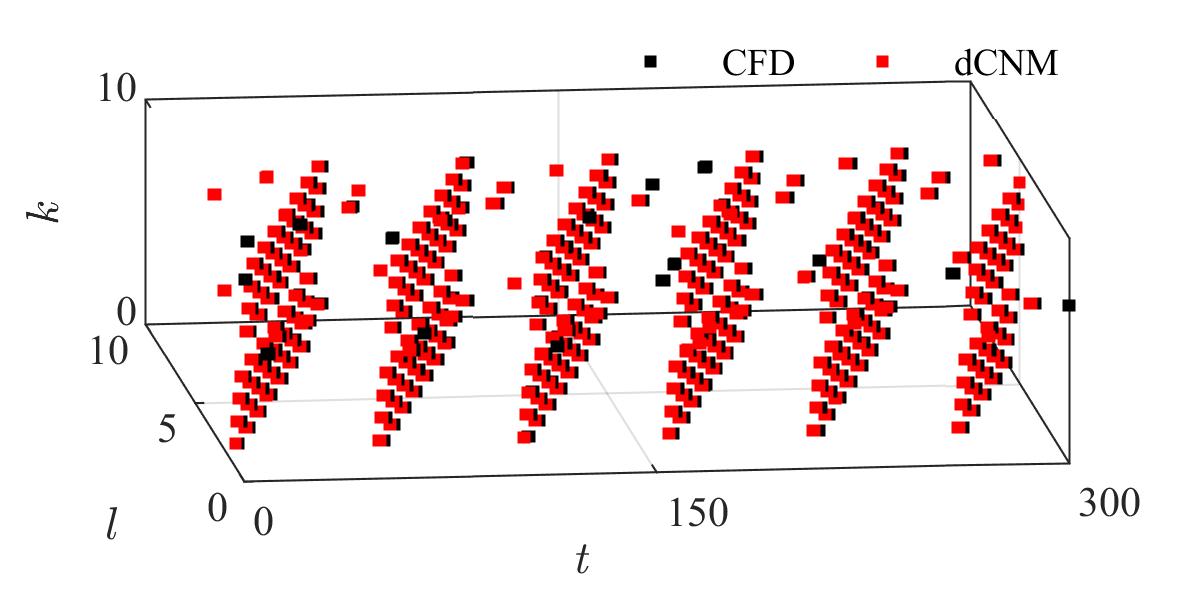}}
    \caption{Transition illustrated with the temporal evolution of the cluster and trajectory segment affiliation of the quasi-periodic flow at $\Rey = 330$.
    The vertical direction represents the cluster-level transition and the horizontal direction represents the trajectory segments inside this cluster.
    The transition with black markers represents the CFD data, and the transition with red markers represents the reconstructed dynamics by the dCNM.
    The $x$ axis is the non-dimensionalized time $t$, the $y$ axis is the trajectory segment affiliation $l$, and the $z$ axis is the cluster affiliation $k$. 
    The reconstruction is achieved from the same parameters as in figure~\ref{Re330_clustering}.}
\label{Re330_idx}
\end{figure}

In the following sections, we shift our focus to the temporal aspects.
The CNM uses the transition matrices to predict the next destination state for each step.
The quasi-periodic feature will be obscured by the stochastic transition probability matrix and the missing historical information.
In contrast, the dCNM preserves the transition sequence by embedding the sub-clusters (see Appendix~\ref{app:d}).
Initially, we explore the cluster and trajectory segment affiliation for each snapshot in both the original data set and the dCNM reconstruction to illustrate the accuracy of transition dynamics, as depicted in figure~\ref{Re330_idx}.
We maintain the same parameters as those used in figure~\ref{Re330_clustering} for the reconstruction.
The affiliation of the original data reveals that the dual clustering effectively represents the quasi-periodic dynamics.
The transition dynamics exhibit significant regularity, with centroids being sequentially and periodically visited, confirming deterministic transitions. 
Each period of centroid visits corresponds to an extended cluster chain, encompassing multiple centroid orbits and capturing cycle-to-cycle variations.
The periodic visits of these extended cluster chains are instrumental in determining the long-timescale periodicity.
These transition characteristics are fully preserved by the dCNM due to the dual indexing constraint. 
In this case, each departure centroid corresponds to only one destination centroid, eliminating the stochastic transition in the model and mitigating the transition errors.

\begin{figure}
    \centerline{\includegraphics[width=8cm]{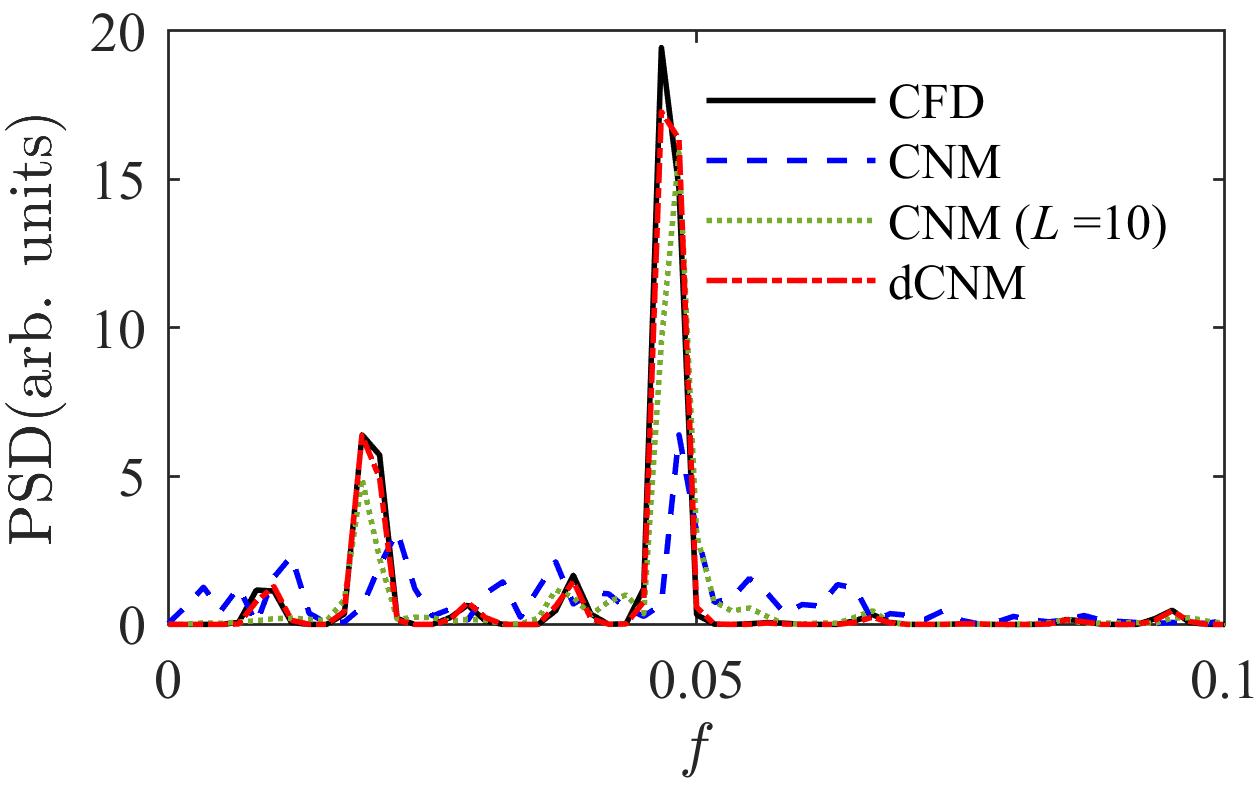}}
    \caption{The envelope spectrum of the streamwise fluctuation velocity $u'_{x}$, obtained by the surface average in $x=5D$.
    The dCNM reconstruction is achieved with the same parameters as in figure~\ref{Re330_clustering}, and the CNM reconstruction is achieved with the same number of centroids as the dCNM}
%    The envelope spectrum of the streamwise fluctuation velocity $u'_{x}$, obtained by the surface average in $x=5D$.
%    (a) CFD.
%    (b) The CNM reconstruction with the same number of centroids as the dCNM.
%    (c) The high-order CNM reconstruction with $L = 10$ and the same number of centroids as the dCNM.
%    (d) The dCNM reconstruction achieved with the same parameters as in figure~\ref{Re330_clustering}.}
\label{Re330_envelop}
\end{figure}

Envelope demodulation can clearly reveal the long-timescale behaviour and is more efficient in reflecting the quasi-periodic dynamics.
We analyse the envelope spectrum of the streamwise fluctuation velocity $u'_{x}$ from the data set, CNM, high-order CNM, and dCNM, as depicted in figure~\ref{Re330_envelop}.
The spectrum of the data set exhibits a dominant frequency $f=0.05$, representing long-timescale periodicity.
However, the CNM spectrum shows significant noise and lacks a clear dominant frequency due to frequent transition errors.
This observation supports the CNM's limitation in capturing multifrequency dynamics effectively.
By incorporating the historical information, the high-order CNM demonstrates superior performance by producing a cleaner spectrum closely aligned with the CFD data's dominant frequency.
Remarkably, the dCNM outperforms all other models by precisely reconstructing both the frequency and amplitude while minimising noise.

%%%%%%%%%%%%%%%%%%%%%%%%%%%%%%%%%%%%%%%%%%
\subsection{The chaotic flow regime at \texorpdfstring{$\Rey = 450$}{Lg}}
\label{sec4.4}
\begin{figure}
    \centerline{\includegraphics[width=12cm]{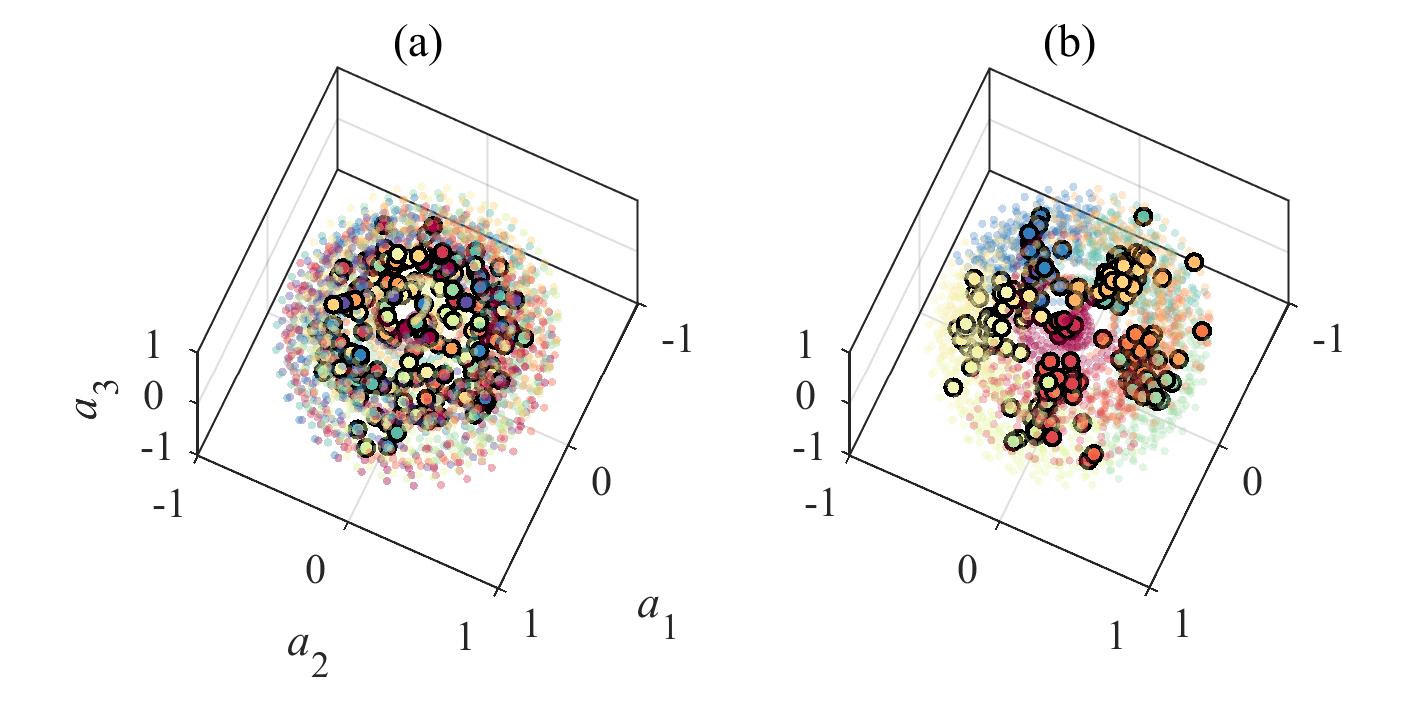}}
    \caption{Same as figure~\ref{Re300_clustering}, but for the chaotic flow of the sphere wake at $\Rey = 450$.
    (a) The CNM result with the same number of centroids as the dCNM.
    (b) The dCNM result with $K = 10$ and $\beta = 0.40$.}
\label{Re450_clustering}
\end{figure}

The comparison between the CNM and dCNM clustering with $K = 10$ and $\beta = 0.40$ for the chaotic flow are illustrated in figure~\ref{Re450_clustering}.
This value of $\beta$ is the sweet point between model complexity and model accuracy for this test case. Discussions with different values of $\beta$ for this flow regime are presented in Appendix~\ref{app:c}.
The normalized transverse cluster size vertor $\boldsymbol{\hat{R}}^{\mathcal{T}} = [0.1068,0.1073,0.1063,0.1068,0.1061,0.0966,0.1001,0.0954,0.0984,0.0763]^{\intercal}$, corresponds to the number of sub-clusters $\boldsymbol{L} = [22,22,22,22,22,20,21,20,20,8]^{\intercal}$.
%
%Similar to the quasi-periodic flow regime, the clustering results of the CNM and dCNM with $K = 10$ and $\beta = 0.5$  for the chaotic flow regime are illustrated in figure~\ref{Re450_clustering}.
%The results with different $\beta$ for this flow regime are also presented in appendix~\ref{app:c}.
As the dynamics become more complex, the snapshots form a chaotic cloud, which is driven by numerous cyclic behaviours of different scales and indicates irregular three-dimensional vortex shedding. 
The CNM continues to cluster the data set primarily based on spatial properties, essentially dividing the chaotic cloud into different segments in an evenly distributed manner. 
Figure~\ref{Re450_clustering} (a) illustrates this process, with the uniformly spread centroids capturing only part of one whole cyclic behaviour, limiting their ability to resolve the multiscale dynamics.
The dCNM centroids concentrate in regions of rich dynamics, enabling a more comprehensive resolution of the cyclic behaviours. 
These centroids, in various combinations, form the basis of multi-frequency and multiscale cyclic behaviour. 
Even after sparsification, the dCNM centroids can encompass a significant amount of scale diversity by merging only those that are spatially close to each other.

\begin{figure}
    \centerline{\includegraphics[width=12cm]{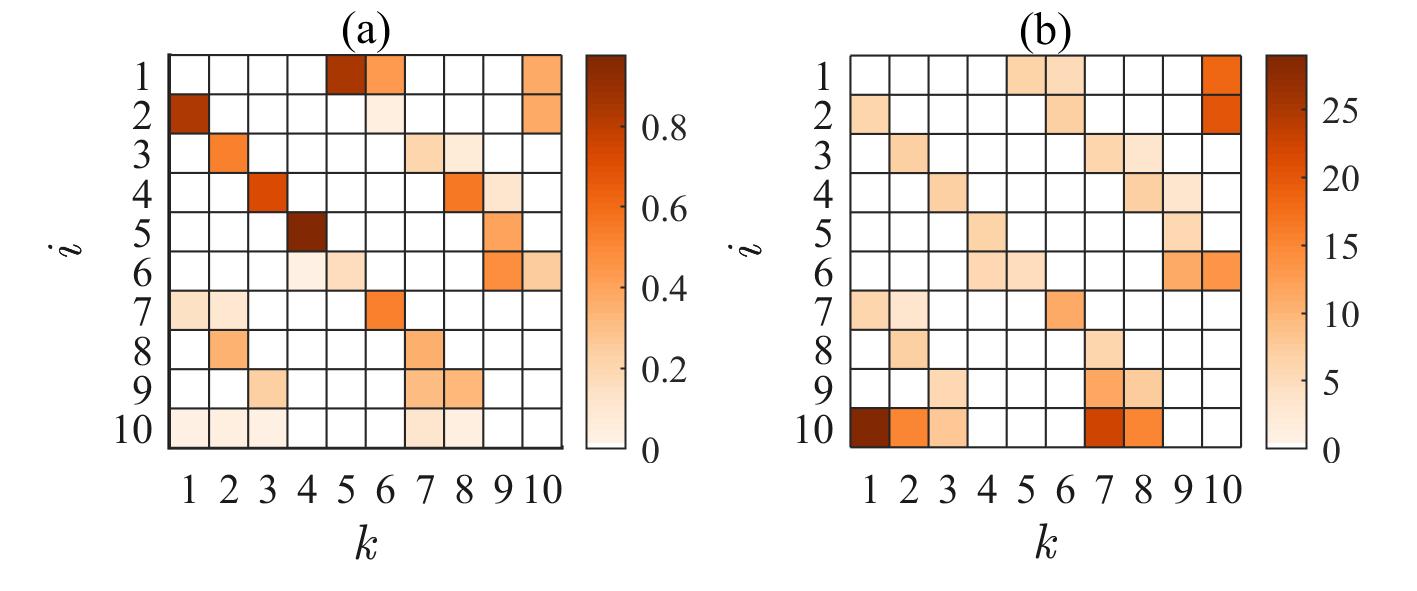}}
    \caption{Same as figure~\ref{Lorenz_matrixes}, but for the chaotic flow at $\Rey = 450$.
    (a) Transition probability matrix $\mathbf{Q}$.
    (b) Transition time matrix $\mathbf{T}$.}
\label{Re450_matrixes}
\end{figure}

The cluster transition matrices of the chaotic flow are illustrated in figure~\ref{Re450_matrixes}.
The probability matrix $\mathbf{Q}$ in figure~\ref{Re450_matrixes} (a) shows that most of the clusters have three or more destination clusters, indicating complex transition dynamics among them. 
Several dominant transition loops are identifiable, such as the large-size cluster chain: $\mathcal{C}_{1} \to \mathcal{C}_{2} \to \mathcal{C}_{3} \to \mathcal{C}_{4} \to \mathcal{C}_{5} \to \mathcal{C}_{6} \to \mathcal{C}_{1}$, the mid-size cluster chains: $\mathcal{C}_{1} \to \mathcal{C}_{2} \to \mathcal{C}_{3} \to \mathcal{C}_{4} \to \mathcal{C}_{5} \to \mathcal{C}_{1}$ and $\mathcal{C}_{3} \to \mathcal{C}_{4} \to \mathcal{C}_{5} \to \mathcal{C}_{6} \to \mathcal{C}_{7} \to \mathcal{C}_{3}$, and the small-size cluster chain: $\mathcal{C}_{6} \to \mathcal{C}_{7} \to \mathcal{C}_{8} \to \mathcal{C}_{6}$. 
These cluster chains with different lengths represent cyclic loops at different scales. 
The small number of chains facilitates human understanding of the transition dynamics but is insufficient for accurately capturing the dynamics. 
The dominant loops have their key transition clusters inside, from which they can randomly jump into each other by choosing periodic or stochastic routes.
This is where the transition error often occurs. 
The time matrix $\mathbf{T}$ in figure~\ref{Re450_matrixes} (b) shows the difference in the transition times between different types of transitions. 
Regarding the main loops, the time scale changes smoothly within its transitions.
However, for the jumps between the loops, the time scale fluctuates considerably, and some transitions can be very large, showing the diversity of the dynamics.
Moreover, this observation implies that the distribution density of snapshots differs among clusters. 
In other words, the distribution of the trajectory segments in different clusters also exhibits significant variations. 
This explains the necessity for determining the number of sub-clusters in the second-stage clustering based on the deviation $\boldsymbol{R}^{\mathcal{T}}$.
The refined transition matrices between the centroids are shown in Appendix~\ref{app:d}.

\begin{figure}
    \centerline{\includegraphics[width=12cm]{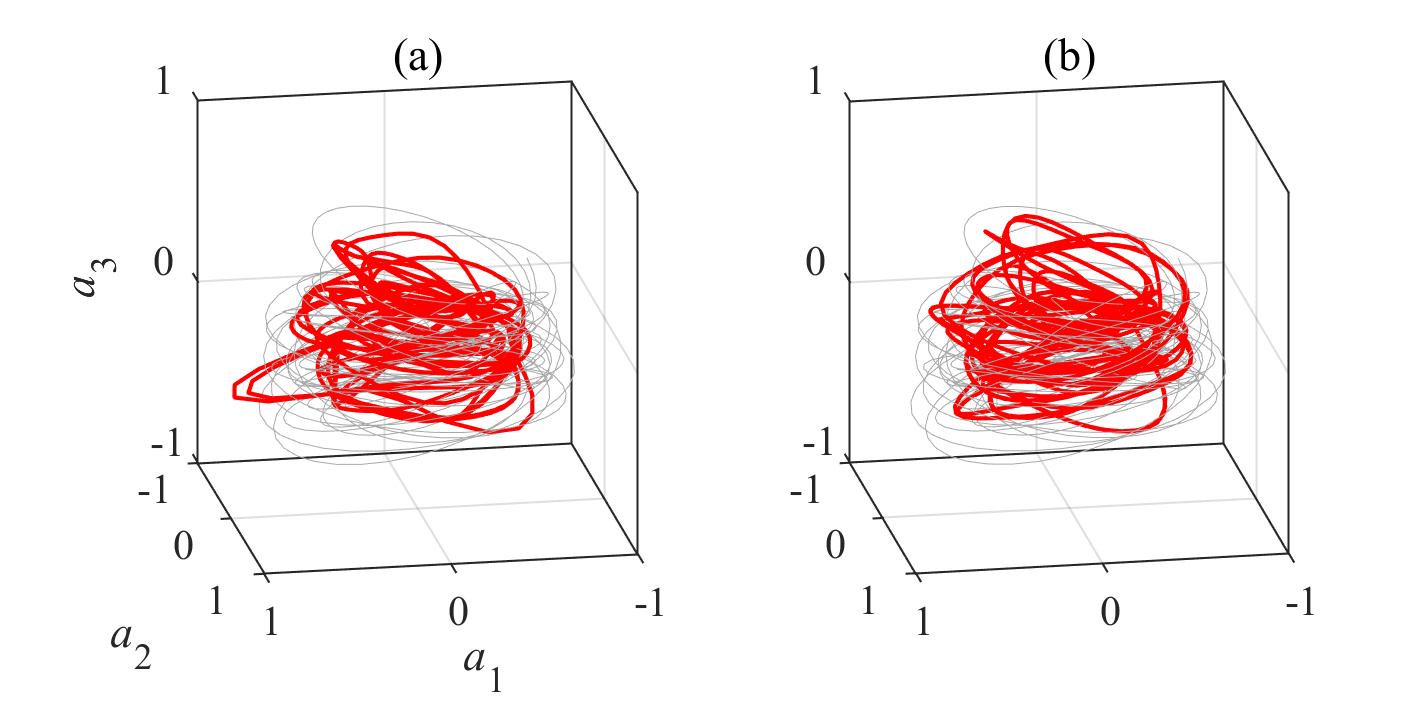}}
        \caption{Same as figure~\ref{Re300_reconstruction}, but for the chaotic flow at $\Rey = 450$.
        (a) The CNM reconstruction and (b) the dCNM reconstruction are obtained from the same parameters as in figure~\ref{Re450_clustering}.}
\label{Re450_reconstruction}
\end{figure}

The original and reconstructed trajectories by the CNM and dCNM for this flow regime are shown in figure~\ref{Re450_reconstruction}.
The reconstruction is achieved with the same parameters as in figure~\ref{Re450_clustering}.
For a clear visualisation, only the trajectories from the first half of the entire time window are plotted.
This selection suffices to analyse the precision of the current trajectory, as it contains ample dynamics. 
We exclude trajectory discrepancies triggered by phase mismatch and focus exclusively on the accuracy of the present trajectory.
In the case of the CNM, noticeable disparities exist between the original trajectory and the reconstructed trajectory. 
These differences include variations in the shape, spatial location, and inclination angle of the cyclic loops. 
These disparities can be attributed to the elimination of small-scale structures and the blending of certain large-scale structures due to the uniform distribution of centroids.
Regarding the dCNM, the reconstructed trajectory nearly occupies the entire chaotic cloud, closely resembling the original trajectory. 
The external and internal geometries are accurately reproduced, capturing both large-scale and small-scale structures. 
However, despite the improved accuracy, some deformations persist. 
These deformations arise from the interpolations between the limited centroids during one single cyclic loop. 
Notably, due to its complexity, achieving a superior reconstruction of a chaotic system often requires more refined centroids compared to a quasi-periodic system.
The kinematic comparison with the POD reconstruction is also introduced in Appendix~\ref{app:e}.

Figure~\ref{Re450_autocorrelation} shows the auto-correlation function of the CNM, high-order CNM, and dCNM.
\begin{figure}
    \centerline{\includegraphics[width=14cm]{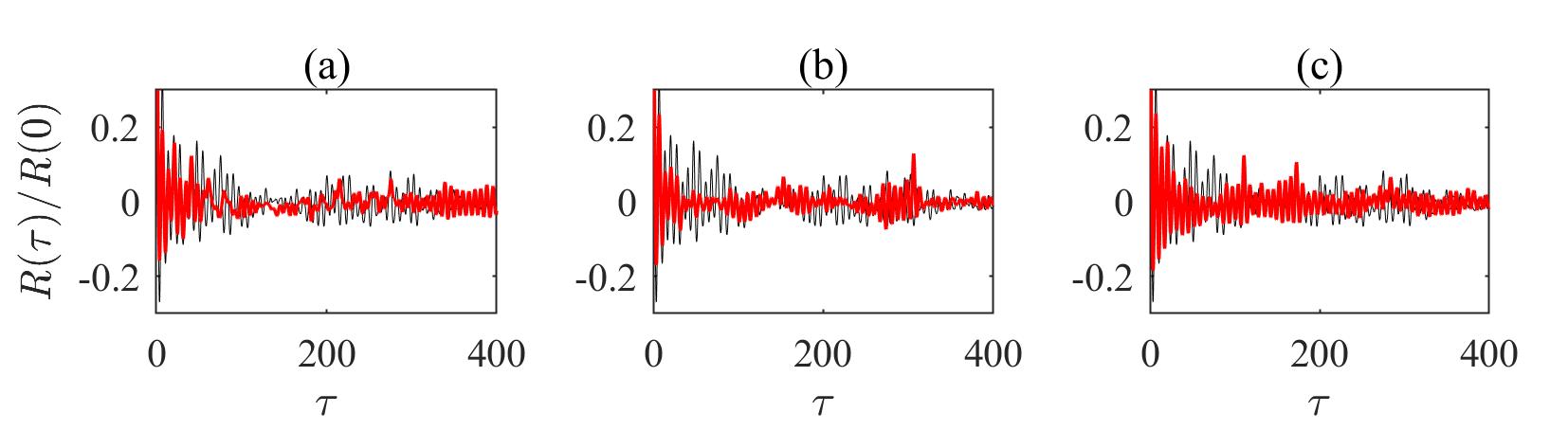}}
    \caption{Auto-correlation function of the chaotic flow at $\Rey = 450$; here $R$ is normalised by $R(0)$.
    The thin black curve represents the CFD data, and the thick red curve represents the reconstruction from different models.
    (a) The CNM reconstruction with the same number of centroids as the dCNM.
    (b) The high-order CNM reconstruction with $L = 10$ and the same number of centroids as the dCNM.
    (c) The dCNM reconstruction achieved with the same parameters as in figure~\ref{Re450_clustering}.}
\label{Re450_autocorrelation}
\end{figure}
We still normalise this function by $R(0)$, and the time window is chosen from $t = 0$ to $t = 400$, which is sufficient for comparison.
For the chaotic flow regime, $R(\tau) / R(0)$ denotes the kinetic energy level of the time window. 
Nonetheless, a notable discrepancy arises in the CNM, where the amplitude experiences a distinct decay after the initial few periods. 
It eventually stabilises with minimal variation, primarily due to the distorted reconstructed trajectory and transition errors. 
This limited variance is indicative of inaccuracies in capturing short-term dynamics, consistent with the absence of historical information.
The high-order CNM, which incorporates this historical information, outperforms the CNM in this regard. 
Its amplitude decays gradually and exhibits variance akin to that of the data set. 
Additionally, it reveals some peaks with similar time delays, due to the potential introduction of unnecessary long-timescale periodicity into the reconstruction via the high-order cluster chain.
The dCNM also surpasses the CNM with regard to accuracy. 
Both the amplitude and phase are faithfully retained, with a gradual amplitude decay and more pronounced variation.
Eventually, the amplitude diminishes, similar to the original data set.
As $\tau$ increases, all three models exhibit some degree of phase delay or lead. 
This is a consequence of averaged transition times introducing some errors into the model \citep{li2021cluster}.

%%%%%%%%%%%%%%%%%%%%%%%%%%%%%%%%%%%%%%%%%%
\subsection{Physical interpretation}
\label{sec4.5}
One of the major advances of the cluster-based model is its strong physical interpretability. 
The dCNM also maintains and even enhances this nature while improving the model accuracy.
In this section, we discuss the physical interpretation of the CROM exemplified for the sphere wake, with particular emphasis on the dCNM.

The cluster-based model spatially coarse-grains the snapshots into groups and represents them by centroids to reduce dimensionality.  
In contrast to the projection-based methodology, such as the POD-Galerkin model, the cluster-based model uses cluster centroids which are linear combinations of several snapshots, and thus reflects the representative patterns.
This feature contributes to its high interpretability.
The snapshot dynamics are mapped into the pattern dynamics, followed by the construction of a probabilistic mechanism to reduce temporal dimensionality. 
The network model, with centroids as nodes and centroid transitions as edges, converts the complex dynamics into pure data analysis.
The centroids act as a bridge between the data-driven model and its underlying physical background.
Furthermore, it is conceivable that the same model can be easily transferred to analogous pattern dynamics, even with distinct backgrounds, through adjustments to the centroids' backstory. 

The sphere wake offers a concise physical interpretation based on the centroids. 
The coherent structure evolution governs the flow field and manifests as vortex shedding events with diverse dynamics. 
These shedding events can be captured well by a limit cycle, with a set of centroids representing flow patterns at different shedding phase as foundational elements. 
The cyclic transitions between these specific flow patterns collectively characterise the entire shedding process. 
The deterministic-stochastic transitions between different shedding events contribute to the overall periodic-chaotic dynamics.

To explain the physical mechanisms of the flow regimes, we propose a chord transition diagram for the cluster transitions and sub-cluster transitions along with centroid visualisation, which provides a comprehensive view of the flow regime.
%To explain the physical mechanisms of the flow regimes, we propose a chord transition diagram along with centroid visualisation, which provides a comprehensive view of the flow regime.
{We start with the periodic flow, as shown in figure~\ref{Re300_Chord}.
\begin{figure}
\centering
    \includegraphics[width=6.5cm]{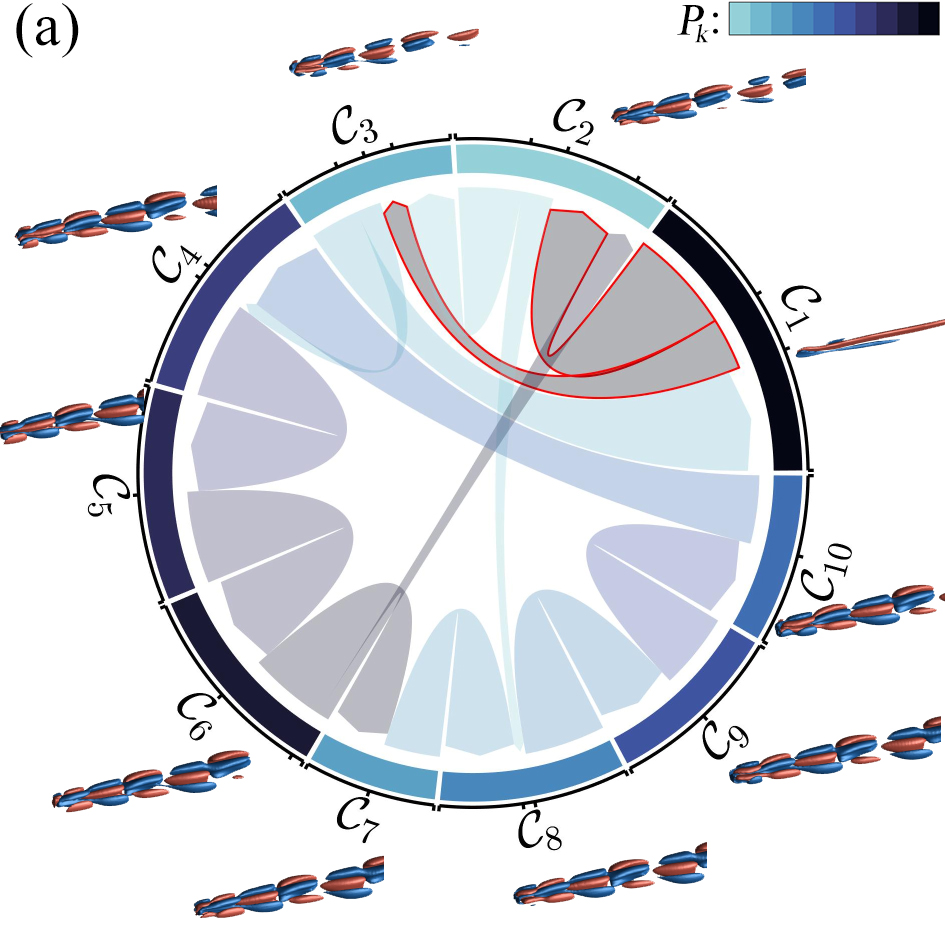}
    \includegraphics[width=6.5cm]{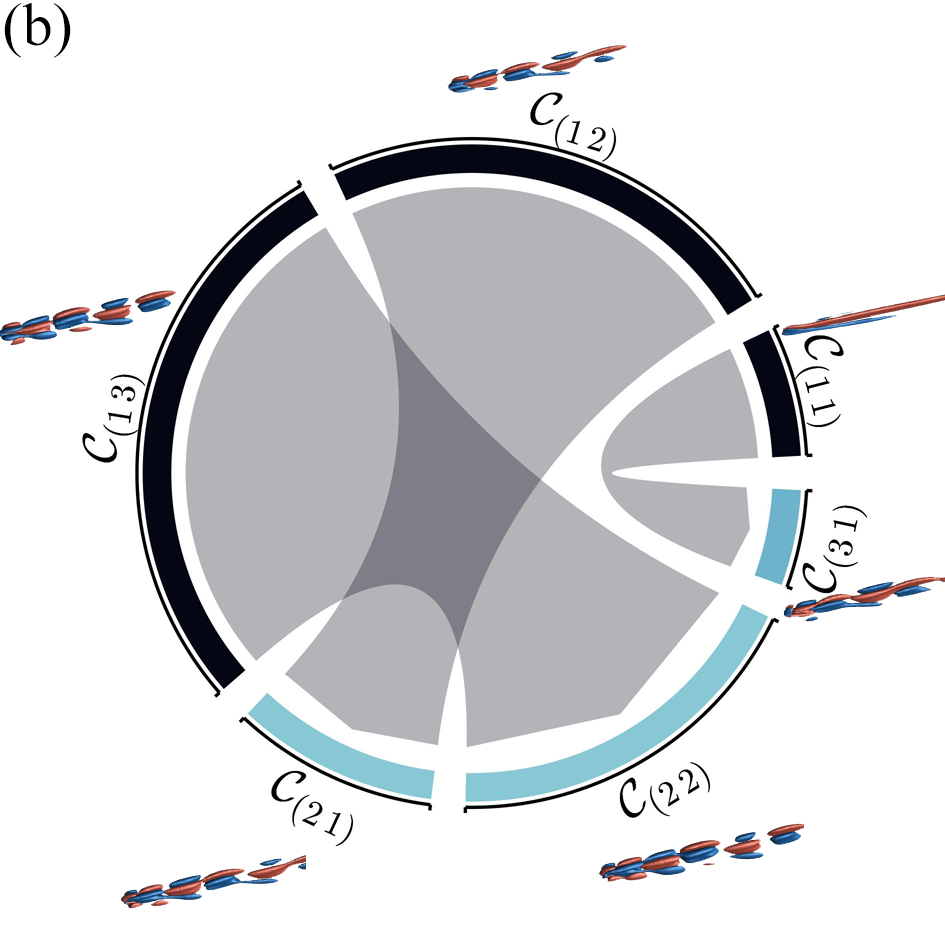}  
    \caption{Transition diagram of the quasi-periodic flow at $\Rey = 300$.
    The centroids are depicted by the vortex distribution.
    The vortices are identified by the iso-surfaces of $z$-vorticity, with $-1$ for the negative vortices coloured in blue and $1$ for the positive vortices coloured in red.
    The transition dynamics are depicted by the directed arrows, the size of the arrow tail represents the transition probability and the colour is consistent with the departure block.
    (a) Cluster transitions. Different blocks represent different clusters, the colour of the block represents the corresponding cluster probability distribution $P_{k}$, and the size of the block represents the cluster size $\boldsymbol{R^u}$.
    (b) Sub-cluster transitions of $\beta = 0.50$, with transitions specifically departing from $\mathcal{C}_{1}$, corresponding to the red-bordered cluster transition in (a). 
    Blocks with the same colour belong to the same cluster, the colour still represents the cluster probability distribution $P_{k}$, and the size of each block represents the sub-cluster size $\boldsymbol{R}^{\boldsymbol{u}}_{\mathrm{sub}}$.
    }
\label{Re300_Chord}
\end{figure}
The cluster probability distribution $P_{k}$ and the cluster size $\boldsymbol{R^u}$ used for visualising the blocks in figure~\ref{Re300_Chord} (a) are shown in figure~\ref{Re300_property} (a) and (b).
The blocks in figure~\ref{Re300_Chord} (b) are split based on the sub-clusters, the transverse cluster size is shown in figure~\ref{Re300_property} (c) and the sub-cluster size is shown in figure~\ref{Re300_property} (d).
% Note that in this section $\beta$ is chosen to have fewer sub-clusters, which is more suitable for visualisation.
%}
%
\begin{figure}
    \centerline{\includegraphics[width=10cm]{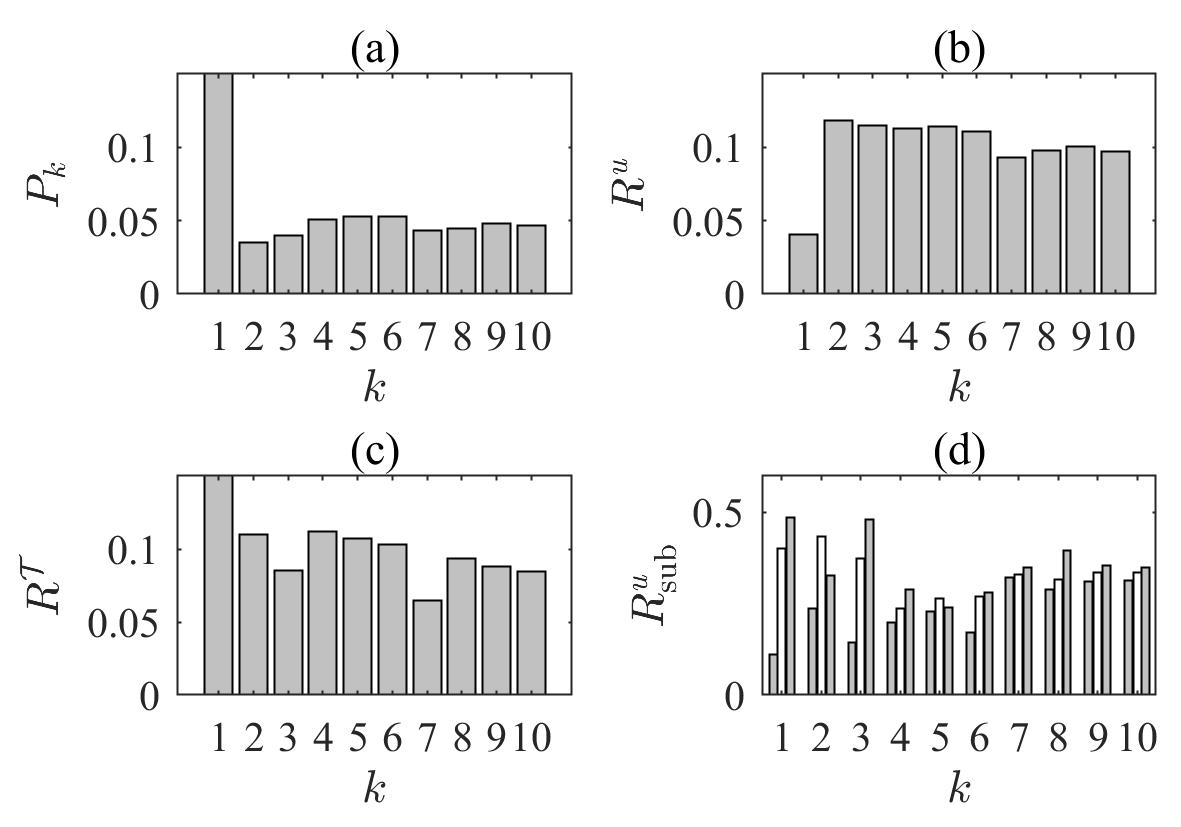}}
    \caption{Cluster and centroid properties of the periodic flow at $\Rey = 300$.
    (a) Cluster probability distribution.
    (b) Normalised cluster size.
    (c) Normalised transverse cluster size.
    (d) Normalised sub-cluster size, where the elements from the same cluster sum to unity.}
\label{Re300_property}
\end{figure}
The cluster transition diagram is capable of clearly distinguishing the dynamic behaviour categories.
The circumferential arrows along the boundary represent the cyclic behaviours, with centroids transitioning to adjacent destination centroids.
The radial arrows crossing the graph signify cycle-to-cycle transitions, with the centroids transitioning to non-adjacent destinations.
These arrows usually originate from the transition clusters.
The number of radial arrows indicates the dynamic characteristics, with more arrows indicating more chaotic features.

The cluster dynamics is illustrated in figure~\ref{Re300_Chord} (a).
The limit cycle is captured by the clusters $\mathcal{C}_4$ to $\mathcal{C}_{10}$, shown as the deterministic transitions between adjacent clusters.
The transient phase is resolved by clusters from $\mathcal{C}_1$ to $\mathcal{C}_3$.
However, the stochastic transitions between the first three clusters are in contrast to the slowly varying amplitude in the transient state and are insufficient to represent this deterministic process.
The distinct vortex structures of the three centroids also indicate a need for higher resolution.
The sub-cluster transitions leaving from $\mathcal{C}_1$ are shown in figure~\ref{Re300_Chord} (b).
The sub-cluster centroids manifest as varying vortex structures, corresponding to the growing amplitudes.
Each sub-cluster has only one destination, resulting in deterministic transitions.
The stochastic cluster transitions from $\mathcal{C}_1$ to $\mathcal{C}_2$ and $\mathcal{C}_3$ are now terminated into a chain of deterministic sub-cluster transitions, effectively reducing the prediction error.

\begin{figure}
    \centering
    \includegraphics[width=6.5cm]{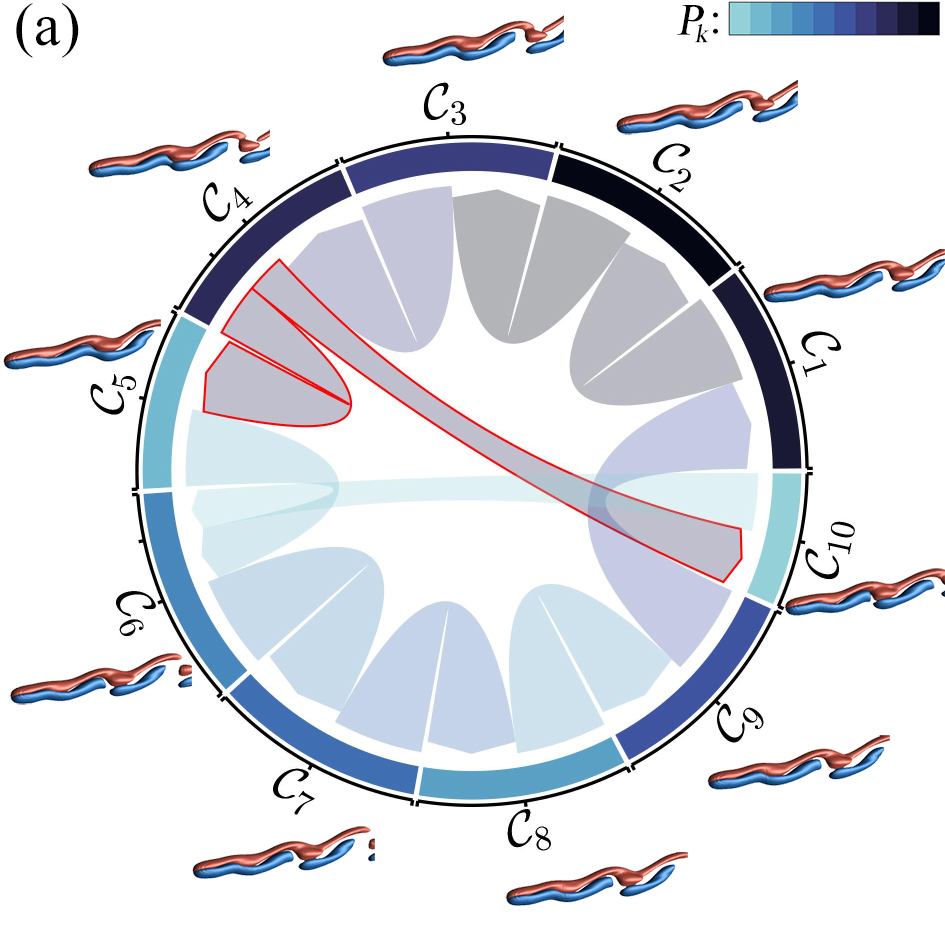}
    \includegraphics[width=6.5cm]{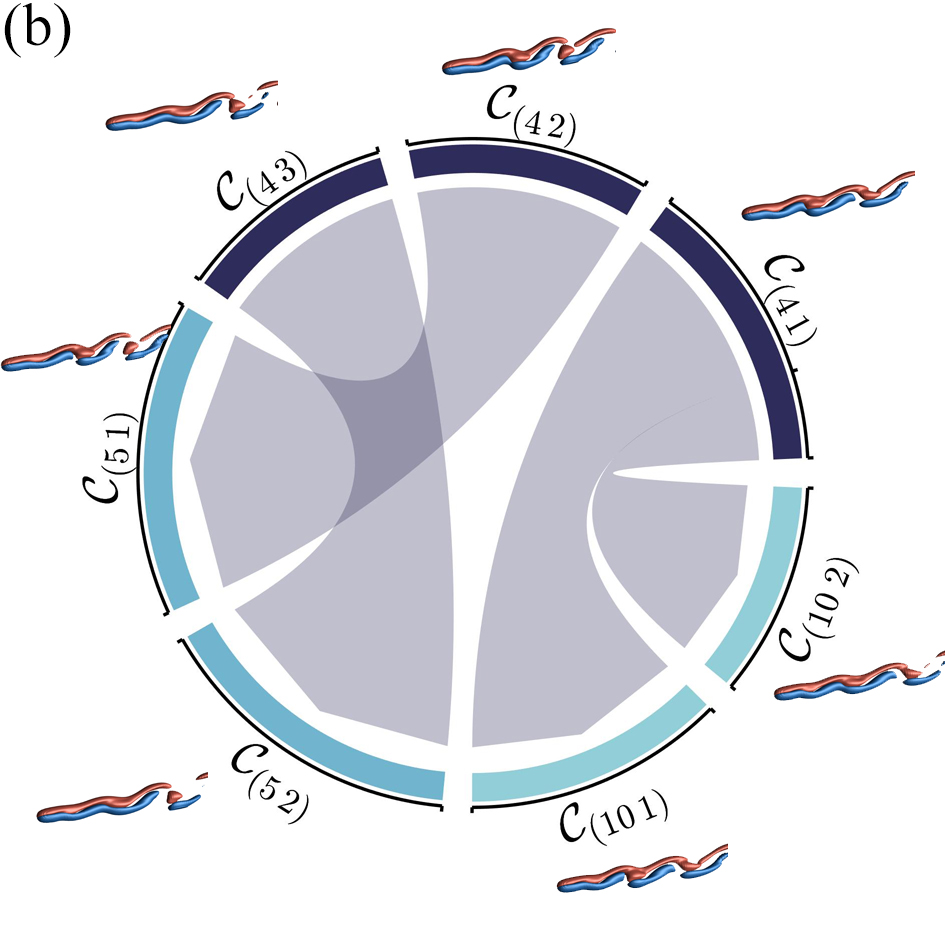}
    \caption{Same as figure~\ref{Re300_Chord}, but for the quasi-periodic flow at $\Rey = 330$.
    (a) Cluster transitions.
    (b) Sub-cluster transitions of $\beta = 0.95$, with transitions specifically departing from $\mathcal{C}_{4}$, corresponding to the marked cluster transition in (a). 
%    Transition diagram of the quasi-periodic flow at $\Rey = 330$.
%    The cluster centroids are depicted by the vortex distribution.
%    The vortices are identified by the iso-surfaces of $z$-Vorticity, with $-1$ for the negative vortices coloured in blue and $1$ for the positive vortices coloured in red.
%    The transition dynamics are depicted by the directed arrows, the size of the arrows represents the transition probability and the colour is consistent with the departure block.
%    (a) Cluster transitions. Different blocks represent different clusters, the colour of the block represents the corresponding cluster probability distribution $P_{k}$, and the size of the block represents the cluster deviation on the snapshots $\boldsymbol{R^u}$.
%    (b) Sub-cluster transitions with $\beta = 0.95$, departuring from $\mathcal{C}_{4}$. 
%    Here the blocks are split based on the second-stage clustering results.
%    Blocks with the same colour belong to the same cluster, the colour still represents the cluster probability distribution $P_{k}$, and the size of the block represents the sub-cluster deviation on the snapshots $\boldsymbol{R}^{\boldsymbol{u}}_{\mathrm{sub}}$.
%    The transition dynamics of dCNM are more detailed while still exhibiting the same rhythm as that of the CNM, which is intuitive for understanding.
    }
\label{Re330_Chord}
\end{figure}
\begin{figure}
    \centerline{\includegraphics[width=10cm]{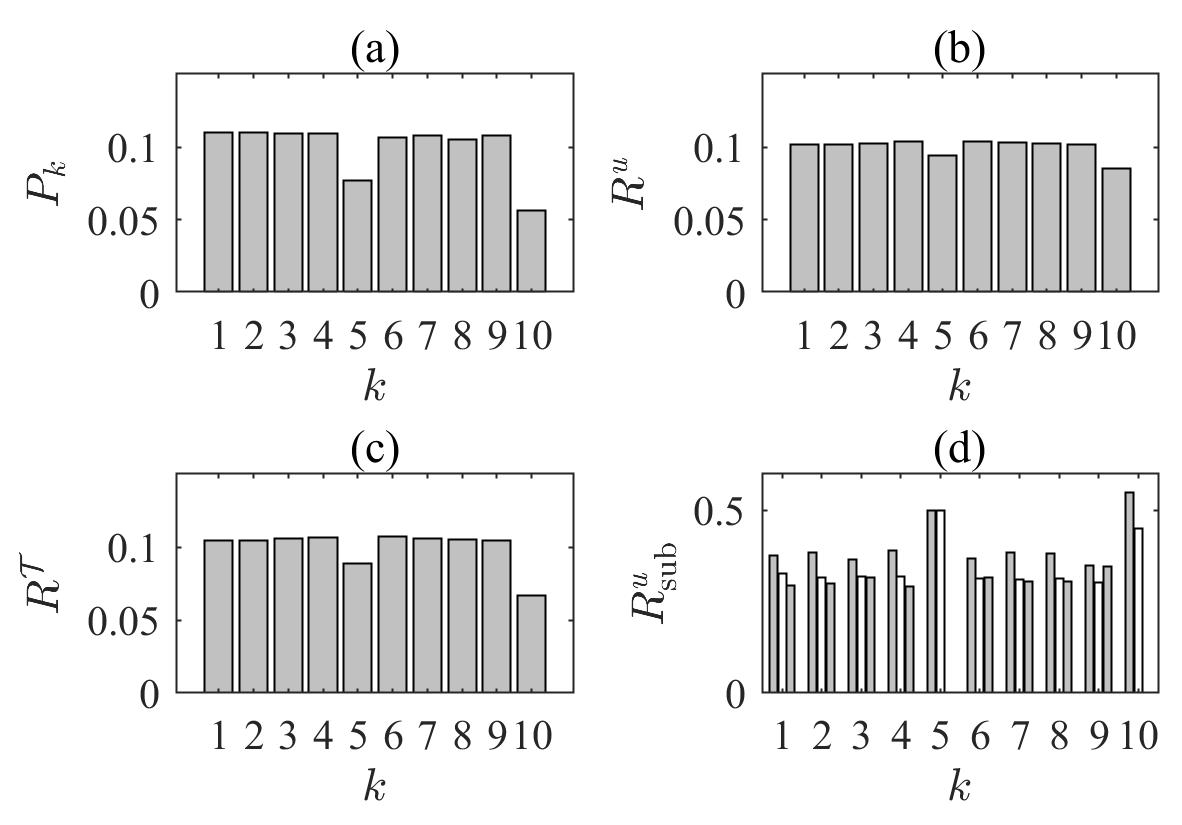}}
    \caption{Same as figure~\ref{Re300_property}, but for the quasi-periodic flow at $\Rey = 330$.
    (a) Cluster probability distribution.
    (b) Normalised cluster size.
    (c) Normalised transverse cluster size.
    (d) Normalised sub-cluster size, where the elements from the same cluster sum to unity.} 
\label{Re330_property}
\end{figure}
The transition graph of the quasi-periodic flow regime is shown in figure~\ref{Re330_Chord}, with the corresponding cluster and centroid properties given in figure~\ref{Re330_property}.
In figure~\ref{Re330_Chord} (a), the flow regime is characterised by the cyclic cluster transitions,  
with only three non-adjacent transitions, i.e., $\mathcal{C}_{4}$ to $\mathcal{C}_{10}$, $\mathcal{C}_{10}$ to $\mathcal{C}_{6}$, and $\mathcal{C}_{9}$ to $\mathcal{C}_{1}$.
The clusters involved in the cyclic transitions exhibit varying vortex structures within one shedding period. 
Their relatively higher probability distribution, as shown in figure~\ref{Re330_property} (a), suggests dominant flow patterns. 
For the bifurcating cluster $\mathcal{C}_{4}$, its destination clusters $\mathcal{C}_{5}$ and $\mathcal{C}_{10}$ manifest visible differences in the far wake.
Further distinction of transitions from $\mathcal{C}_{4}$ is provided by the sub-clusters, as shown in figure~\ref{Re330_Chord} (b).
The centroids belonging to the same cluster are roughly at the same shedding phase, but exhibit different vortex structures, exemplified by $\mathcal{C}_{4\, 1}$ and $\mathcal{C}_{4\, 2}$. 
This difference leads to distinct sheddings, such as $\mathcal{C}_{10\, 1}$, and $\mathcal{C}_{5\, 1}$.
Consequently, finer dynamic resolution originating from $\mathcal{C}_{4}$ is captured, enabling the deterministic transitions to $\mathcal{C}_{5}$ and $\mathcal{C}_{10}$, respectively.

The chaotic flow regime exhibits a more complex transition graph, as shown in figure~\ref{Re450_Chord}.
\begin{figure}
\centering
    \includegraphics[width=6.5cm]{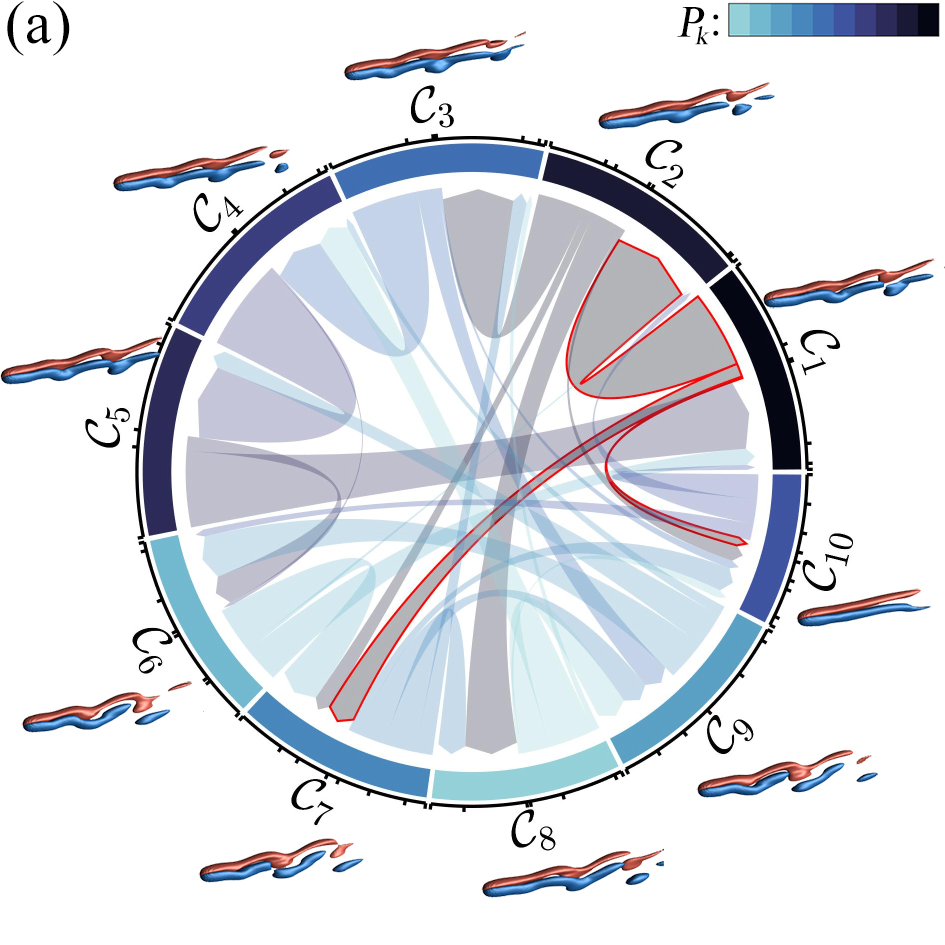}
    \includegraphics[width=6.5cm]{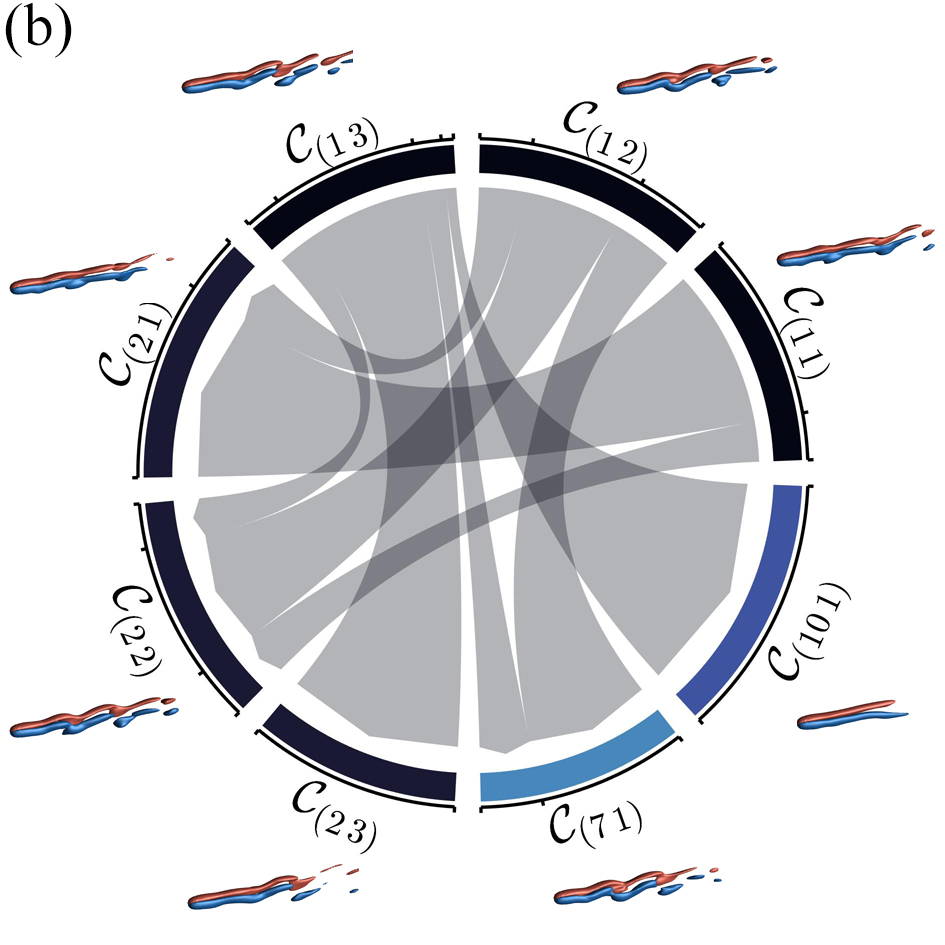}  
    \caption{Same as figure~\ref{Re300_Chord}, but for the chaotic flow at $\Rey = 450$.
    (a) Cluster transitions.
    (b) Sub-cluster transitions of $\beta = 0.95$, with transitions specifically departing from $\mathcal{C}_{1}$.}
\label{Re450_Chord}
\end{figure}
The relative information is illustrated in figure~\ref{Re450_property}.
\begin{figure}
    \centerline{\includegraphics[width=10cm]{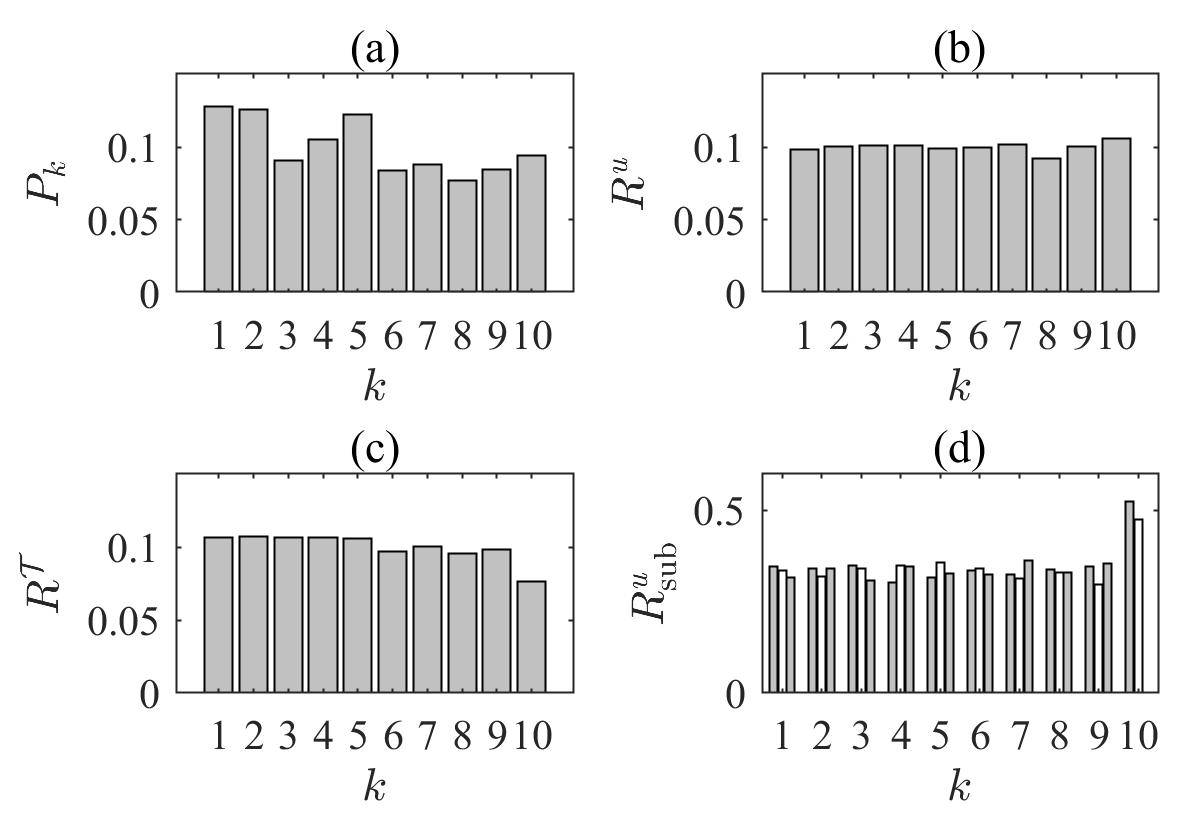}}
    \caption{Same as figure~\ref{Re300_property}, but for the chaotic flow at $\Rey = 450$.
    (a) Cluster probability distribution.
    (b) Normalised cluster size.
    (c) Normalised transverse cluster size.
    (d) Normalised sub-cluster size, where the elements from the same cluster sum to unity.}
\label{Re450_property}
\end{figure}
In figure~\ref{Re450_Chord} (a), similar to other flow regimes, adjacent cluster transitions continue to dominate the flow field, reflecting cyclic behaviours.
However, the increasing number of radial arrows with varying transition probabilities indicates chaotic features. 
Each centroid represents a distinct flow field with different scales of vortex structures, even within the same cyclic cluster transition.
This discrepancy indicates that the current flow patterns are inadequate for capturing the entire shedding dynamic.
Concerning the sub-cluster transitions in figure~\ref{Re450_Chord} (b), the increased arrows maintain the transition rhythm but offer more specificity. 
The flow states can be inferred from the vortex structures surrounding the cyclic diagram. 
The centroids within the same cluster also represent vortex structures sharing the same shedding phase but exhibiting different scales.
Only centroids with similar scales of vortex structures are connected by sub-cluster transitions.
The departing sub-clusters are thus restricted, for instance, $\mathcal{C}_{(2\,3)}$ and $\mathcal{C}_{(10\,1)}$ each have only one departing sub-cluster.
The diversity of the centroid transitions guarantees diverse flow scales, while simultaneously maintaining a consistent scale within the same cluster loop.
Therefore, the dCNM facilitates multi-scale fidelity and smoother cyclic behaviours, significantly enhancing the representation capacity of the model.

When comparing the cluster transition and the sub-cluster transition in dCNM, it is evident that the state space clustering can be seen to automatically introduce prior knowledge into the model.
%When comparing the CNM and dCNM, it is evident that with the trajectory segments, the state space clustering of the dCNM can be seen to automatically introduce prior knowledge into the model. 
This prior knowledge includes the inner-state kinematic information, as resolved by the trajectory segments, and the inter-state dynamic information, as resolved by the transitions between the trajectory segments.
The incorporation aids in the automatic assignment of refined centroids within each cluster and constrains the probabilistic transition dynamics.
In essence, the dCNM can be regarded as having a built-in unsupervised physics-informing process, which results in superior model accuracy.

%%%%%%%%%%%%%%%%%%%%%%%%%%%%%%%%%%%%%%%%%%%%%%%%%%%%%%%%%%%%%%%
\section{Conclusions and outlook}
\label{sec:conclusion}
We propose an automatable data-driven reduced-order model for nonlinear dynamics. This model can resolve the periodic, quasi-periodic and chaotic dynamics of the sphere wake featuring multi-frequency and multiscale behaviours.
%We propose an automatable data-driven reduced-order model for nonlinear dynamics. This model can resolve the quasi-periodic and chaotic dynamics of a three-dimensional sphere wake featuring multi-frequency and multiscale behaviours.
%
The starting point is the cluster-based network model (CNM) \citep{fernex2021cluster,li2021cluster}, which is an automated framework employing clustering and network science.
The dynamics within the CNM are described using a deterministic-stochastic approach on a network, where centroids act as nodes, and transitions serve as edges.
However, the clustering process in the CNM relies on a uniform geometric coverage of the snapshot data, agnostic of the temporal dynamic relevance.
For multi-frequency dynamics, this can result in large prediction errors. 
One example is the long transient to a limit cycle.
Here, the slow increase in the radius requires a finer resolution
than the robust angular dynamics.
Hence, the CNM can be expected to be more accurate
if the centroids are much denser in the radial direction than in the angular motion.
This idea is incorporated in the proposed dynamics-augmented CNM (dCNM).
The model can automatically stratify the state space along the trajectory direction.

The dCNM was applied to the Lorenz system (in \S~\ref{sec:Lorenz}) and the three-dimensional sphere wake (in \S~\ref{sec:sphere}), with $K=10$ clusters for the coarse-graining of the state space.
The Lorenz system features oscillatory dynamics, presented as two ``ears'' consisting of many unstable orbits, and stochastic dynamics, presented as random switching between the ``ears''.
For the future state, the phase can be accurately predicted, but the amplitude requires a higher resolution.
The CNM is only capable of reconstructing limited loops of the cyclic behaviours and their related transitions in the branching area. 
Non-physical radial jumps also occur due to transition errors.
On the other hand, the dCNM coarsely resolves the deterministic phases but accurately resolves the slowly varying amplitude.
The attractor oscillations are distinctly defined, and the transitions in the branching region are subsequently constrained.
For the transient and post-transient dynamics of the periodic sphere wake, the dCNM accurately resolves the slowly growing amplitude between the cyclic behaviours.
Regarding the quasi-periodic sphere wake, the dCNM successfully captures both the periodic behaviour and cycle-to-cycle variations. 
Notably, it discerns intrinsic deterministic transition behaviours, which are often misinterpreted as stochastic transitions by the CNM.
For the chaotic flow dominated by unstable periodic orbits with varying scales, the dCNM accurately distinguishes between these orbits and captures their transitions.
Even after sparsification, chaotic features remain preserved, with transition dynamics demonstrating stochastic characteristics.
Overall, these findings underscore the notable improvement of the dCNM in capturing and accurately representing multi-frequency and multiscale dynamics.

The dCNM offers several advantages over other reduced-order modelling strategies. 
It preserves the advantages of previous cluster-based approaches 
and adds new noteworthy features.
\begin{enumerate}
\item The prediction error is minimized.
The slow evolution of amplitude oscillations, the deterministic quasi-periodic dynamics, and the stochastic chaotic dynamics can be automatically resolved without any prior knowledge.
\item The model complexity is significantly reduced as the number of non-trivial transitions is mitigated by design.
The CNM often requires more clusters and a higher order to achieve similar accuracy, with more complex cluster transition relationships.
\item The physical interpretability of the model is enhanced.
%\item The model is more flexible.
%The strong dependence of the model accuracy on the number of clusters is mitigated, and the model resolution can be arbitrarily changed based on the needs.
\end{enumerate}

Our results suggest entropy as a guiding principle
of future cluster-based models.
We characterize the prediction accuracy of the cluster-based network model
with the Kullback-Leibler entropy of the transition matrix,
called \emph{transition entropy} for brevity.
This transition entropy is significantly reduced
for the dCNM as compared to the CNM
for the same number of centroids.
Further improvements may be expected
by optimizing the $\beta$ parameter.
Thus, the dCNM development from snapshots can be fully automated.
The results even inspire a new clustering
based on the prediction uncertainty expressed
with the transition entropy.
An intrusive framework with Navier-Stokes propagator
may be a further avenue for improvement.

The dCNM may be compared with the POD-based Galerkin method.
By construction, centroids are physically interpretable 
as coherent structures.
In contrast, POD models have no intrinsic meaning
and typically mix different frequencies.
However, in select cases, 
the Galerkin method may yield deep insights into 
linear and nonlinear dynamics.
Examples are the Galerkin mean-field models
for the effect of forcing on a vortex shedding \citep{Semaan2016jfm},
for a single  oscillator \citep{noack2003hierarchy} and 
for frequency cross-talk  \citep{luchtenburg2009generalized}.
The authors work on combining the advantages of clustering and POD 
in human-interpretable dynamic models.

%%%%%%%%%%%%%%%%%%%%%%%%%%%%%%%%%%%%%%%%%%%%%%%%%%%%%%%%%%%%%%%%%%%%%%%%%%%%%%%%%%%%%%%%%%%%%%%%%%%%%%%%%%%%%%%%%%%%%%%%%%%%%%%%%%%%%%%%%%%%
% \backsection[Supplementary data]{\label{SupMat}Supplementary material and movies are available at \\https://doi.org/10.1017/jfm.2019...}

% \backsection[Acknowledgements]{Acknowledgements may be included at the end of the paper, before the References section or any appendices. Several anonymous individuals are thanked for contributions to these instructions.}
{\bf Acknowledgements}. 
The authors appreciate the valuable discussions with Steven Brunton,  Antonio Colanera, Guy Yoslan Cornejo Maceda,  Stefano Discetti, Andrea Ianiro, Fran\c{c}ois Lusseyran, Luc R. Pastur and Xin Wang.

% \backsection[Funding]{Please provide details of the sources of financial support for all authors, including grant numbers. Where no specific funding has been provided for research, please provide the following statement: "This research received no specific grant from any funding agency, commercial or not-for-profit sectors." }
{\bf Funding}. 
This work is supported by the National Natural Science Foundation of China under grants 12172109,  12172111,  and 12202121, 
by the China Postdoctoral Science Foundation under grants 2023M730866 and 2023T160166,
by the Guangdong Basic and Applied Basic Research Foundation under grant 2022A1515011492, 
and by the Shenzhen Science and Technology Program under grant JCYJ20220531095605012.

% \backsection[Declaration of interests]{A Competing Interests statement is now mandatory in the manuscript PDF. Please note that if there are no conflicts of interest, the declaration in your PDF should read as follows: {\bf Declaration of Interests}. The authors report no conflict of interest.}
{\bf Declaration of Interests}. The authors report no conflict of interest.

% \backsection[Data availability statement]{The data that support the findings of this study are openly available in [repository name] at http://doi.org/[doi], reference number [reference number]. See JFM's \href{https://www.cambridge.org/core/journals/journal-of-fluid-mechanics/information/journal-policies/research-transparency}{research transparency policy} for more information}

% \backsection[Author ORCIDs]{Authors may include the ORCID identifers as follows.  F. Smith, https://orcid.org/0000-0001-2345-6789; B. Jones, https://orcid.org/0000-0009-8765-4321}
{\bf Author ORCIDs}. 
C. Hou, https://orcid.org/0000-0001-7477-4242;\\
N. Deng, https://orcid.org/0000-0001-6847-2352;\\
B.~R. Noack, https://orcid.org/0000-0001-5935-1962

% \backsection[Author contributions]{Authors may include details of the contributions made by each author to the manuscript'}
{\bf Author contributions}. 
C. Hou: Methodology, Data Curation, Validation, Writing-Original draft preparation.\\
N. Deng: Supervision, Methodology, Validation, Writing-Reviewing and Editing, Funding acquisition.\\
B.~R. Noack: Methodology, Conceptualisation, Supervision, Funding acquisition, Writing-Reviewing and Editing.

%%%%%%%%%%%%%%%%%%%%%%%%%%%%%%%%%%%%%%%%%%
\appendix
%%%%%%%%%%%%%%%%%%%%%%%%%%%%%%%%%%%%%%%%%%
\section{Convergence and validation studies on the simulation of the sphere wake}
\label{app:a}

To determine an optimal grid size for the numerical analysis, grid convergence studies were conducted at a $\Rey = 300$.
For a set of grids with different numbers of grid cells, the values of the typical flow characteristics are compared to obtain grid-independent results, including the time-averaged drag coefficient $\overline{C_D}$ and its standard deviation $C_{D}^{\prime}$, the time-averaged lift coefficient $\overline{C_L}$ and its standard deviation $C_{L}^{\prime}$, and the Strouhal number $St$. 

The grid refinement is specifically applied to the surface of the sphere and the wake region. 
Across all grid configurations, the boundary layer thickness is adjusted to ensure that the $y^+$ value on the sphere's surface remains below $1$.
This adjustment implies that the first layer of the near-wall grid has a thickness of $0.01D$ \citep{pan2018wake} with a spacing ratio of $1.1$.

The related flow characteristics of the simulations using different grids are listed in Table~\ref{tab:grid}.
\begin{table}
\centering
\def~{\hphantom{0}}
  \begin{tabular}{lcccccccc}
        Cases   &   $n_s$   &   $n_r$   &   Grid cells   &   $\overline{C_{D}}$   &   $C_{D}^{\prime}$   &   $\overline{C_{L}}$   &   $C_{L}^{\prime}$   &   $St$\\[4pt]
        Grid (a)   &   32   &   61   &   1.93 million   &   0.6637   &   0.00183   &   0.0674   &   0.00960   &   0.1363\\        
        Grid (b)   &   49   &   61   &   4.61 million   &   0.6615   &   0.00194   &   0.0666   &   0.01062   &   0.1363\\
        Grid (c)   &   49   &   70   &   5.07 million   &   0.6624   &   0.00185   &   0.0674   &   0.01031   &   0.1363\\
        Grid (d)   &   49   &   100  &   6.58 million   &   0.6623   &   0.00175   &   0.0664   &   0.01051   &   0.1363\\   
        Grid (e)   &   64   &   61   &   8.12 million   &   0.6607   &   0.00193   &   0.0661   &   0.01084   &   0.1363
  \end{tabular}
  \caption{Grid independence test at $\Rey = 300$.}
  \label{tab:grid}
\end{table}
Here, $n_s$ refers to the number of nodes along the circumference of the sphere within one of the `O'-blocks.
This parameter is interconnected with the grid elements along the streamwise direction and the circumference of the cylinder.
On the other hand, $n_r$ signifies the number of elements along the radial direction originating from the surface of the sphere.
Consequently, $n_s$ governs the resolution of the wake region, whereas $n_r$ dictates the resolution of the sphere surface region.
Comparing grids $(a)$, $(b)$, and $(e)$ reveals a relatively smaller difference between grids $(b)$ and $(e)$, especially in terms of standard deviations.
As a result, we select $n_s = 49$ for further analysis concerning the sphere surface region.
Examining grids $(b)$, $(c)$, and $(d)$ leads to similar conclusions, given that there is a more significant increase in the number of grid cells from $(c)$ to $(d)$ than from $(b)$ to $(c)$, despite limited variations in the flow characteristics. 
Consequently, based on these comparisons, it can be concluded that grid $(c)$ is suitable for conducting efficient simulations with sufficient accuracy in this study.

To validate the numerical method, we compare our results with available data from related studies. 
Table~\ref{tab:validition} presents a comparison between the time-averaged drag coefficient $\overline{C_D}$, the time-averaged lift coefficient $\overline{C_L}$, and the Strouhal number $St$ obtained in this study and those reported in other work for $\Rey = 300$.
\begin{table}
\centering
\def~{\hphantom{0}}
  \begin{tabular}{lccc}
               &   $\overline{C_{D}}$   &   $\overline{C_{L}}$   &   $St$\\[4pt]
        Present study                   &   0.662   &   0.067   &   0.136\\        
        \citet{johnson1999flow}         &   0.656   &   0.069   &   0.137\\ 
        \cite{kim2001immersed}          &   0.657   &   0.067   &   0.137\\ 
        \cite{giacobello2009wake}       &   0.658   &   0.067   &   0.134\\    
        \cite{rajamuni2018transverse}   &   0.665   &   0.070   &   0.137\\ 
     
  \end{tabular}
\caption{Validation of the numerical method at $\Rey = 300$, compared to the listed literature.}
\label{tab:validition}
\end{table}
The results obtained from various studies exhibit a high degree of similarity. 
This consistency indicates that our study aligns well with these flow characteristics, as the values are all small and sensitive.

The convergence and validation studies presented here instil confidence that our computational grid and selected numerical schemes are adequate for the wake simulations and for testing the reduced-order modelling method.

%%%%%%%%%%%%%%%%%%%%%%%%%%%%%%%%%%%%%%%%%%
\section{Optional POD before clustering}
\label{app:b}
The computational burden of clustering algorithms becomes a concern when dealing with high-dimensional flow field data. 
Utilising a \emph{lossless} proper orthogonal decomposition (POD) can effectively compress the dataset. 
Implementing the clustering algorithm on the compressed data rather than the high-dimensional velocity fields can significantly reduce the computational time.

Here we introduce the snapshot POD methodology for the completeness of our work.
The $M$ snapshots of the flow field can be decomposed into spatial POD modes with temporal amplitudes, where the $m$-th snapshot can be expressed as:
\begin{equation}
\boldsymbol{u}^{m}(\boldsymbol{x}) \approx \boldsymbol{u}_{0}(\boldsymbol{x}) + \sum_{i=1}^{M-1} a_{i}^{m} \boldsymbol{u}_{i}(\boldsymbol{x}),
\label{eqa1}
\end{equation}
where $\boldsymbol{u}_{0}$ is the mean flow, $a_{i}$ is the mode amplitude and $\boldsymbol{u}_{i}$ is the related mode.
For the three-dimensional sphere flow in this work, we maintain the leading $500$ POD modes for a \emph{loseless} POD, which can resolve more than 99.9\% of the fluctuation energy from all the flow regimes.

The distance between the snapshots translates into the distance between the corresponding mode amplitudes as follows:
\begin{equation}
D(\boldsymbol{u}^{m}, \boldsymbol{u}^{n}) = D(\boldsymbol{a}^{m}, \boldsymbol{a}^{n}).
\label{eqa2}
\end{equation}
With this transformation, the reduction in the computational time can be one or two orders of magnitude, and the statistical description has been formulated in \citet{fernex2021cluster} and \citet{li2021cluster}.

%%%%%%%%%%%%%%%%%%%%%%%%%%%%%%%%%%%%%%%%%%
\section{Modelling with different values of $\beta$}
\label{app:c}

In the dCNM framework, a sparsification controller $\beta \in [0, 1]$ is set to determine the number of sub-clusters in each cluster.
As in Eq.~\eqref{eq18}, the number of sub-clusters is decided by the transverse cluster size and the number of trajectory segments in each cluster.
With a given $\beta$, the number of sub-clusters in each cluster can be decided, then the second-stage clustering algorithm will automatically search the centroids.
Decreasing $\beta$ will lead to more sub-clusters, which means higher model complexity and also higher accuracy in the dynamic reconstruction.
The optimal choice of $\beta$ can be determined by searching for a sweet point which balances the model complexity and the model accuracy.
% The number of sub-clusters can be different for different data sets, while the sparsification controller $\beta$ can be a generalised standard for all datasets.
% The optimal choice of $\beta$ can be determined from the representation error, where a sweet point of $\beta$ can be found considering the model complexity and the model accuracy.

We demonstrate the impact of $\beta$ on the modelling of the post-transient sphere wake. 
Figure~\ref{APP_Re330_clustering} displays the clustering results for the quasi-periodic flow, where $\beta$ takes values of $1$, $0.95$, $0.80$, and $0$. 
Figure~\ref{APP_Re450_clustering} illustrates the results for the chaotic flow, with $\beta$ values of $1$, $0.95$, $0.40$, and $0$.
\begin{figure}
    \centerline{\includegraphics[width=12cm]{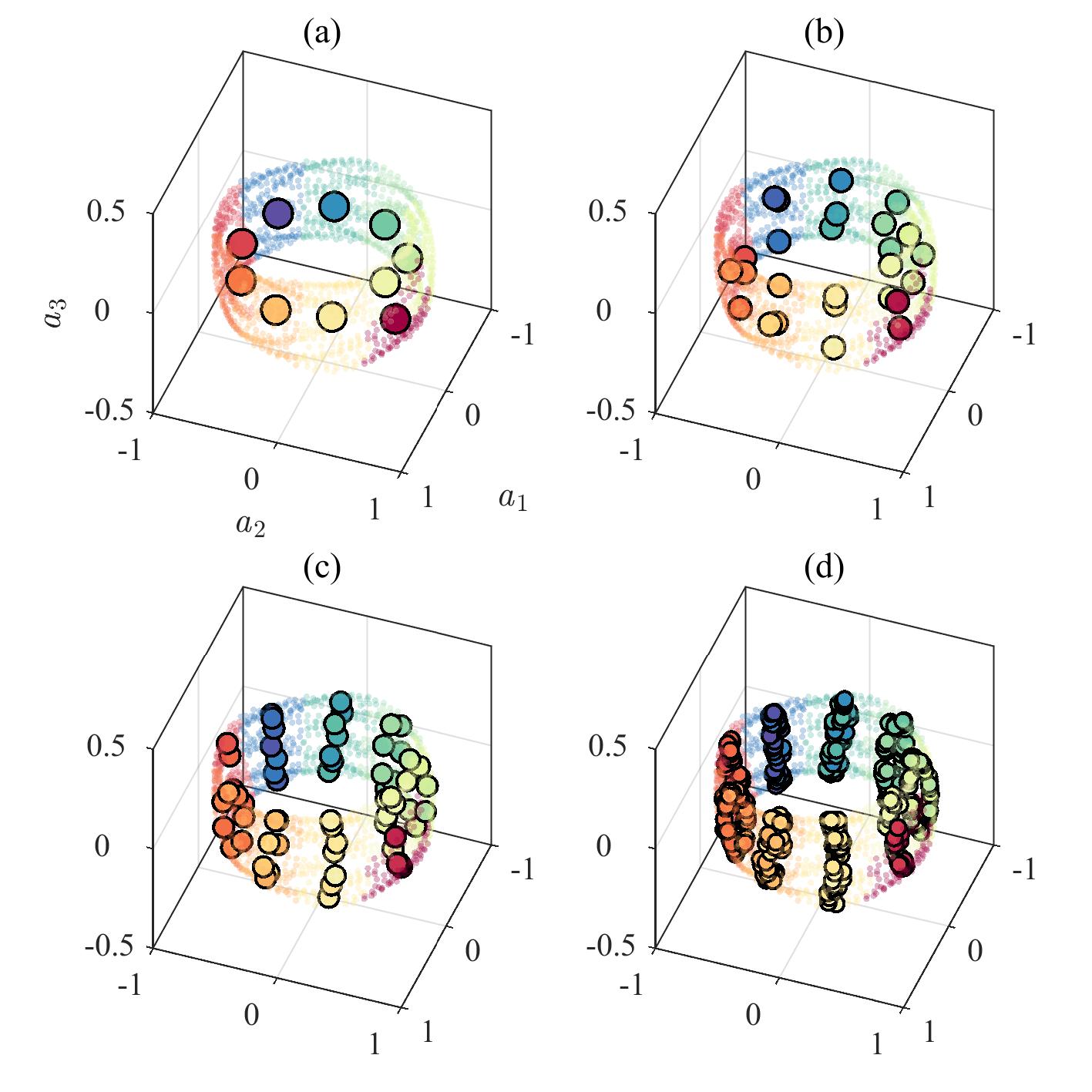}}
    \caption{Clustering results with different $\beta$ on the quasi-periodic flow.
    (a) $\beta = 1$.
    (b) $\beta = 0.95$.
    (c) $\beta = 0.80$.
    (d) $\beta = 0$.}
\label{APP_Re330_clustering}
\end{figure}
\begin{figure}
    \centerline{\includegraphics[width=12cm]{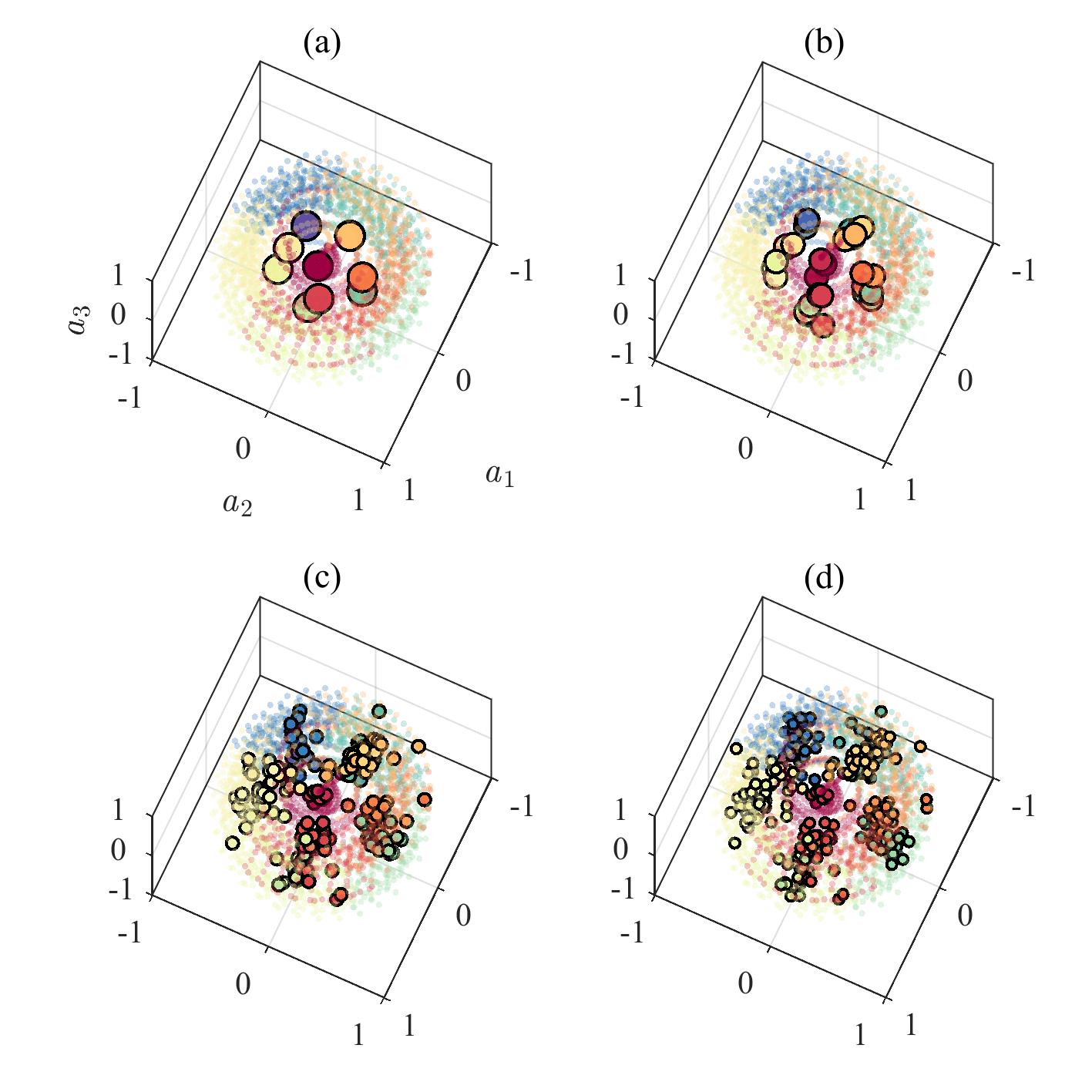}}
    \caption{Clustering results with different $\beta$ on the chaotic flow.
    (a) $\beta = 1$.
    (b) $\beta = 0.95$.
    (c) $\beta = 0.40$.
    (d) $\beta = 0$.}
\label{APP_Re450_clustering}
\end{figure}
$\beta = 1$ means fully sparse, and the centroids are equivalent to the cluster averages, yielding results identical to those of the CNM. 
Conversely, when $\beta = 0$, the model is minimally sparse, resulting in the highest model accuracy, but also the highest model complexity.
As $\beta$ decreases, the centroids try to cover more cyclic behaviours, gradually outlining the entire structure. 
This expansion involves more trajectory segments and, consequently, increases the model resolution. 
For the quasi-periodic flow regime, there are limitations to this enhancement. 
Due to the finite axial length of the cylinder, the trajectory segments often overlap. 
Consequently, increasing the number of sub-clusters to a certain extent results in extensive centroid overlap, offering minimal contributions to the resolution improvement.
This situation is evident when comparing figure~\ref{APP_Re330_clustering} (c) and (d), where the centroid distributions are very similar, and where centroid overlap is prevalent. 
In contrast, centroid overlap is rare in chaotic flows, allowing for noticeable accuracy improvements with smaller $\beta$ values. 
However, using a small $\beta$ will result in lengthy and complex centroid transition information.
Therefore, for chaotic dynamics, it is advisable to strike a balance between the model accuracy and complexity by adjusting $\beta$ based on specific purposes.

From a spatial perspective, we evaluated the representation error using different $\beta$ for the two flow regimes, which is also relevant to determine the appropriate $\beta$, where a sweet point of $\beta$ can be found considering the model complexity and the model accuracy, as shown in figure~\ref{Er}.
%From a spatial perspective, we evaluated the representation error using different $\beta$ for the two flow regimes, which is also %relevant to determine the appropriate $\beta$, as shown in figure~\ref{Er}.
%
\begin{figure}
    \centerline{\includegraphics[width=10cm]{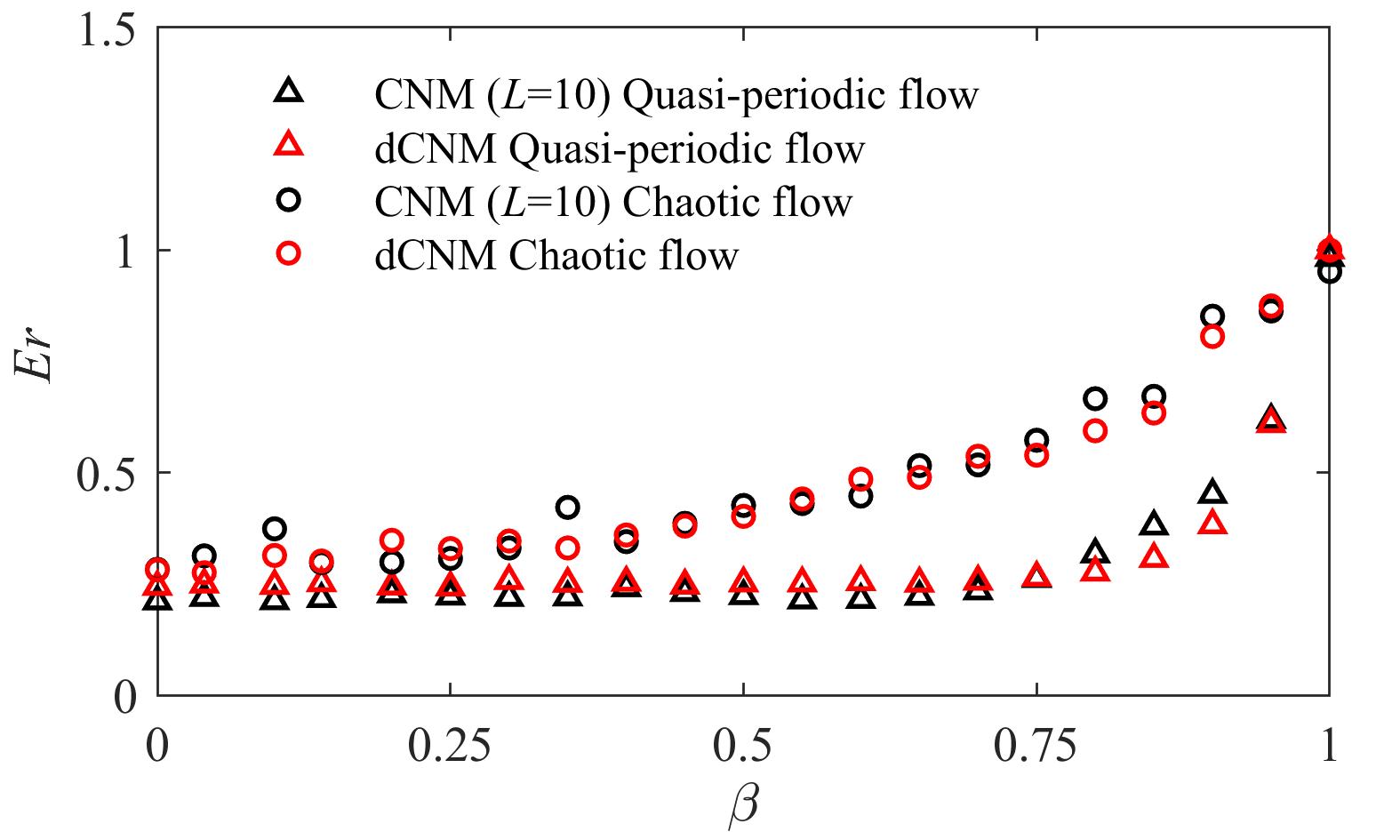}}
    \caption{Representation error versus the sparsification index $\beta$ for the quasi-periodic flow and the chaotic flow.
    The results of dCNM are marked with red, and the corresponding results of high-order CNM with the same number of centroids are marked with black.
    All values have been normalised using the representation error of classical CNM with 10 clusters.
    The marginally lower error of the high-order CNM for the quasi-periodic case is due to the more numerous distribution of the centroids in one limit cycle, which constitutes a smoother cyclic trajectory.
    The dCNM centroids also focus on the variation between loops, thus with fewer centroids in each loop.}
\label{Er}
\end{figure}
The representation error exhibits different trends for the two flow regimes as $\beta$ increases.
In the quasi-periodic flow, the representation error remains relatively constant over a wide range of $\beta$ values and then sharply increases near $\beta=1$. 
This abrupt rise suggests that sparsification eliminates the cycle-to-cycle variations.
%An acceptable reconstruction of this flow regime can be achieved with $\beta < 0.80$.
For the chaotic flow, the representation error changes smoothly from $\beta = 0$ to $\beta = 1$, indicating the loss of diversity of the main loop. 
%For an accurate reconstruction of this flow regime, a relatively smaller value of $\beta$ is needed.
We can expect a Pareto Optimality from the spatial representation error for these two cases, i.e. $\beta = 0.80$ for the quasi-periodic case and $\beta = 0.40$ for the chaotic case.

From a temporal perspective, the assignments of each snapshot to the clusters and centroids of the chaotic flow are illustrated in figure~\ref{APP_Re450_idx}.
\begin{figure}
    \centerline{\includegraphics[width=14cm]{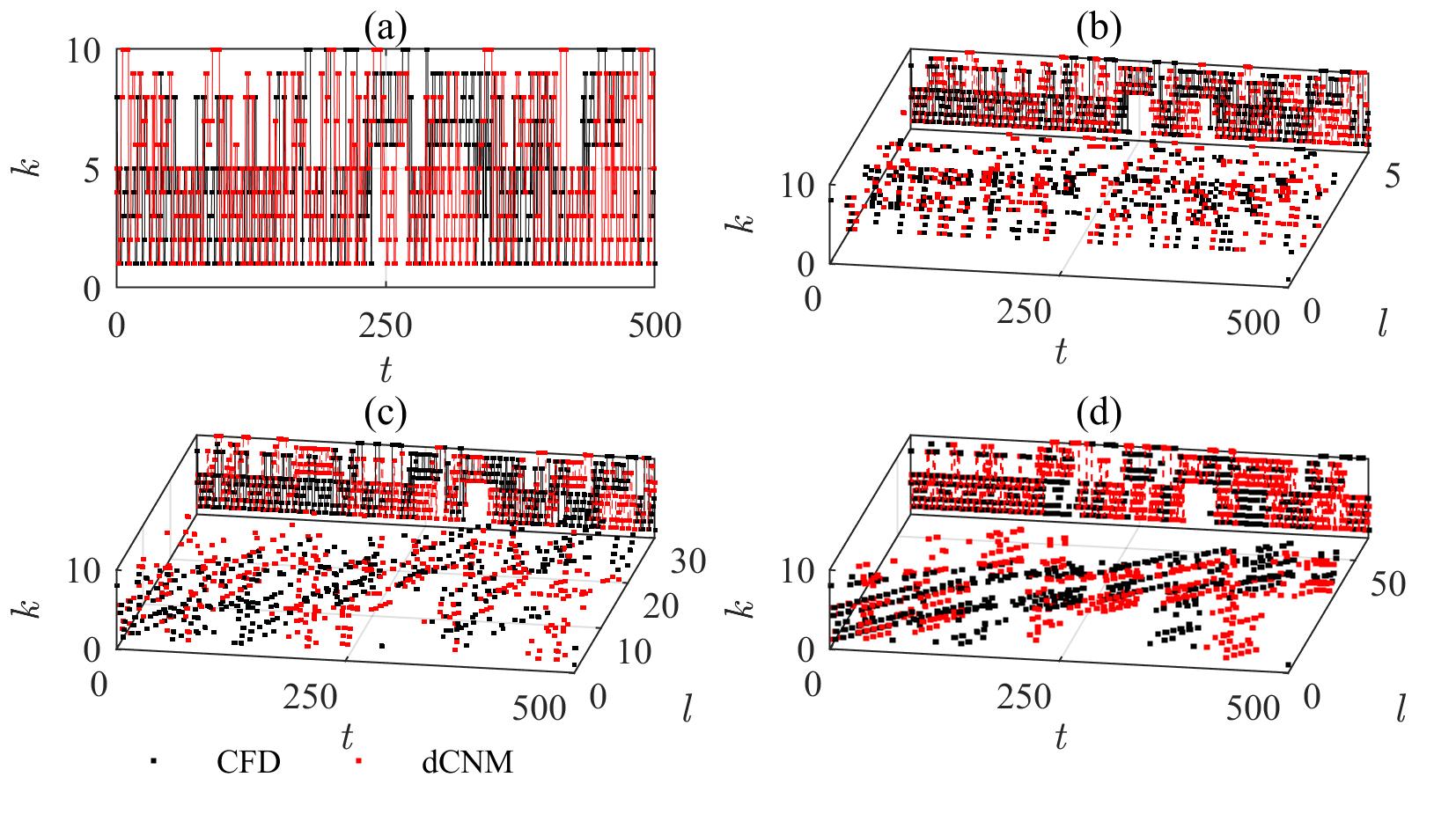}}
    \caption{Temporal evolution of cluster and trajectory segment affiliation with different $\beta$ for the chaotic flow.
    (a) $\beta = 1$,
    (b) $\beta = 0.95$,
    (c) $\beta = 0.40$,
    and (d) $\beta = 0.$}
\label{APP_Re450_idx}
\end{figure}
When $\beta = 1$, the centroid affiliation is disregarded, and only the cluster-level transitions can be observed, this is the same with the CNM.
The temporal evolution of centroids relies solely on the stochastic cluster transition probabilities, with each centroid visited multiple times, as shown in figure~\ref{APP_Re450_idx} (a). 
Conversely, for $\beta = 0$, most centroids are visited only once, leading to the minimum transition error, as seen in figure~\ref{APP_Re450_idx} (d).
From figure~\ref{APP_Re450_idx} (b) and (c), we can conclude that even with sparsification, varied cyclic behaviours can still be effectively captured by the dCNM.
This is because different centroid combinations in the dCNM reconstruction constitute extended cluster chains mentioned in \S~\ref{sec4.3}, and the occurrence of extended cluster chains affirms the capability of the dCNM to effectively resolve the multiscale dynamics.
The generally similar visiting sequences in the extended cluster chains from the dCNM reconstruction and the data set ensure the model accuracy and the difference highlights that the stochastic transition characteristics of chaotic dynamics are also reserved.

%%%%%%%%%%%%%%%%%%%%%%%%%%%%%%%%%%%%%%%%%%
\section{The centroid transition matrix}
\label{app:d}

For the nonzero terms in the cluster transition probability matrix, we can embed a corresponding centroid transition matrix based on the sub-clusters and then record all the dual indexing centroid transitions by the transition tensors $\mathcal{Q}$.
%For the nonzero terms in the cluster transition probability matrix, we can embed a corresponding centroid transition matrix and then record all the dual indexing centroid transitions by the transition tensors $\mathcal{Q}$.

The centroid transition matrices of the quasi-periodic flow, as discussed in \S~\ref{sec4.2}, departing from $\mathcal{C}_4$ are shown in figure~\ref{APP_Re330_matrices}.
The matrices of the chaotic flow, as discussed in \S~\ref{sec4.3}, departing from $\mathcal{C}_1$ are shown in figure~\ref{APP_Re450_matrices}.
\begin{figure}
    \centerline{\includegraphics[width=12cm]{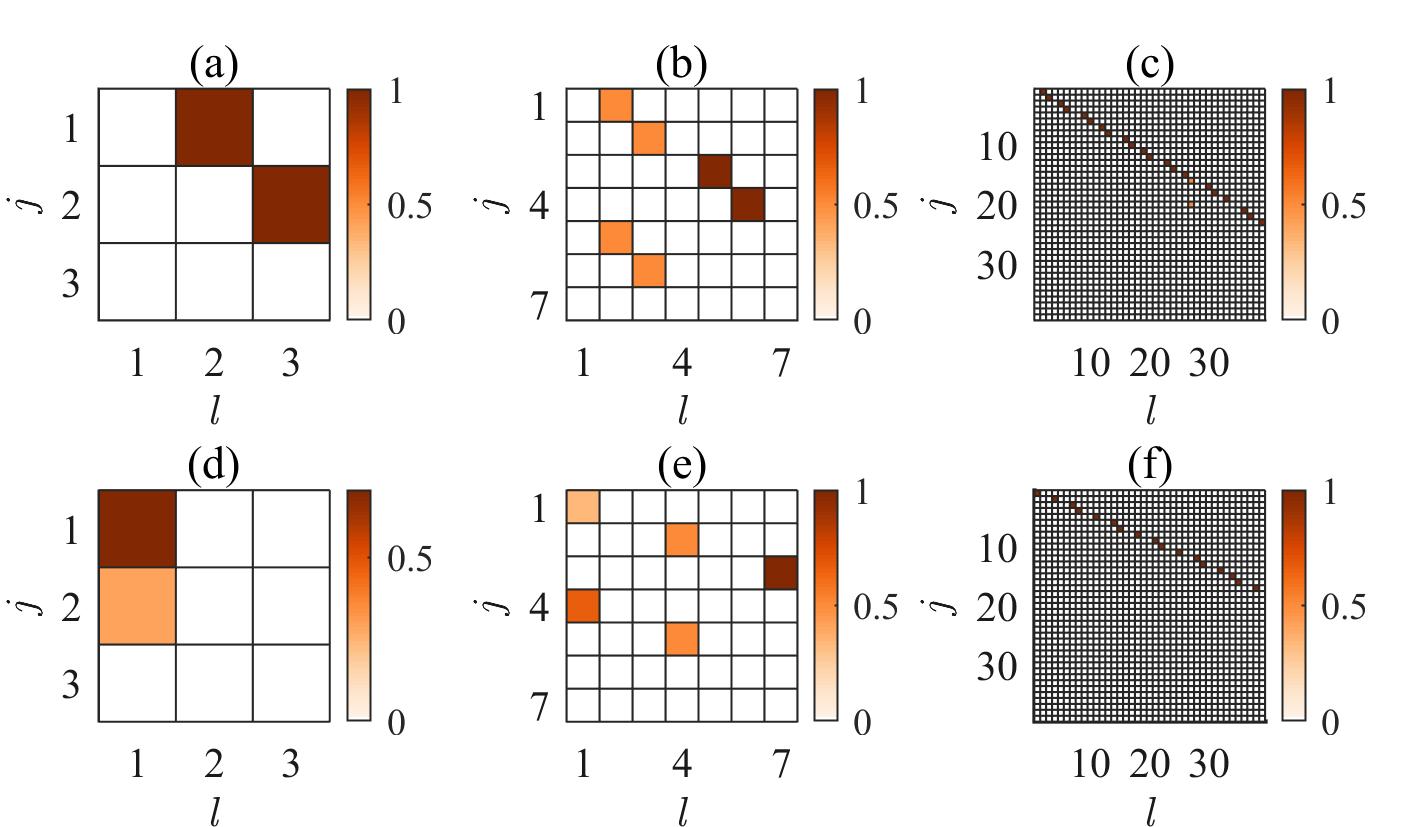}}
    \caption{Centroid transition matrices departing from $\mathcal{C}_4$ with different $\beta$ for the quasi-periodic flow:  
    (a) $\beta = 0.95$, (b) $0.80$, and (c) $0$ for the cluster transition $\mathcal{C}_4 \to \mathcal{C}_5$; 
    and (d) $\beta = 0.95$, (e) $0.80$, and (f) $0$ for $\mathcal{C}_4 \to \mathcal{C}_{10}$.}
\label{APP_Re330_matrices}
\end{figure}
\begin{figure}
    \centerline{\includegraphics[width=12cm]{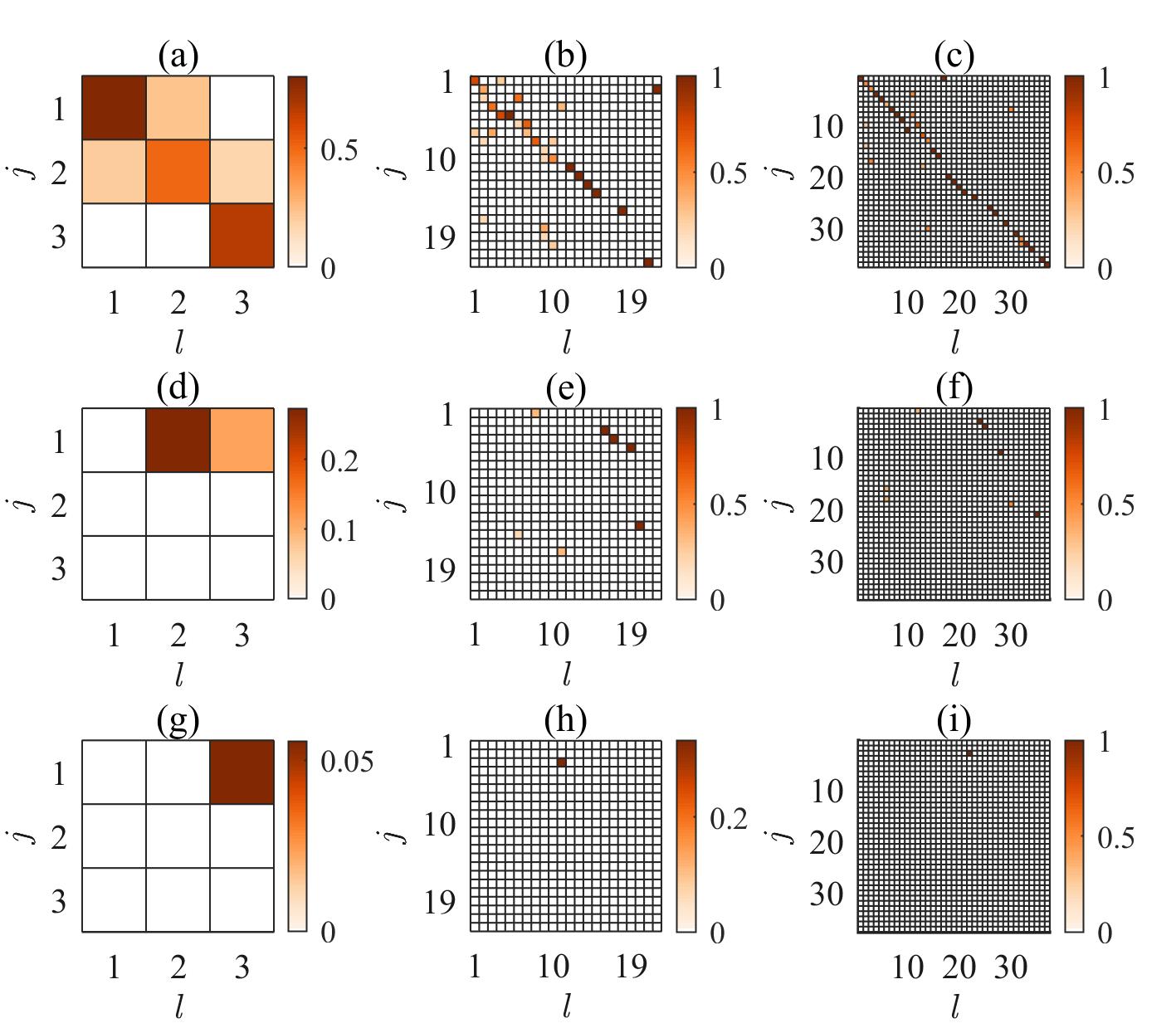}}
    \caption{Centroid transition matrices departing from $\mathcal{C}_1$ with different $\beta$ for the chaotic flow:
    (a) $\beta = 0.95$, (b) $0.40$, and (c) $0$ for the cluster transition $\mathcal{C}_1 \to \mathcal{C}_2$; 
    (d) $\beta = 0.95$, (e) $0.40$, and (f) $0$ for $\mathcal{C}_1 \to \mathcal{C}_7$;
    and (g) $\beta = 0.95$, (h) $0.40$, and (i) $0$ for $\mathcal{C}_1 \to \mathcal{C}_{10}$.}
\label{APP_Re450_matrices}
\end{figure}
For the quasi-periodic flow regime, the matrices are sparse and clear, with centroids having only one destination, indicating deterministic transitions.
Moreover, these transitions impose specific constraints on the quasi-stochastic dynamics. 
Once the departing centroid is determined, all destination centroids belong to the same destination cluster, and the nonzero terms in this column appear only in one matrix.
The stochastic cluster transition can therefore become deterministic.
Compared to the quasi-periodic flow, the transition probabilities in the matrices of the chaotic flow exhibit stochastic centroid transitions.
Some departing centroids have destination centroids within the same cluster, while others do not. 
Consequently, some centroids participate solely in deterministic cluster loops, while others also engage in random jumps between cluster loops.
This distinction separates the cluster transitions from periodic and stochastic routes and serves as a constraint that distinguishes multiscale loops and their associated cycle-to-cycle transitions.

%%%%%%%%%%%%%%%%%%%%%%%%%%%%%%%%%%%%%%%%%%
\section{The POD reconstruction and the dCNM reconstruction}
\label{app:e}
In this section, we compared the flow kinematics reconstructed by POD and the dCNM.
% To evaluate the POD reconstruction and the dCNM reconstruction up to the same standard, 
The POD reconstruction uses the leading POD modes and their mode amplitudes to reconstruct the flow.
The number of POD modes is chosen so that the resolved fluctuation energy is equal to the value of $1-\beta$ from the dCNM, i.e. $90\%$ of the fluctuation energy resolved by the POD reconstruction is comparable to the dCNM reconstruction with $\beta = 0.1$.
For the quasi-periodic flow at $\Rey = 330$, the POD reconstruction with $50\%$ of the fluctuation energy needs the $5$ leading POD modes. 
For the chaotic flow at $\Rey = 450$, the POD reconstruction with $60\%$ of the fluctuation energy takes the $19$ leading modes.
The dCNM outperforms the POD in resolving the key features of the data under the same standard.
For the quasi-periodic flow in figure~\ref{APP_Re330_reconstruction_POD}, the POD results better resolve the tiny variation between trajectories while exhibiting a larger deformation of the overall geometry.
\begin{figure}
    \centerline{\includegraphics[width=12cm]{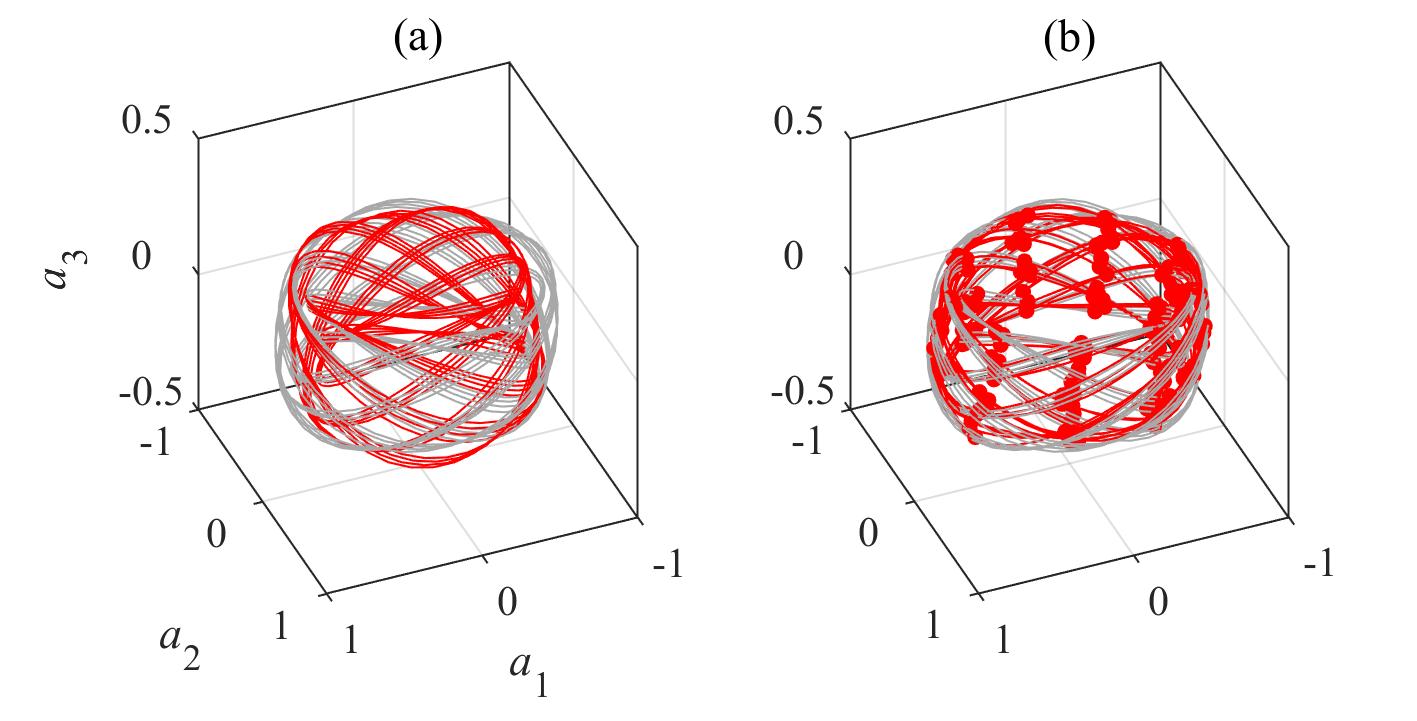}}
        \caption{Comparison between the POD reconstruction and the dCNM reconstruction for the quasi-periodic flow at $\Rey = 330$: (a) The POD reconstruction resolving $50\%$ of the fluctuation energy and (b) the dCNM reconstruction with $\beta = 0.5$.}
\label{APP_Re330_reconstruction_POD}
\end{figure}
The dCNM results cover the whole geometry better, while the tiny variation between trajectories is averaged by the centroids.
This elaborates the dCNM with the advantage of being more robust to noise.
This trend is similarly observed in the chaotic flow in figure~\ref{APP_Re450_reconstruction_POD}, where the dCNM better outlines the geometry and resolves the prominent features.
\begin{figure}
    \centerline{\includegraphics[width=12cm]{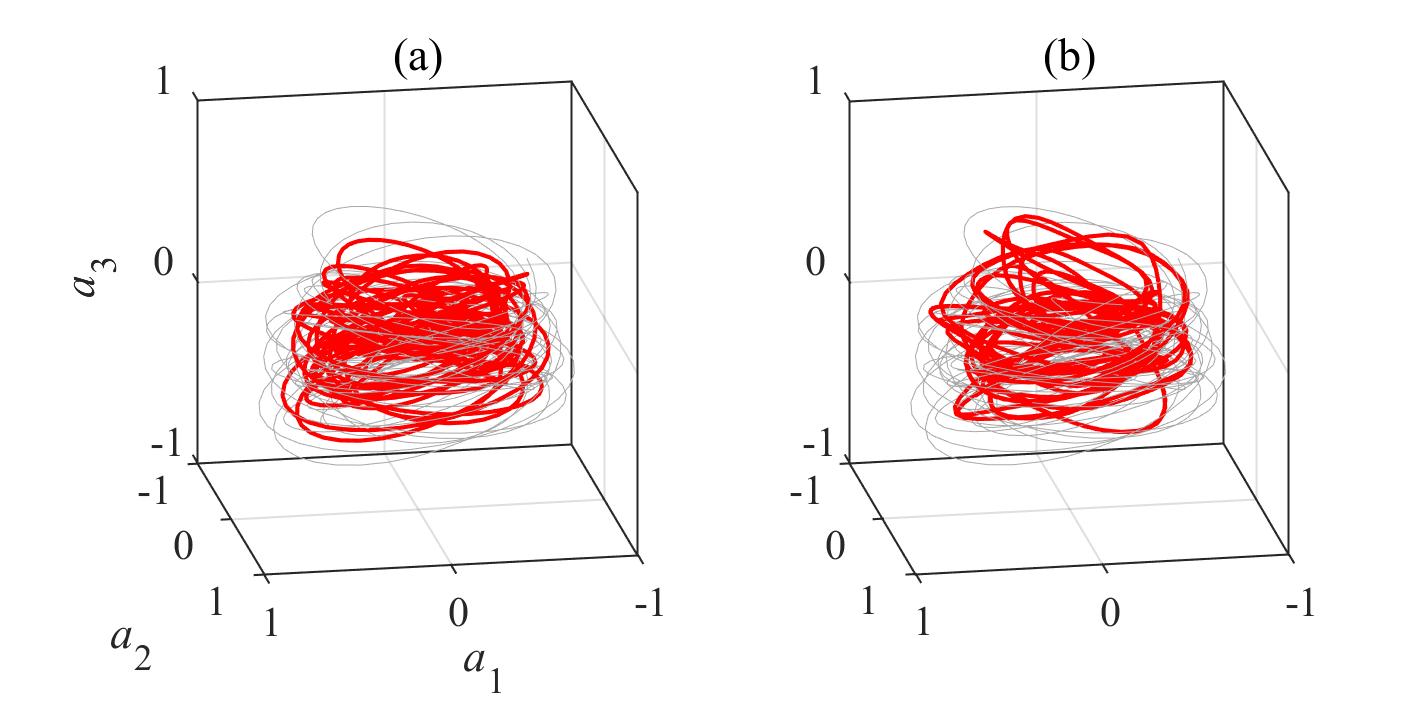}}
        \caption{Same as figure~\ref{APP_Re330_reconstruction_POD}, but for the chaotic flow at $\Rey = 450$.
        (a) The POD reconstruction resolving $60\%$ of the fluctuation energy and (b) the dCNM reconstruction with $\beta = 0.4$.}
\label{APP_Re450_reconstruction_POD}
\end{figure}
%

%%%%%%%%%%%%%%%%%%%%%%%%%%%%%%%%%%%%%%%%%%%%%%%%%%%%%%%%%%%%%%%%%%%%%%%%%%%%%%%%%%%%%%%%%%%%%%%%%%%%%%%%%%%%%%%%%%%%%%%%%%%%%%%%%%%%%%%%%%%%
%\bibliographystyle{jfm}
%\bibliography{jfm}
%Use of the above commands will create a bibliography using the .bib file. Shown below is a bibliography built from individual items.

\bibliographystyle{jfm}
\bibliography{ref}

%% End of file `jfm2esam.bib'.

\end{document}